\documentclass[cernpreprint,USenglish]{na61doc}
\usepackage[utf8]{inputenc}
\usepackage[T1]{fontenc}
\usepackage{amsmath}
\usepackage{url}
\usepackage{multirow}
\usepackage{colortbl}
\usepackage[colorlinks=true,linkcolor=firebrick,citecolor=darkgreen,urlcolor=darkblue]{hyperref}
\usepackage{mathptmx} 
\usepackage{enumerate}
\usepackage{lineno}
\usepackage{xspace}
\usepackage{color}
\usepackage{tikz}
\usetikzlibrary{patterns}
\usepackage{varwidth}
\usepackage{placeins}
\usepackage{bbding}
\usepackage{enumerate}
\usepackage{afterpage}
\usepackage{longtable}
\usepackage{varwidth}
\usepackage{flafter}
\usepackage{cite}

\renewcommand{\NASixtyOne}{\mbox{NA61/SHINE}\xspace}

\newcommand{\pp}{\mbox{\textit{p}+\textit{p}}\xspace}
\newcommand{\aGevNumerical}{\textit{A} GeV/$c$\xspace}
\newcommand{\GeantThree}{{\scshape Geant3}\xspace}
\newcommand{\Epos}{{\scshape Epos}\xspace}
\newcommand{\EposLong}{{\scshape Epos1.99}\xspace}

\newcommand{\coordinate}[1]{{\fontfamily{lmss}\selectfont#1}}

\newcommand{\MeV}{\mbox{Me\kern-0.1em V}\xspace}
\newcommand{\GeV}{\mbox{Ge\kern-0.1em V}\xspace}
\newcommand{\A}{\textit{A}\xspace}
\newcommand{\GeVc}{\mbox{Ge\kern-0.1em V\kern-0.15em /\kern-0.05em\textit{c}}\xspace}
\newcommand{\MeVc}{\mbox{Me\kern-0.1em V\kern-0.15em /\kern-0.05em\textit{c}}\xspace}

\newcommand{\AGeVc}{\A~\GeVc}
\newcommand{\Ar}{\textsuperscript{40}Ar\xspace}
\newcommand{\Sc}{\textsuperscript{45}Sc\xspace}
\newcommand{\ArSc}{\mbox{\Ar~+~\Sc}\xspace}
\newcommand{\dedx}{\mbox{\ensuremath{\textrm{d}E\!/\!\textrm{d}x}}\xspace}
\newcommand{\sNN}{\ensuremath{\sqrt{s_{\mathrm{NN}}}}\xspace}
\newcommand{\AVG}[1]{\ensuremath{\langle #1\rangle}\xspace}
\newcommand{\aIt}{\textit{A}\xspace}

\definecolor{darkred}{rgb}{0.5,0,0}
\definecolor{darkblue}{rgb}{0,0,0.5}
\definecolor{firebrick}{rgb}{0.75,0.125,0.125}
\definecolor{darkgreen}{rgb}{0,0.5,0}
\definecolor{redShading}{RGB}{229,127,127}
\definecolor{colorPSDCentral}{RGB}{66,129,164}
\definecolor{colorPSD150}{RGB}{72,169,166}
\definecolor{colorPSD75}{RGB}{212,180,131}
\definecolor{colorPSD19}{RGB}{193,102,107}

\ShineTitle{
Search for a critical point of strongly-interacting matter in central \ArSc
collisions at 13\A--75\AGeVc beam momentum
}

\PreprintIdNumber{CERN-EP-2023-305}

\ShineJournal{}

\ShineJournalRef{}
\ShineDOI{}

\ShineAbstract{
The critical point of strongly interacting matter is searched for at
the CERN SPS by the \NASixtyOne experiment in central \ArSc collisions at 13\A, 19\A, 30\A, 40\A, and 75\AGeVc.
The dependence of the second-order scaled factorial moments of proton multiplicity distributions on the number of subdivisions in transverse momentum space is measured.
The intermittency analysis uses statistically independent data sets for every subdivision in transverse and cumulative-transverse momentum variables.

The results obtained do not indicate the searched intermittent pattern. An upper limit on the fraction of correlated protons and the intermittency index is obtained based on a comparison with the Power-law Model.
}

\begin{document}

\maketitle

\section{Introduction}

The experimental results are presented on intermittency analysis using second-order scaled factorial moments of mid-rapidity protons produced in central \ArSc collisions at 13\A, 19\A, 30\A, 40\A, and 75\AGeVc beam momentum (\sNN \footnote{collision energy per nucleon pair in the center-of-mass system} $\approx$ 5.1--11.9 \GeV). The measurements were performed by the multi-purpose \NASixtyOne~\cite{Abgrall:2014xwa} apparatus being operated at the CERN Super Proton Synchrotron (SPS).
They are part of the strong interactions program of \NASixtyOne devoted to studying the properties of strongly interacting matter, such as the onset of deconfinement and critical endpoint (CP). Within this program, a two-dimensional scan of collision energy and colliding nuclei size was conducted~\cite{Aduszkiewicz:2642286}.

In the QCD phase diagram, CP is a hypothetical endpoint of the first-order phase transition line with properties of second-order phase transition. In the proximity of CP, self-similar dynamics~\cite{Antoniou:2006zb} is expected to lead to the corresponding fluctuations of the chiral order parameter, 
belonging to the 3D-Ising universality class \cite{Antoniou:2000nh} and can be detected in transverse momentum space
within the framework of intermittency analysis of proton density fluctuations using scaled factorial moments. 

 The idea of "intermittency", which was first introduced to the study of turbulent flow~\cite{Mandelbrot:1974xfn}, later became important in the physics of particle production, especially as a way to study fluctuations. In the pioneering article of Bialas and Peschanski~\cite{Bialas:1985jb} introducing intermittency analysis to high-energy physics, it was proposed to study the scaled factorial moments of the multiplicity of particles produced in high-energy collisions as a function of the resolution size of rapidity interval.


This paper follows the proton intermittency analysis of~ \ArSc collisions at 150\AGeVc~\cite{NA61SHINE:2023gez} by the \NASixtyOne Collaboration. This analysis was performed in intervals of transverse momentum and cumulative transverse momentum distributions. Using the approach of intermittency analysis introduced in Ref.~\cite{NA61SHINE:2023gez}, statistically independent data sets were used to obtain results for the different numbers of intervals.
 
The paper is organized as follows. Section~\ref{sec:sfm} introduces quantities exploited for the CP search using the intermittency analysis.
In Sec.~\ref{sec:detector}, the characteristics
of the \NASixtyOne detector, relevant for the current study, are briefly presented.
The details of data selection and the analysis procedure are presented in Sec.~\ref{sec:analysis}.
Results obtained are shown in Sec.~\ref{sec:results} and compared with models in Sec.~\ref{sec:models}.
A summary in Sec.~\ref{sec:summary} closes the paper.

Throughout this paper, the rapidity, $y = \text{atanh}\left(\beta_{L}\right)$, is calculated in the collision center-of-mass frame by shifting rapidity in laboratory frame by rapidity of the center-of-mass, assuming proton mass.
The $\beta_L = p_L/E$ ($c\equiv1$) is the longitudinal (\coordinate{z}) component of the velocity, while $p_L$ and $E$ are particle longitudinal momentum and energy in the collision center-of-mass system. The transverse component of the momentum is denoted as $p_T = \sqrt{p_x^2 + p_y^2}$, where $p_x$ and $p_y$ are its horizontal and vertical components. The azimuthal angle is the angle between the transverse momentum vector and the horizontal (\coordinate{x}) axis. Total momentum in the laboratory system is denoted as $p_{\textit{lab}}$.

The \ArSc collisions are selected by requiring a low energy value measured by the forward calorimeter, the Projectile Spectator Detector (PSD). This is the energy emitted into the region populated mostly by projectile spectators. These collisions are called PSD-central collisions, and a selection of collisions based on the PSD energy is called a PSD-centrality selection.
\section{Scaled factorial moments}
\label{sec:sfm}

\subsection{Critical point and intermittency in heavy-ion collisions}
\label{sec:sfm_CP}

A second-order phase transition leads to the divergence of the correlation length ($\xi$).
The infinite system becomes scale-invariant with the particle density-density correlation function exhibiting power-law scaling, which induces intermittent behavior of particle multiplicity fluctuations~\cite{Wosiek:1988}.


The maximum critical signal is expected when the freeze-out occurs close to the CP. On the other hand, the energy density at the freeze-out is lower than at the early stage of the collision. Clearly, the critical point should be experimentally searched for in nuclear collisions at energies higher than that of the onset of deconfinement -- when quark-gluon plasma creation sets in.

The intermittent multiplicity fluctuations~\cite{Bialas:1985jb}  were
discussed as the signal of CP by Satz~\cite{Satz:1989vj}, Antoniou et al.~\cite{Antoniou:1990vf} and Bialas, Hwa~\cite{Bialas:1990xd}. This initiated experimental studies of the structure of the phase
transition region via analyses of particle multiplicity fluctuations using scaled factorial moments~\cite{NA49:2012ebu}. Later, additional measures of fluctuations were also proposed as probes of the critical behavior ~\cite{Stephanov:1998dy,Stephanov:1999zu}. The \NASixtyOne experiment has performed a systematic scan of collision energy and system size. To date none of the anticipated signals have been observed for the critical behavior~\cite{Andronov:2018ccl,NA61SHINE:2015uhh,Mackowiak-Pawlowska:2019nmk}. The new measurements may answer the question about the nature of the transition region and, in particular, whether or not the critical point of strongly interacting matter exists.

The scaled factorial moments $F_r(M)$~\cite{Bialas:1985jb} of order $r$ are defined as:
\begin{equation}
  F_r(M) = \frac
    {\bigg<{\displaystyle{\frac{1}{M^{D}}\sum_{i=1}^{M^{D}}} n_i\;...\;(n_i-r+1) }\bigg>}
    {\bigg<{\displaystyle{\frac{1}{M^{D}}\sum_{i=1}^{M^{D}}} n_i }\bigg>^r }\:,
  \label{eq:scaled-factorial-moments}
\end{equation}
\begin{figure}[!ht]
  \centering
    \includegraphics[scale=0.75]{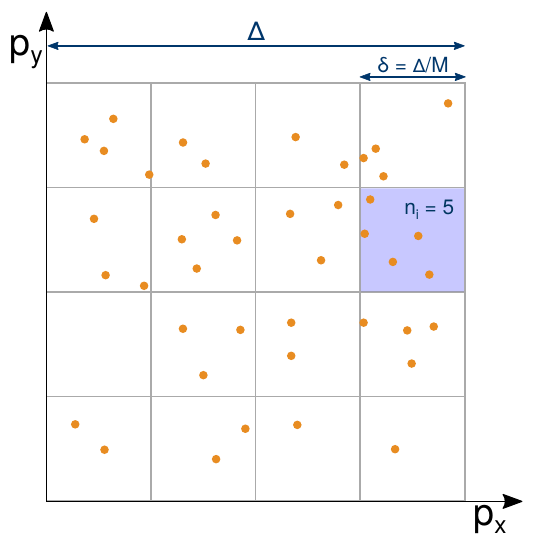}
    \rule{35em}{0.5pt}
  \caption{Two-dimensional transverse momentum space is sub-divided into $M \times M$ number of equally sized bins. The $n_{i}$ is the particle multiplicity in a given sub-interval, $\Delta$ is the momentum region, and $\delta$ is bin width (in this example $n_{i}$ equals 5 in the (4,3) sub-interval).} 
  \label{fig:F2explained}
\end{figure}
where $M$ is the number of subdivision intervals in each of the $D$ dimensions of the selected range $\Delta$ (see Fig.~\ref{fig:F2explained}), $n_{i}$ is the particle multiplicity in a given sub-interval and angle brackets denote
averaging over the analyzed events. In the presented analysis, $\Delta$ is divided into two-dimensional ($D=2$) cells in $p_{x}$ and $p_{y}$.
In case the mean particle multiplicity, $\left<n_{i}\right>$, is proportional to the subdivision interval size and for a Poissonian multiplicity distribution, $F_r(M)$ is equal to 1 for all values of $r$ and $M^{D}$.
This condition is satisfied in the configuration space when the particle density is uniform. The momentum distribution is, in general, non-uniform, and thus in the momentum space, it is more convenient to use the cumulative variables~\cite{Bialas:1990dk} which, for a small cell size, leave a power-law behavior unaffected and at the same time lead to a uniformly
distributed particle density.

If the system freezes out near CP, its properties are expected to be different from those of an ideal gas~\cite{Becattini:2003wp}. Such a system represents a simple fractal and $F_r(M)$ follows a power-law dependence:
\begin{equation}
\label{eq:cp_1}
  F_{r}(M) = F_{r}(\Delta) \cdot (M^{D})^{\varphi_{r}}~.
\end{equation}
Moreover, the exponent (intermittency index) $\varphi_{r}$ obeys the relation:
\begin{equation}
\label{eq:cp_2}
  \varphi_{r} = ( r - 1 ) \cdot (d_{r}/D)\:,
\end{equation}
where $d_{r}$, the anomalous fractal dimension, is independent of $r$~\cite{Bialas:1990xd}.
Such behavior is the analogue of critical opalescence in electromagnetically interacting matter~\cite{Antoniou:2006zb}.
Importantly the critical properties given by Eqs.~\ref{eq:cp_1} and~\ref{eq:cp_2} are approximately preserved for a small cell size (large $M$) under transformation to the cumulative
variables~\cite{Bialas:1990dk,Samanta:2021dxq}.

The ideal CP signal, Eqs.~\ref{eq:cp_1} and~\ref{eq:cp_2}, derived for the infinite system in equilibrium, are generally deteriorated by numerous experimental effects present in high-energy nuclear collisions.
This includes the system's finite size and evolution time, other dynamical correlations between particles, and limited acceptance and finite resolution of measurements. Moreover, to experimentally search for CP in high-energy collisions, the dimension and size of the momentum interval
must be chosen. Note that unbiased results can be obtained only by analyzing variables and dimensions in which the singular behavior appears~\cite{Bialas:1990gu,Ochs:1988ky,Ochs:1990mg}. Any other procedure is likely to distort the critical-fluctuation signal.


Search for the CP via the study of proton fluctuations was suggested in several publications~\cite{Stephanov:1999zu,Hatta:2003wn,Stephanov:2004wx,Fukushima:2010bq, Hatta:2002sj,Antoniou:2008vv,Karsch:2010ck,Skokov:2010uh,Morita:2012kt}. In the case of a pure system exhibiting critical fluctuations, for proton, $d = \varphi_2 = 5/6$ is expected~\cite{Antoniou:2006zb}.

\subsection{Cumulative transformation}
\label{sec:sfm_cumulative}

Scaled factorial moments are sensitive to the shape of the single-particle momentum distribution.
This dependence biases the signal of critical fluctuations.
To remove this dependence, one has two possibilities. The first possibility is to exploit the mixed events, where each event is constructed using particles from different experimental events, thereby removing all possible correlations.
The objective is to measure the following quantity:
\begin{equation}
    \Delta F_{2}(M) = F_{2}^\text{data}(M) - F_{2}^{\text{mixed}}(M)~.
\end{equation}

It was already verified~\cite{NA49:2012ebu} that this procedure, to a large extent, removes the dependence of
$\Delta F_{r}(M)$ on the shape of single-particle distribution, at least for $r = 2$.

The second possibility is to use cumulative transformation~\cite{Bialas:1990dk}, which for a
one-dimensional single-particle distribution $f(x')$ reads:
\begin{equation}
    Q_{x} = \int\limits_{a}^{x} f(x')dx' \Bigg/ \int\limits_{a}^{b} f(x')dx'~,
\end{equation}
where $a$ and $b$ are lower and upper limits of the variable $x'$.
For a two-dimensional distribution $f(x',y')$ and a given $x'=x$ the transformation reads
\begin{equation}
    Q_{y}(x) = \int\limits_{a}^{y} f(x,y')dy' \Bigg/ \int\limits_{a}^{b} f(x,y')dy'.
\end{equation}
After the cumulative transformation, any single-particle distribution becomes flat, ranging from 0 to 1, see example distributions in Ref.~\cite{NA61SHINE:2023gez}, and therefore, it removes the dependence on the shape of the single-particle distribution
for uncorrelated particles~\cite{Bialas:1990dk}. It also distorts all non-scale-invariant correlations.
On the other hand, the transformation is proven to preserve the critical behaviour~\cite{Samanta:2021dxq} given by Eq.~\ref{eq:cp_1}, at least for the second-order scaled factorial moments.

Both methods are approximate. Subtracting moments for mixed data set may introduce negative $\Delta F_{2}(M)$ values~\cite{NA49:2012ebu} and using cumulative quantities mixes the scales of the momentum differences and therefore may distort eventual power-law behavior due to the finite size of the momentum interval, $\Delta$.

\section{The \NASixtyOne detector}
\label{sec:detector}

\begin{figure*}[ht]
  \centering
  \includegraphics[width=\textwidth]{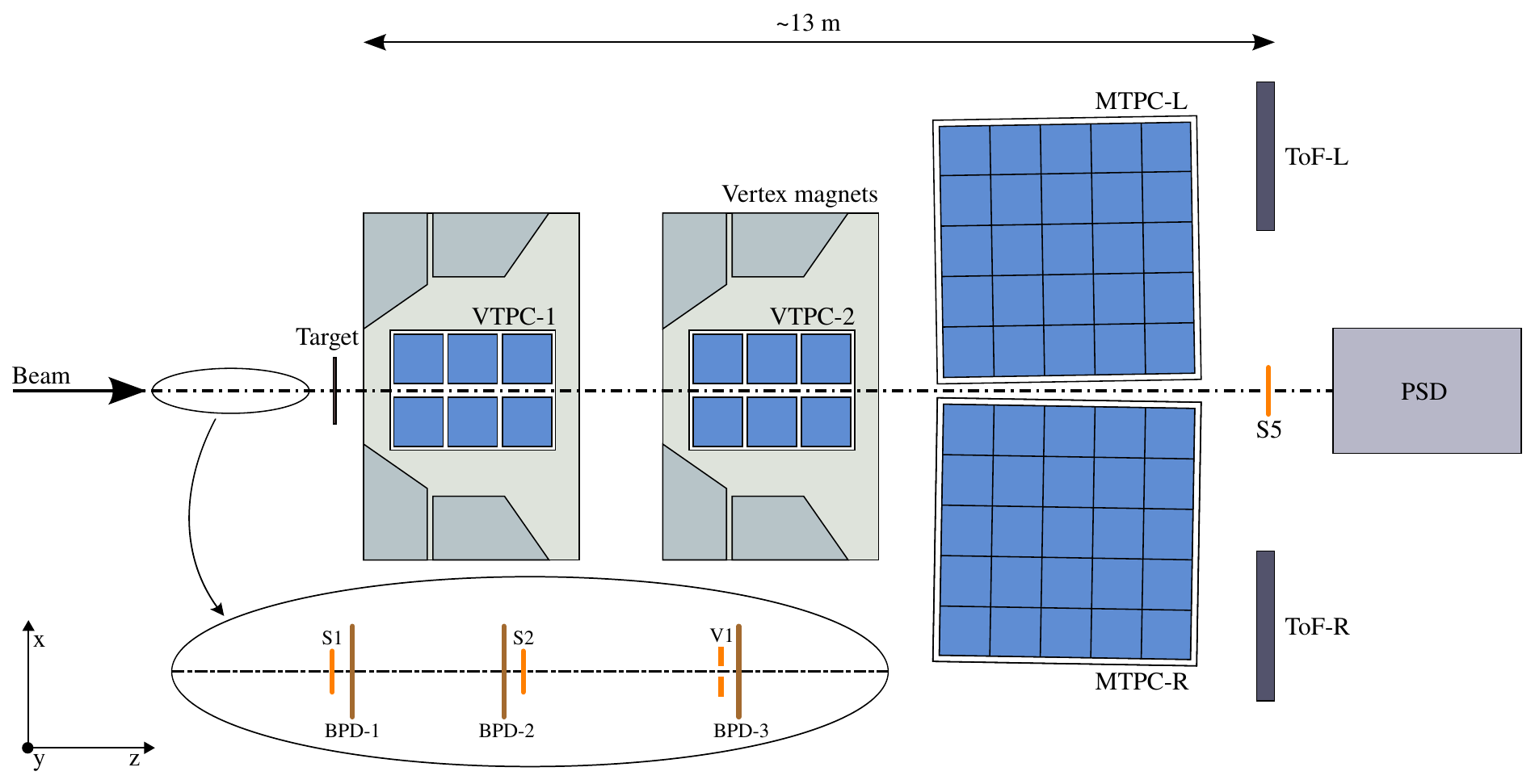}
  \rule{42em}{0.5pt}
  \caption{
    The schematic layout of the \NASixtyOne experiment at the CERN SPS ~\cite{Abgrall:2014xwa}
    showing the components used for the Ar+Sc energy scan (horizontal cut, not to scale).
    The detector configuration upstream of the target is shown in the inset.
    The alignment of the chosen coordinate system is shown on the plot; its origin (\coordinate{x}=\coordinate{y}=\coordinate{z}=0) lies in the middle of VTPC-2,
    on the beam axis. The nominal beam direction is along the \coordinate{z}-axis.
    Target is placed at \mbox{\coordinate{z} = -580.00~cm}.
    The magnetic field bends charged particle trajectories in the \coordinate{x}--\coordinate{z} (horizontal) plane.
    The drift direction in the TPCs is along the (vertical) \coordinate{y}-axis.
  }
  \label{fig:setup}
\end{figure*}

The \NASixtyOne detector (see Fig.~\ref{fig:setup}) is a large-acceptance hadron spectrometer situated in the North Area H2 beam-line of the CERN SPS~\cite{Abgrall:2014xwa}.
The main components of the detection system used in the analysis are four large-volume Time Projection Chambers (TPC).
Two Vertex TPCs (VTPC-1/2) are located downstream of the target inside superconducting magnets with a maximum combined bending power of 9~Tm. The magnetic field was scaled in proportion to the beam momentum in order to obtain similar $y-p_{T}$ acceptance at all beam momenta.
The main TPCs (MTPC-L/R) and two walls of pixel Time-of-Flight (ToF-L/R) detectors are placed symmetrically on either side of the beamline downstream of the magnets. The TPCs were filled with Ar:CO\textsubscript{2} gas mixture in proportions 90:10 for the VTPCs and 95:5 for the MTPCs. The Projectile Spectator Detector (PSD), a zero-degree hadronic calorimeter, is positioned 20.5 m (16.7 m) downstream of the MTPCs at beam momenta of 75\A (13\A, 19\A, 30\A, and 40\A), centered in the transverse plane on the deflected position of the beam.

The PSD consists of 44 modules that cover a transverse area of almost 2.5 m\textsuperscript{2}. The central part of the PSD consists of
16 small modules with transverse dimensions of 10 x 10 cm\textsuperscript{2} and its outer part consists of
28 large 20 x 20 cm\textsuperscript{2} modules. Moreover, a brass cylinder of 10 cm (30\A–75\AGeVc) or \mbox{5 cm (19\AGeVc)} length and 5 cm diameter (degrader) was placed in front of the center of the PSD to reduce electronic saturation effects and shower leakage from the
downstream side caused by the Ar beam and its heavy fragments. No degrader was used at 13\AGeVc.

Primary beams of fully ionized \Ar nuclei were extracted from the SPS accelerator at beam momenta of 13\A, 19\A, 30\A, 40\A, and 75\AGeVc.
Two scintillation counters, S1 and S2, provide beam signal, and a veto counter V1, with a 1~cm diameter hole, defines the beam before the target.
The S1 counter also provides the timing reference (start time for all counters). Beam particles are selected by the trigger system requiring the coincidence T1 = $\textrm{S1} \wedge\textrm{S2} \wedge\overline{\textrm{V1}}$. The three beam position detectors (BPDs) placed upstream of the target~\cite{Abgrall:2014xwa} precisely measure individual beam particle trajectories.  Collimators in the beam line were adjusted to obtain beam rates of $\approx10^4$/s during the 10.4 s spill and a super-cycle \mbox{time of 32.4 s}.

The target was a stack of 2.5 x 2.5~cm\textsuperscript{2} area and 1 mm thick \Sc plates of 6~mm total thickness \mbox{placed $\approx$ 80~cm} upstream of VTPC-1. Impurities due to other isotopes and elements were measured to be 0.3\%~\cite{Banas:2018sak}. No correction was applied for this negligible contamination.

Interactions in the target are selected with the trigger system by requiring an incoming \Ar ion and a signal below that of beam ions from S5, a small 2~cm diameter scintillation counter placed on the beam trajectory behind the MTPCs. This minimum bias trigger is based on the breakup of the beam ion due to interactions in and downstream of the target. In addition, central collisions were selected by requiring an energy signal below a set threshold from the 16 central modules of the PSD, which measure mainly the energy carried by projectile spectators. The cut was set to retain only the events with the $\approx$~30\% smallest energies in the PSD. The event trigger condition thus was
T2 = T1$\wedge\overline{\textrm{S5}}\wedge\overline{\textrm{PSD}}$. 
\section{Analysis}
\label{sec:analysis}

The goal of the analysis was to search for the critical point of the strongly interacting matter by measuring the second-order scaled factorial moments for a selection of protons produced in central \ArSc collisions at 13\A--75\AGeVc, using statistically independent points and cumulative variables.

\subsection{Event selection}
\label{sec:event-selection}

The \NASixtyOne detector recorded events using 13\A--75\AGeVc \Ar beam impinging
on a stationary \Sc target. However, not all of those events contain well-reconstructed central Ar+Sc interactions. Therefore the following criteria were used to select data for further analysis:
\begin{enumerate}[(i)]
   \item T2 trigger set to select central and semi-central collisions,
   \item beam particle detected in at least three planes out of four of BPD-1 and BPD-2 and in both planes of BPD-3,
   \item no off-time beam particle detected within a time window of $\pm$ 4 $\mu$s around the trigger particle,
   \item no interaction-event trigger detected within a time window of $\pm$ 25 $\mu$s around the trigger particle,
   \item a high-precision interaction vertex with \coordinate{z} position (fitted using  the beam trajectory and TPC tracks) no further than 2~cm away from the center of the Sc target.
\end{enumerate}

\subsection{Centrality selection}
\begin{figure}
    \centering
    \includegraphics[width=0.45\textwidth]{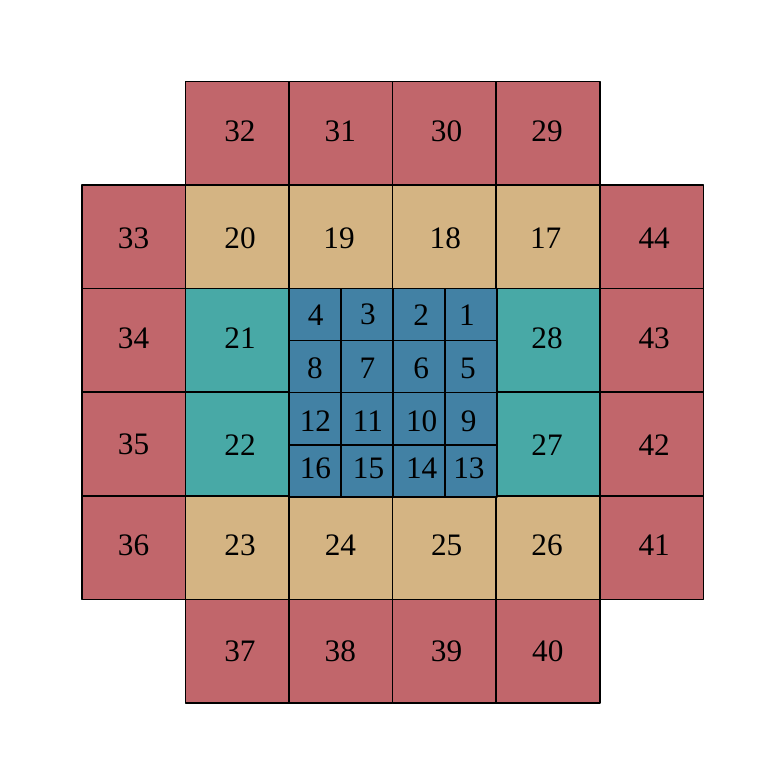}\\
    \begin{varwidth}{\textwidth}
    \begin{itemize}
        \item[\textcolor{colorPSDCentral}{\SquareSolid}] T2 trigger,
        \item[\textcolor{colorPSDCentral}{\SquareSolid} + \textcolor{colorPSD150}{\SquareSolid} + \textcolor{colorPSD75}{\SquareSolid}] 30\aIt, 40\aIt and 75\aIt,
        \item[\textcolor{colorPSDCentral}{\SquareSolid} + \textcolor{colorPSD150}{\SquareSolid} + \textcolor{colorPSD75}{\SquareSolid} + \textcolor{colorPSD19}{\SquareSolid}] 13\aIt and 19\aGevNumerical.
	\end{itemize}
    \end{varwidth}\\
	\vspace{0.5cm}
   \rule{42em}{0.5pt}
    \caption{Event centrality selection using the PSD energy. Modules used in the centrality class determination were chosen based on the anti-correlation between the measured energy and the track multiplicity in a given event. All modules were used for $13$\aIt and $19$\aGevNumerical. For $30$\aIt, $40$\aIt, and $75$\aGevNumerical 28 central modules were chosen. For the T2 trigger, 16 central modules were used.}
    \label{fig:centralityModules}
\end{figure}
The final results presented in this paper refer to the 0--10\% of \ArSc collisions with the lowest energy value measured by a subset of PSD modules (see Fig.~\ref{fig:centralityModules}). The selection of the modules was to optimize the sensitivity to projectile spectators. For more details see Ref.~\cite{NA61SHINE:2020ggt}.
Online event selection by the hardware trigger (T2) used a threshold on the sum of electronic signals from the 16 central modules of the PSD set to accept $\approx$ 30\%, 35\%, 30\%, 35\%, and 20\% of the inelastic \ArSc collisions at 13\A--75\AGeVc. To select 0--10\% central collision events, an upper limit of energy values measured by a subset of modules are 143, 264, 446, 666, and 1290.6 \GeV corresponding to \ArSc collisions \mbox{at 13\A--75\AGeVc.}

The event statistics after applying the selection criteria are summarized in Table~\ref{tab:statbeam}.
\begin{table}[!ht]
	\caption{The statistics of selected events for \ArSc collisions at beam momentum ($p_{beam}$) of 13\A--75\AGeVc.}
	\vspace{0.5cm}
	 \centering
	\begin{tabular}{ c || c | c | c | c | c }
    $p_{beam}$ & \multicolumn{5}{c}{number of events after cuts (10\textsuperscript{6})}\\
    \cline{2-6}
    (\GeVc) & T2 trigger& beam quality  &beam off-time  &vertex z position   & 0--10\% of most central\\
   \hline
   \hline
   13\A & 2.14 & 1.60 & 1.56 &1.48 & 0.50 \\
   19\A & 2.51 & 2.00 & 1.93 &1.83 & 0.52 \\
   30\A & 3.71 & 2.93 & 2.85 &2.74 & 0.91 \\
   40\A & 5.71 & 4.87 & 4.74 &4.53 & 1.29 \\
   75\A & 2.89 & 2.44 & 2.37 &2.32 & 1.16 \\
  	\end{tabular}
	\label{tab:statbeam}
\end{table}

\subsection{Single-track selection}
\label{sec:track_selection}

To select tracks of primary charged hadrons and to reduce the contamination by particles from secondary interactions, weak decays, and off-time interactions, the following track selection criteria were applied:
\begin{enumerate}[(i)]
    \item track momentum fit including the interaction vertex is required to have converged,
    \item total number of reconstructed points on the track is required to be greater than 30,
    \item the sum of the number of reconstructed points in VTPC-1 and VTPC-2 is required to be greater than 15,
    \item the ratio of the number of reconstructed points to the number of potential (maximum possible) points is required to be greater than 0.5 and less than 1.1,
    \item number of points used to calculate energy loss (\dedx) is required to be greater than 30,
    \item the distance between the track extrapolated to the interaction plane and the vertex (track impact parameter)
    is required to be smaller than 4~cm in the horizontal (bending) plane and 2~cm in the vertical (drift) plane.
\end{enumerate}

As the analysis concerns mid-rapidity protons, only particles with center-of-mass rapidity (assuming proton mass) between 0 and 0.75 were considered. Only particles with transverse momentum components, $p_{x}$ and $p_{y}$, values lower than 1.5 \GeVc were accepted for the analysis.

\subsubsection{Proton selection}
\label{sec:protonselection}
\begin{figure}[!ht]
    \centering
    \includegraphics[width=.33\textwidth]{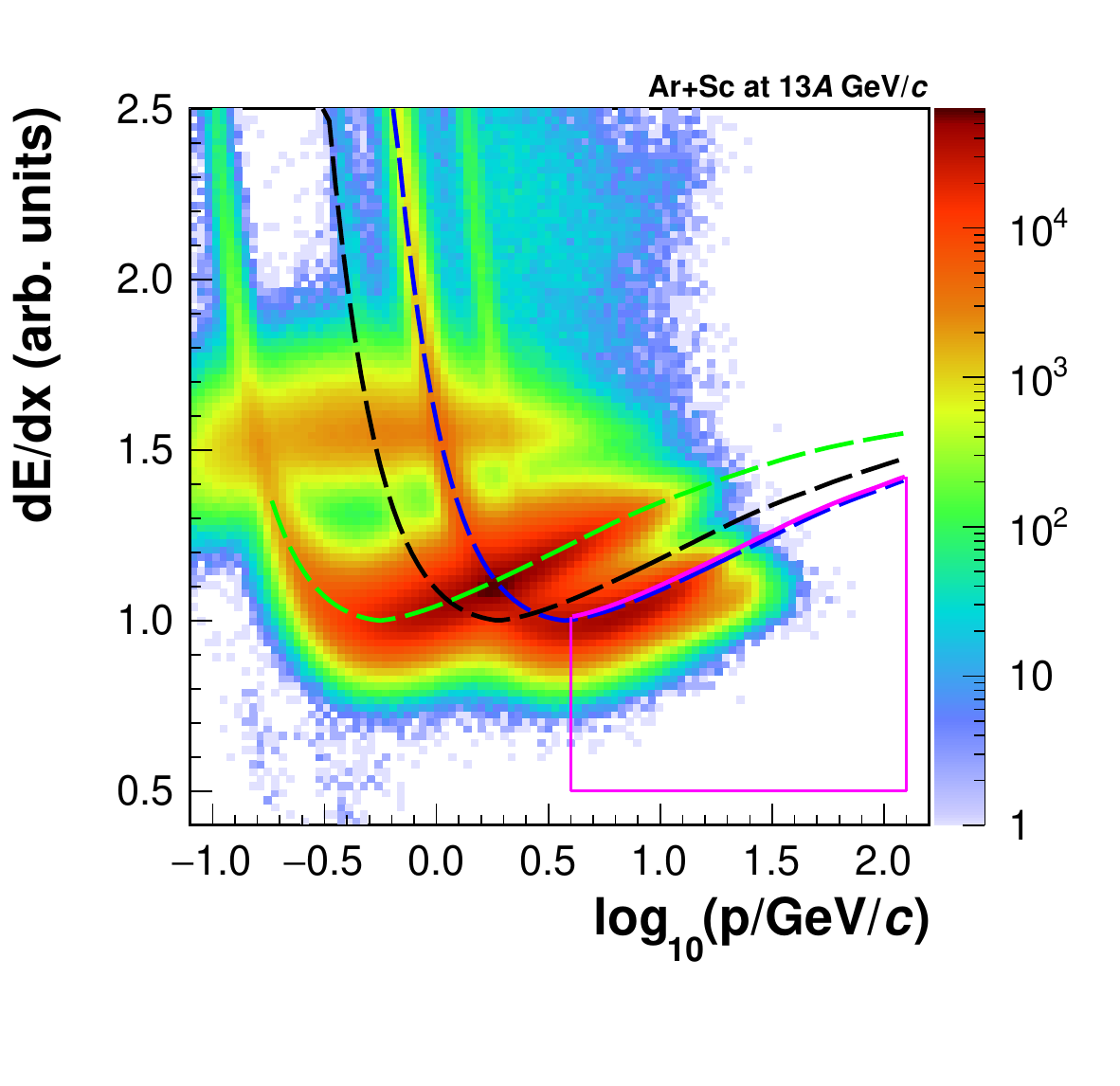}\hfill
    \includegraphics[width=.33\textwidth]{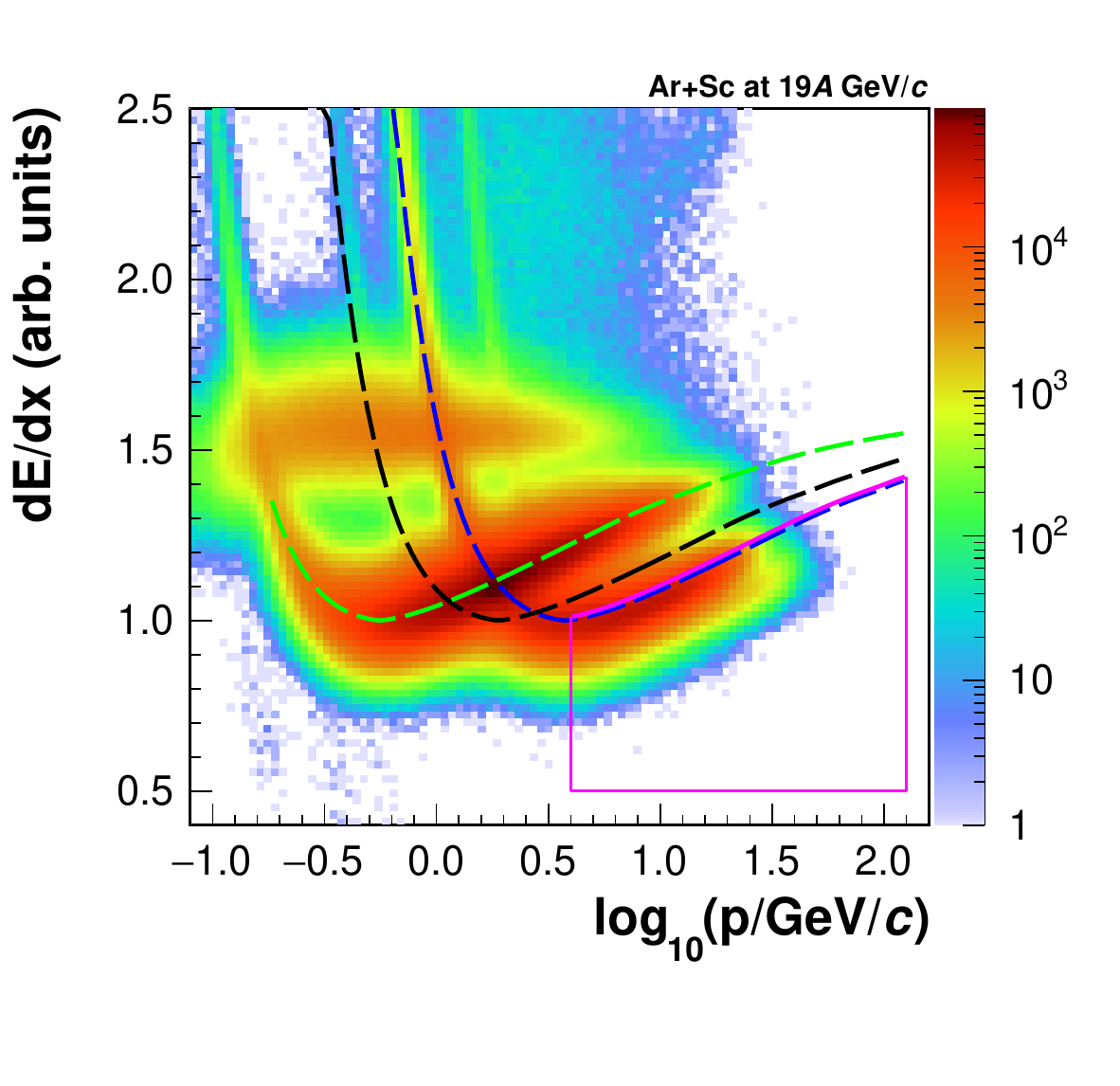}\hfill
    \includegraphics[width=.33\textwidth]{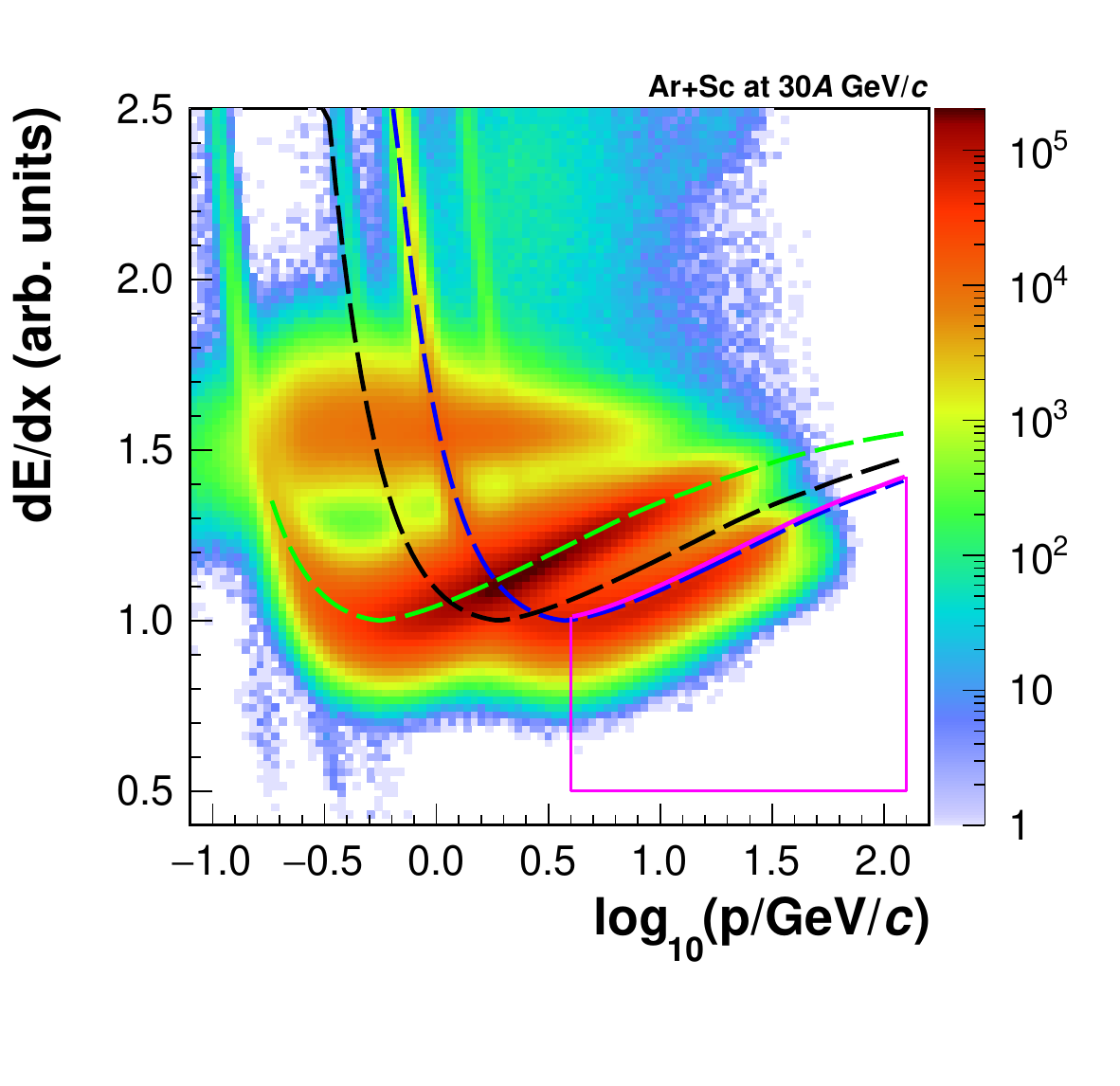}\\
    \includegraphics[width=.33\textwidth]{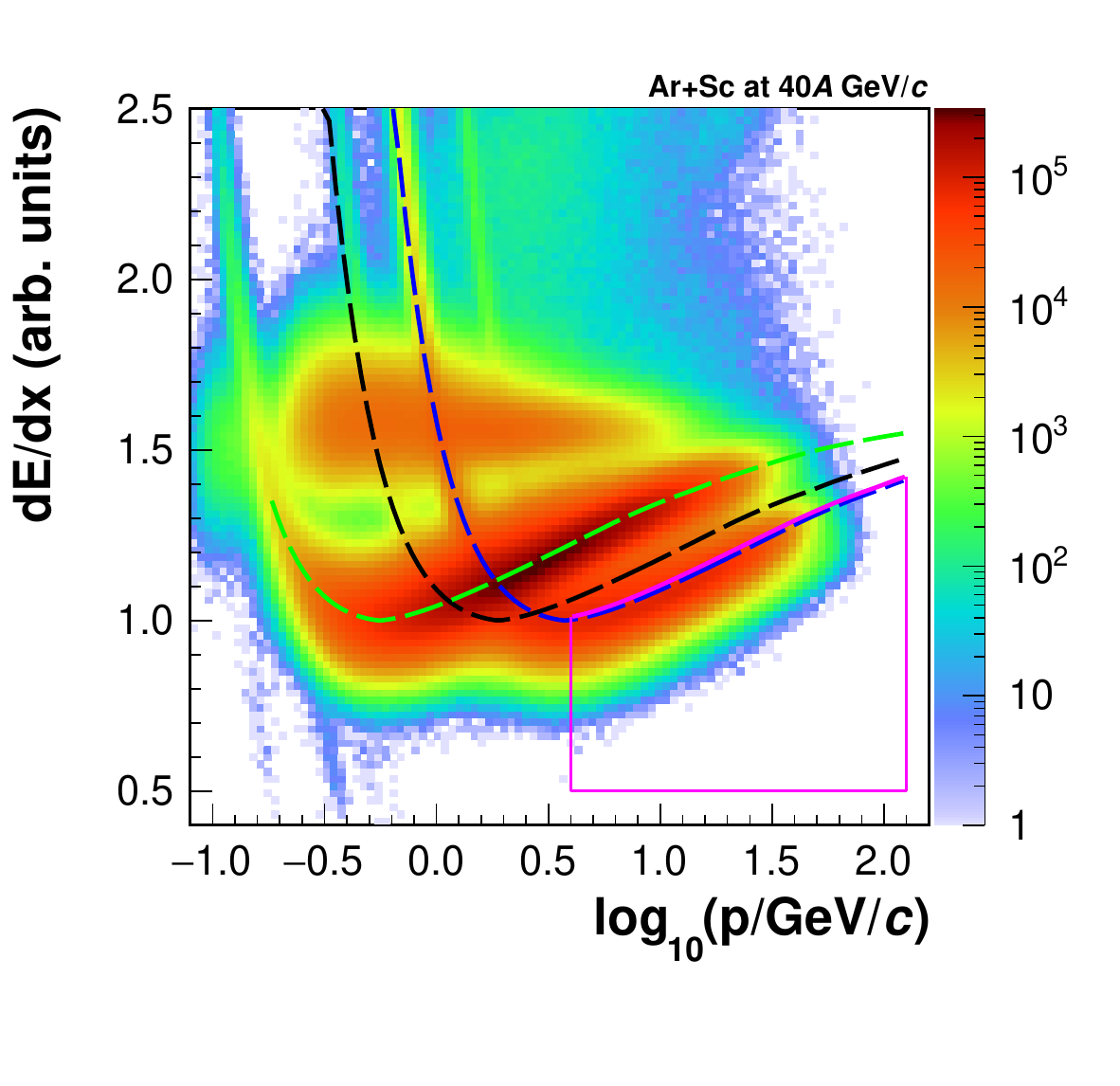}\qquad
    \includegraphics[width=.33\textwidth]{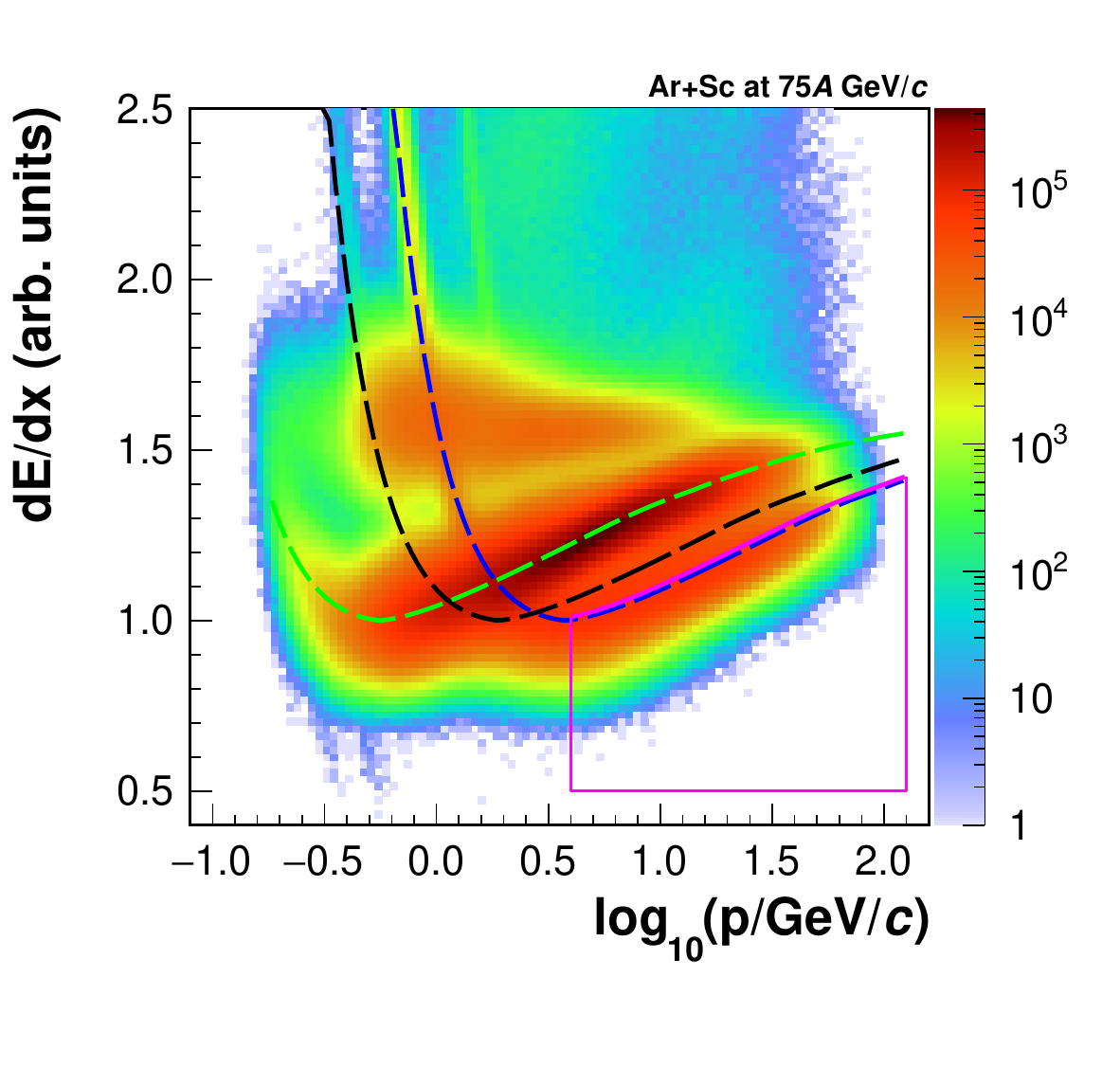}
    \rule{42em}{0.5pt}
    \caption{
        Energy loss versus the logarithm of the total momentum of positively charged particles measured with the \NASixtyOne
        Time Projection Chambers in the selected \ArSc events at 13\A--75\AGeVc.
        The picture's dashed blue, black, and green lines are nominal Bethe-Bloch lines for protons, kaons, and pions, respectively.
        The graphical cut selecting proton candidates is marked with a magenta line.
    }
    \label{fig:dEdx}
\end{figure}

To identify proton candidates, positively charged particles were selected. Their ionization \mbox{energy loss (d$E$/d$x$)} in TPCs is taken to be greater than 0.5 and less than the proton Bethe-Bloch value increased by the 15\% difference between the values for kaons and protons while the momentum is in the relativistic-rise region (from 4 to 125 \GeVc). The energy loss versus the logarithm of the total momentum of the selected positive particles for \ArSc collisions at 13\A--75\AGeVc is shown in Fig.~\ref{fig:dEdx}.
The selected region is marked with a magenta line.

This procedure was found to select, on average, about 60\% of protons. The remaining average kaon contamination is of the order of a few percent, depending on collision energy, see Ref.~\cite{Adhikary:2023xmk}.

The corresponding random proton losses do not bias the final results
of the independent production of protons in the transverse momentum space.
The results for correlated protons will be biased by the selection (see Sec.~\ref{sec:results}). Thus, the random proton selection should be considered when calculating model predictions (see Sec.~\ref{sec:models}).

\clearpage
\subsection{Acceptance maps}
\label{sec:maps}

\subsubsection{Single-particle acceptance map}

A three-dimensional ($p_{x}$, $p_{y}$ and center-of-mass rapidity) acceptance maps~\cite{na61AccMapProtinIntermittencyArSc150} for central \ArSc collisions at 13\A--75\AGeVc were created to describe the momentum region selected for this analysis. These maps were calculated by comparing the number of Monte Carlo-generated mid-rapidity protons before and after detector simulation and reconstruction.
Only particles from the regions with at least 70\% reconstructed particles are analyzed. The single-particle acceptance maps have to be used for calculating model predictions and they are given in Ref.~\cite{na61AccMapProtinIntermittencyArSc150}.

\subsubsection{Two-particle acceptance map}
\label{sec:track-pair-selection}

The Time Projection Chambers (the main tracking devices of \NASixtyOne) cannot distinguish tracks that are too close to each other in space. At a small distance, their clusters overlap, and signals are merged. Consequently, the TPC cluster finder frequently rejects overlapping clusters, and the tracks can be lost. Moreover, the TPC track reconstruction may fail to merge two track fragments. This can generate split tracks out of a single track.

The mixed data set is constructed by randomly swapping particles from different events so that each particle in each mixed event comes from different recorded events.
\begin{figure}[!ht]
    \centering
       \includegraphics[height=7.9cm, width=7.8cm]{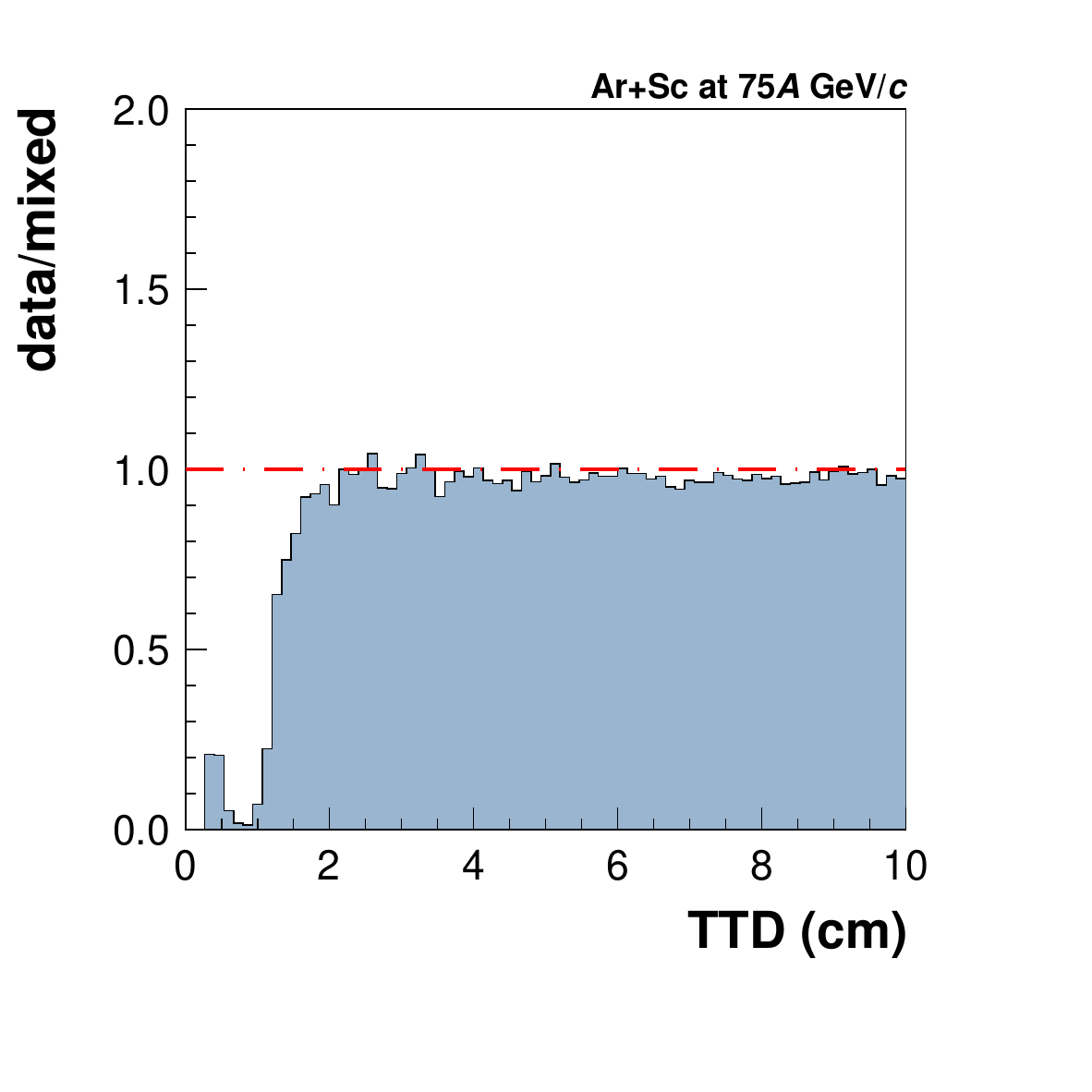}
       \includegraphics[height=7.9cm, width=7.8cm]{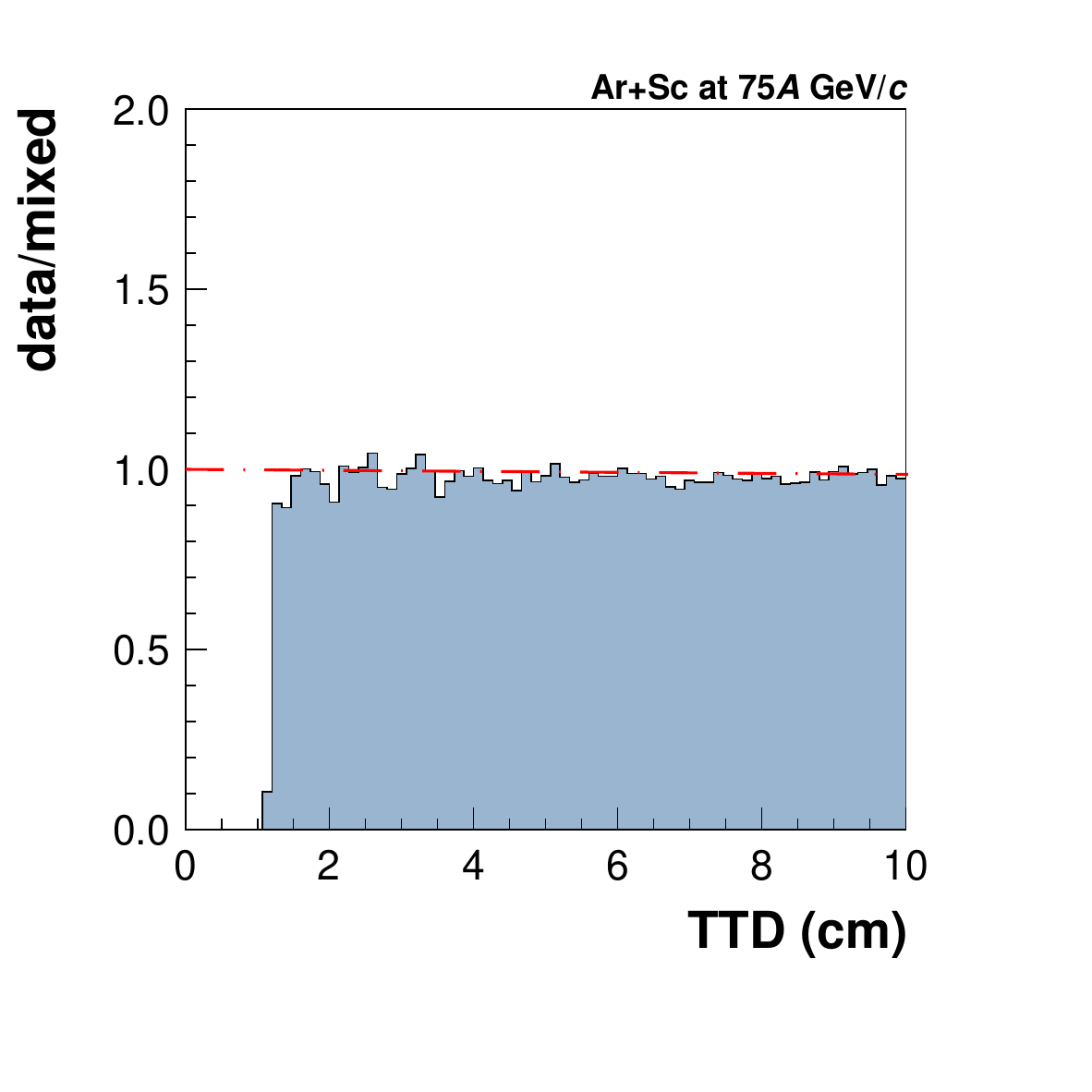}
        \rule{42em}{0.5pt}
    \caption{ The two-track distance distribution of the ratio of the number of track pairs in data and mixed events. The example for central \ArSc collisions at 75\AGeVc is given. The \textit{left} plot shows the distribution before the mTTD cut and the \textit{right} one after. See text for details.}
    \label{fig:TTDdistribution}
\end{figure}

For each pair of particles in both recorded and mixed events, a Two-Track Distance (TTD) is calculated. It is the average distance of their tracks in \coordinate{x}--\coordinate{y} plane at eight different \coordinate{z} planes (-506, -255, -201, -171, \mbox{-125, 125, 352, and 742~cm).}
The TPC’s limitation in recognizing close tracks is illustrated in \mbox{Fig.~\ref{fig:TTDdistribution} (\textit{left})} for \ArSc collisions at 75\AGeVc. The data to mixed ratio is significantly below one for TTD less than $\approx$ 2 cm. The TTD cut values for each data set are listed in Table~\ref{tab:mTTD}.

Calculating TTD requires knowledge of the \NASixtyOne detector geometry and magnetic field. Hence, it is restricted to the Collaboration members. Here, a momentum-based Two-Track Distance (mTTD) cut is introduced.

The mTTD cut removes the remaining split tracks from the data after the potential point ratio cut (see Sec.~\ref{sec:track_selection}), and it provides the precise definition of the biased region in which we do not have good efficiency for measuring two-tracks. Due to its momentum-based definition of the biased region, the mTTD cut can be used for model comparison of the experimental results without having access to the internal \NASixtyOne resources.

The magnetic field bends the trajectory of charged particles in the \coordinate{x}--\coordinate{z} plane. Thus, it is most convenient to express the momentum of each particle in the following momentum coordinates:
\begin{equation}
\label{eqs:momentumcoordinate}
 \begin{split}
    s_{x} & = p_{x}/p_{xz} = \cos(\Psi)~,\\
    s_{y} & = p_{y}/p_{xz} = \sin(\lambda)~,\\
    \rho & = 1/p_{xz}~,
\end{split}
\end{equation}
where $p_{xz} = \sqrt{p_{x}^{2} + p_{z}^{2}}$.

For each pair of particles, a difference in these coordinates is calculated as:
\begin{equation}
\label{eqs:momentumcoordinateforpairs}
 \begin{split}
    \Delta s_{x} & = s_{x,2} - s_{x,1}~,\\
    \Delta s_{y} & = s_{y,2} - s_{y,1}~,\\
    \Delta \rho & = \rho_{2} - \rho_{1}~.
\end{split}
\end{equation}
The distributions of particle pairs’ momentum difference for pairs with TTD less than $\approx$ 2 cm (as an example for Ar+Sc collisions at 75\AGeVc) are parameterized with ellipses in the new momentum coordinates. Such parameterized elliptical cuts are defined as:
\begin{equation}
    \begin{split}
         \left(\frac{\Delta \rho}{r_{\rho}}\right)^{2} + \left(\frac{\Delta s_{y}}{r_{s_{y}}}\right)^{2} & \leq 1~,\\
    \left(\frac{\Delta s_{x}}{r_{s_{x}}}\right)^{2} + \left(\frac{\Delta s_{y}}{r_{s_{y}}}\right)^{2} & \leq 1~,\\
    \left(\frac{\Delta \rho\cos\theta - \Delta s_{x}\sin\theta}{r_{\rho s_{x}}}\right)^{2}  + \left(\frac{\Delta\rho\sin\theta+\Delta s_{x}\cos\theta}{r_{s_{x}\rho}}\right)^{2} & \leq 1~,
    \end{split}
    \label{eq:ellipse}
\end{equation}
where  $r_{\rho s_{x}}$ and $r_{s_{x}\rho}$ is the semi-major and semi-minor axis of an ellipse formed by $\Delta\rho$ and $\Delta s_{x}$, and $\theta$ is the angle from the positive horizontal axis to the ellipse’s major axis.

Proton pairs with momenta difference inside the ellipses (Eqs.~\ref{eq:ellipse}) are rejected. The mTTD cut provides the precise definition of the biased region  (see Fig.~\ref{fig:TTDdistribution} \textit{right}).
The mTTD cut can replace the TTD cut.  The parameters of the mTTD cut (see Eqs.~\ref{eq:ellipse}) are given in Table~\ref{tab:mTTD}, and the mTTD cut is used as a two-particle acceptance map for the data analysis (see Sec.~\ref{sec:results}) and comparison with the models (see Sec.~\ref{sec:models}).

\begin{figure}[!ht]
    \centering
    \includegraphics[height=7.5cm, width=7.7cm]{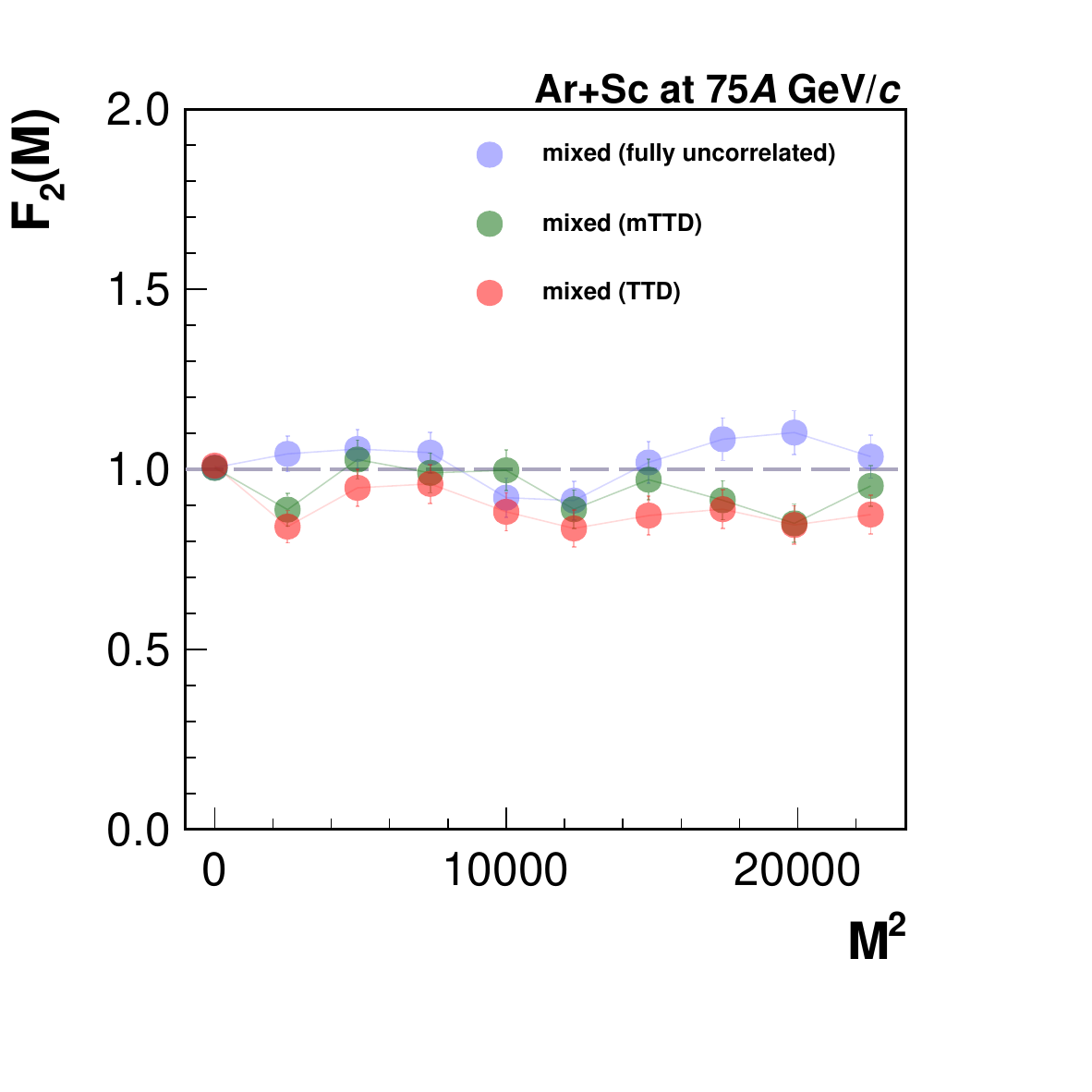}
    \includegraphics[height=7.5cm, width=7.7cm]{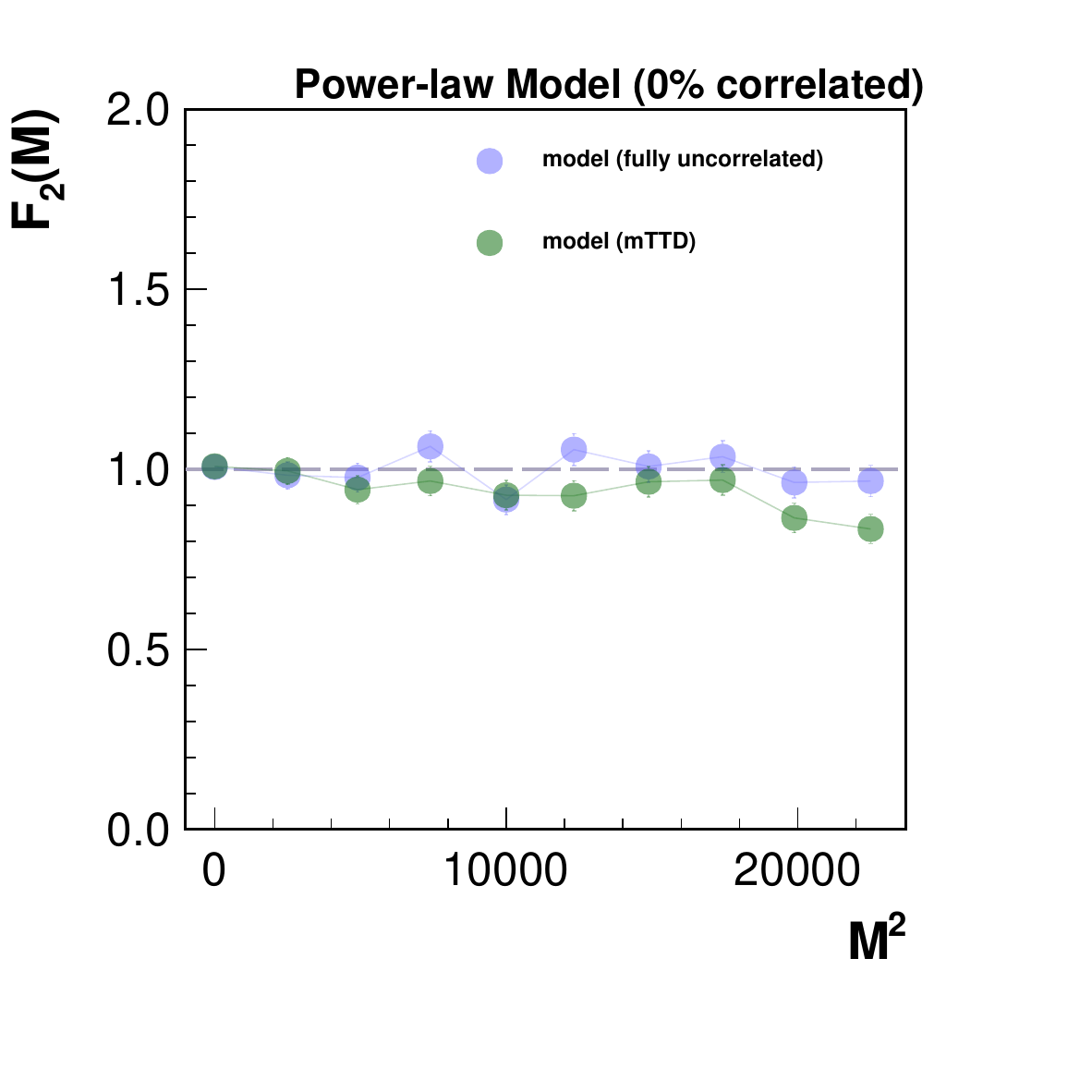}
    \rule{42em}{0.5pt}
    \caption{Example of the TTD or mTTD cut impact on mixed events for central \ArSc collisions \mbox{at 75\AGeVc (\textit{left})} and on the Power-law Model~\cite{Czopowicz:2023xcu} with uncorrelated particles only (\textit{right}) are shown. The blue circles correspond to the dependence of $F_{2}(M)$ on $M^{2}$ where neither TTD nor mTTD cut is applied to the mixed events or the Power-law Model. Green and red points correspond to either TTD or mTTD cut applied.}
    \label{fig:mTTDEFFECT}
\end{figure}

\begin{table}[!ht]
	\caption{Numerical values of the mTTD cut parameters (see Eqs.~\ref{eq:ellipse}) used to analyze central \ArSc collisions at 13\A--75\AGeVc. Particle pairs with momenta inside the ellipses are rejected.
}
	\vspace{0.5cm}
	\centering

    \begin{tabular}{ c || c |c | c | c | c | c | c }

      $p_{beam}$ (\GeVc) & TTD cut (cm) & $r_{\rho} (\GeVc)^{-1}$ & $r_{s_{y}}$ & $r_{s_{x}}$ & $r_{\rho s_{x}} (\GeVc)^{-1}$ & $r_{s_{x}\rho} (\GeVc)^{-1}$ &  $\theta$ \\
      \hline
      \hline
      13\A  & 3.5 &0.470 & 0.004 & 0.047 & 0.470 & 0.004 &  5 \\
      19\A  &2.8 &0.121 & 0.003 & 0.010 & 0.121 & 0.003 &  8 \\
      30\A  &2.8 &0.123 & 0.002 & 0.013 & 0.123 & 0.002 & 13 \\
      40\A  &2.2 &0.043 & 0.002 & 0.010 & 0.043 & 0.002 & 15 \\
      75\A  &2.2 &0.080 & 0.002 & 0.011 & 0.020 & 0.002 & 31 \\

    \end{tabular}
	\label{tab:mTTD}
\end{table}
The effect of the mTTD cut for \ArSc collisions at 75\AGeVc mixed events, and the Power-law Model (see Sec.~\ref{sec:models-power}) is shown in Fig.~\ref{fig:mTTDEFFECT}. The dependence of  $F_{2}(M)$ on $M^{2}$ in cumulative transverse momentum space for $M^{2} > 1$ is systematically below $F_{2}(M=1)$ when TTD or mTTD cut is applied to fully uncorrelated mixed events (\textit{left}) and the Power-law Model with uncorrelated particles only (\textit{right}).

\subsection{Statistically-independent data points}
The intermittency analysis gives the dependence of scaled factorial moments on the number of subdivisions of transverse momentum or cumulative transverse momentum intervals.
In the past intermittency analyses, the same data set was used to obtain results for each number of subdivisions. The results for different $M^{2}$ were statistically correlated. Therefore, the full covariance matrix was required for proper statistical treatment of the results. This is numerically not trivial~\cite{Wosiek:1990kw}. Here, the statistically-independent data subsets were used to obtain results for different subdivision numbers. In this case, the results for different subdivision numbers are statistically independent. Only diagonal elements of the covariance matrix are non-zero, and thus the complete relevant information needed to interpret the results is easy to present graphically and use in the statistical tests. The procedure decreases the number of events used to calculate each data point increasing statistical uncertainties, and therefore forcing to reduce the number of data points.

The number of events used in each subset was selected to obtain similar statistical uncertainties for different subsets. Table~\ref{tab:fraction-of-events} presents the fractions of all selected events used to calculate each of the 10 points.
\begin{table}[!ht]
    \centering
    \caption{
        Fraction of the total number of selected events used to calculate
        second-order scaled factorial moments for the chosen number of cumulative momentum cells.
    }
    \vspace{1ex}
    \begin{tabular}{ l | c | c | c | c | c | c | c | c | c | c }
        number of cells $M^{2}$    & $1^{2}$   & $50^{2}$ & $70^{2}$ & $86^{2}$ & $100^{2}$ & $111^{2}$ & $122^{2}$ & $132^{2}$ & $141^{2}$ & $150^{2}$ \\ \hline
        fraction of all events (\%) & 0.5 &  3.0 &  5.0 &  7.0 &   9.0 &  11.0 &  13.0 & 15.5 & 17.0 &  19.0
    \end{tabular}
    \label{tab:fraction-of-events}
\end{table}

\subsection{Uncertainties and biases}

The standard expression for the scaled factorial moments,
Eq.~\ref{eq:scaled-factorial-moments} can be rewritten as
\begin{equation}
\label{eq:sfm}
    F_{2}(M) = 2M^{2}\frac{\AVG{N_{2}(M)}}{\AVG{N}^{2}},
\end{equation}
where $N_{2}(M)$ denotes a total number of particle pairs in $M^{2}$ bins in an event.
Then the statistical uncertainties can be calculated using the standard error propagation:
\begin{equation}
    \frac{\sigma_{F_{2}}}{|F_{2}|} = \sqrt{\frac{(\sigma_{N_{2}})^{2}}{\AVG{N_{2}}^{2}}
    + 4\frac{(\sigma_{N})^{2}}{\AVG{N}^{2}}
    - 4\frac{(\sigma_{N_{2}N})^{2}}{\AVG{N}\AVG{N_{2}}}}.
\end{equation}

The final results presented in Sec.~\ref{sec:results} are not corrected for possible biases. Their magnitude was estimated by comparing results for pure \EposLong~\cite{Pierog:2006qv} and \EposLong subjected to the detector simulation, reconstruction, and data-like analysis. Figure~\ref{fig:epos} shows the comparison for \ArSc collisions at 13\A--75\AGeVc.
Their differences are significantly smaller than the statistical uncertainties of the experimental data and increase with $M^{2}$ up to about 0.1 at large $M^{2}$ values. Note that protons generated by \EposLong do not show a significant correlation in the transverse momentum space, see Sec.~\ref{sec:models}. In this case, the momentum resolution does not affect the results significantly.

In the case of the critical correlations, the impact of the momentum resolution may be significant, see Ref.~\cite{Samanta:2021usk} and Sec.~\ref{sec:models} for detail.
Thus a comparison with models including short-range correlations in the transverse momentum space requires smearing of momenta according to the experimental resolution, which can be approximately parameterized as:
\begin{equation}\label{eq:smearing}
\begin{split}
    p_{x}^{\text{smeared}} &= p_{x}^{\text{original}} + \delta p_{x} \quad \text{and}\\
    p_{y}^{\text{smeared}} &= p_{y}^{\text{original}} + \delta p_{y},
\end{split}
\end{equation}
where $\delta p_{x}$ and $\delta p_{y}$ are randomly drawn from a Gaussian distribution with $\sigma_{x}$ and $\sigma_{y}$ values for central \ArSc collisions at 13\A--75\AGeVc from the Table.~\ref{tab:resolution}.

\begin{table}[!ht]
	\caption{The $\sigma_{x}$ and $\sigma_{y}$ parameters, see text for details, used for smearing of proton transverse momentum in simulated \ArSc collisions at 13\A--75\AGeVc.
}
	\vspace{0.5cm}
	\centering
	\begin{tabular}{ c || c | c }
     $p_{beam}$ (\GeVc) & $\sigma_{x}$ (\MeVc)& $\sigma_{y}$ (\MeVc) \\
     \hline
    \hline
    13\A  & 4.8 & 3.5\\

    19\A  & 4.6 & 3.4\\

    30\A  & 4.0 & 3.2\\

    40\A  & 3.5 & 3.1\\

    75\A  & 3.1 & 3.1\\

  	\end{tabular}
	\label{tab:resolution}
\end{table}

Uncertainties on the final results presented in Sec.~\ref{sec:results} correspond to statistical uncertainties.

\section{Results}
\label{sec:results}

This section presents results on second-order scaled factorial moments
(Eq.~\ref{eq:scaled-factorial-moments}) of $\approx60\%$ randomly selected protons (losses due to proton misidentification) with momentum smeared due to reconstruction resolution (Eq.~\ref{eq:smearing}) produced within the acceptance maps defined in Sec.~\ref{sec:maps} by strong and electromagnetic processes in 0--10\% central \ArSc interactions at 13\A--75\AGeVc.
The results are shown as a function of the number of subdivisions in
transverse momentum space -- the so-called intermittency analysis.
The analysis was performed for cumulative and original transverse momentum components. Independent data sub-sets were used to calculate the results for each subdivision.

Uncertainties correspond to statistical ones. Biases estimated using the \EposLong~\cite{Werner:2008zza} model (see Sec.~\ref{sec:models})
are significantly smaller than the statistical uncertainties of the experimental data.

\subsection{Two-particle correlation function}
\begin{figure}[!hbt]
\centering
\includegraphics[height=5cm, width=5cm]{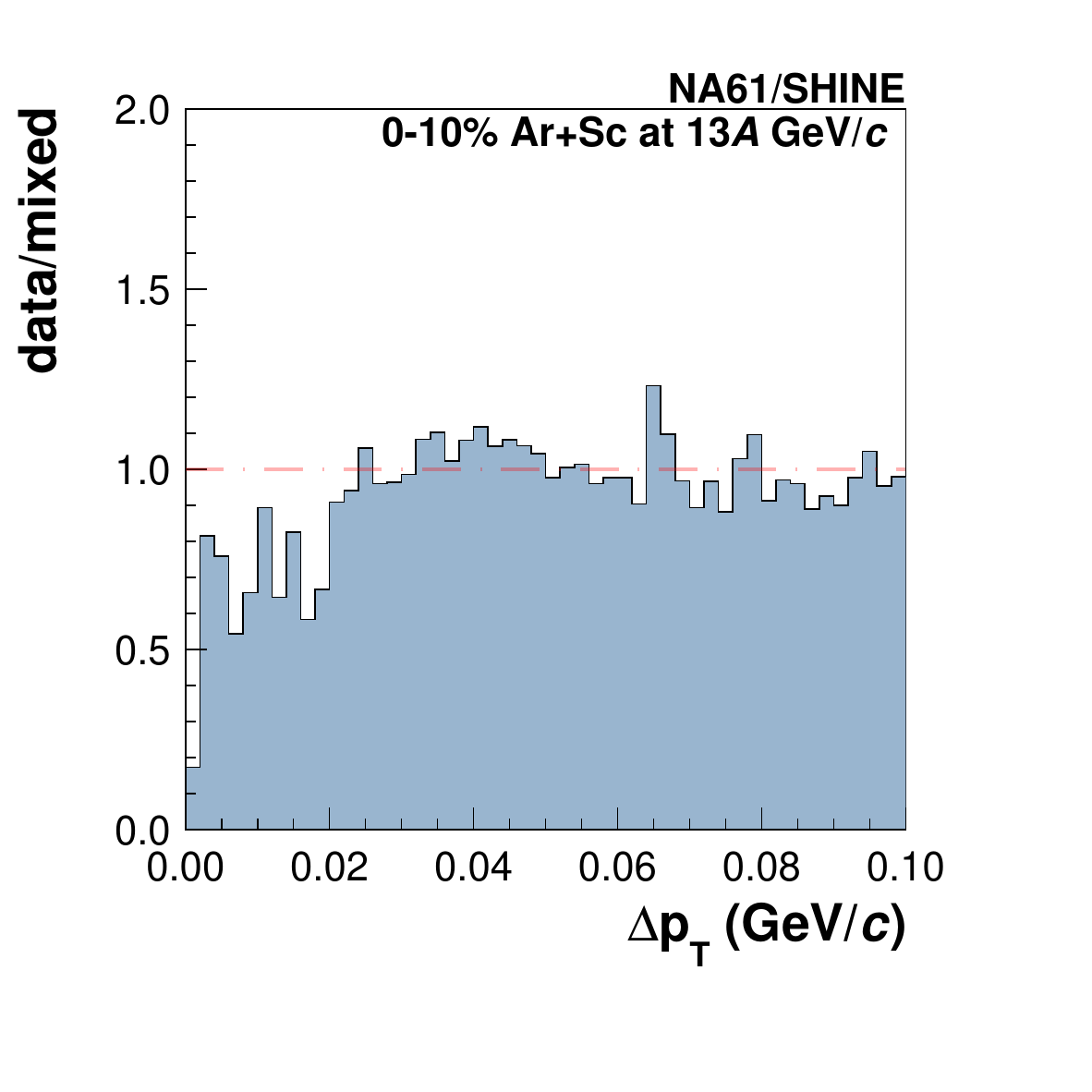}
\includegraphics[height=5cm, width=5cm]{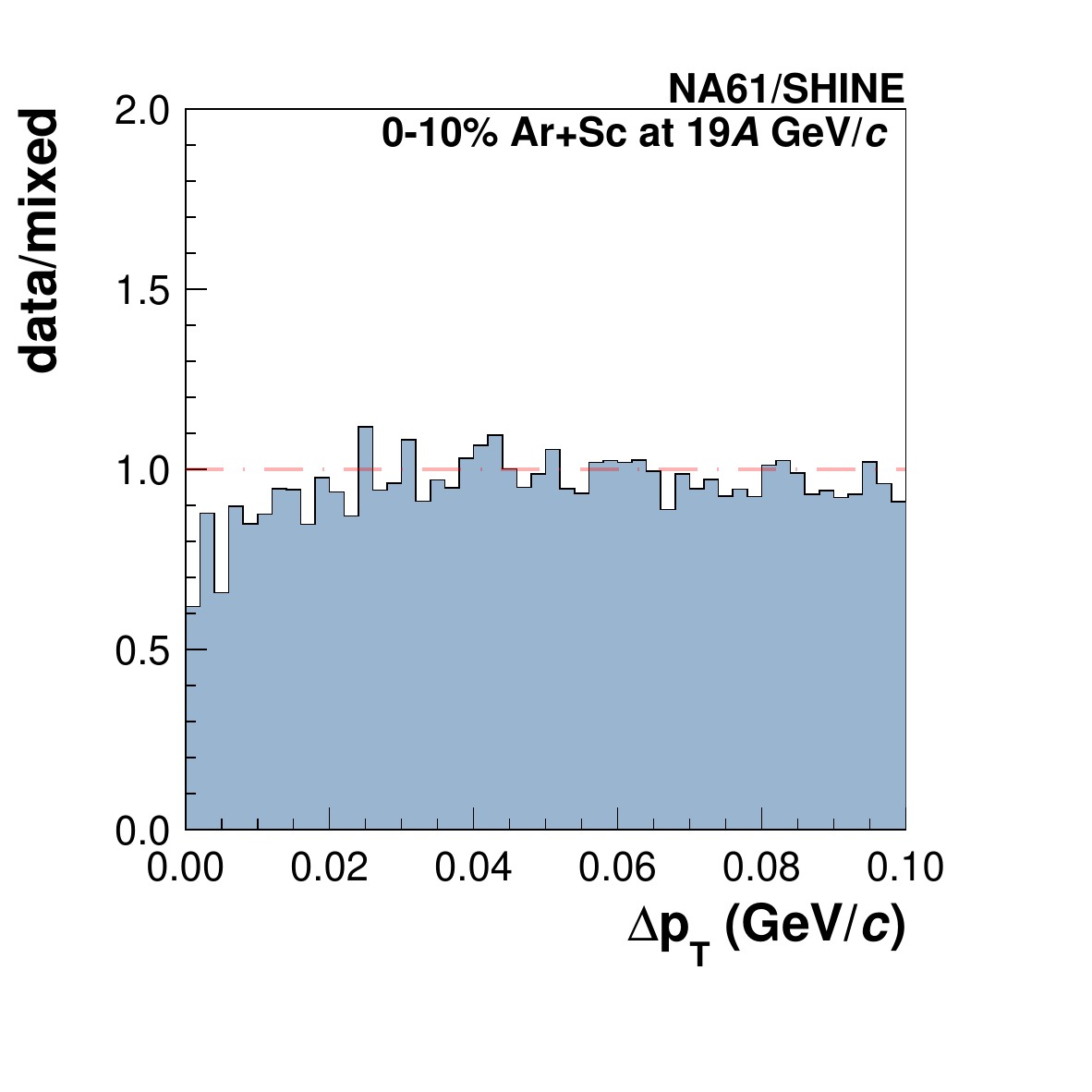}
\includegraphics[height=5cm, width=5cm]{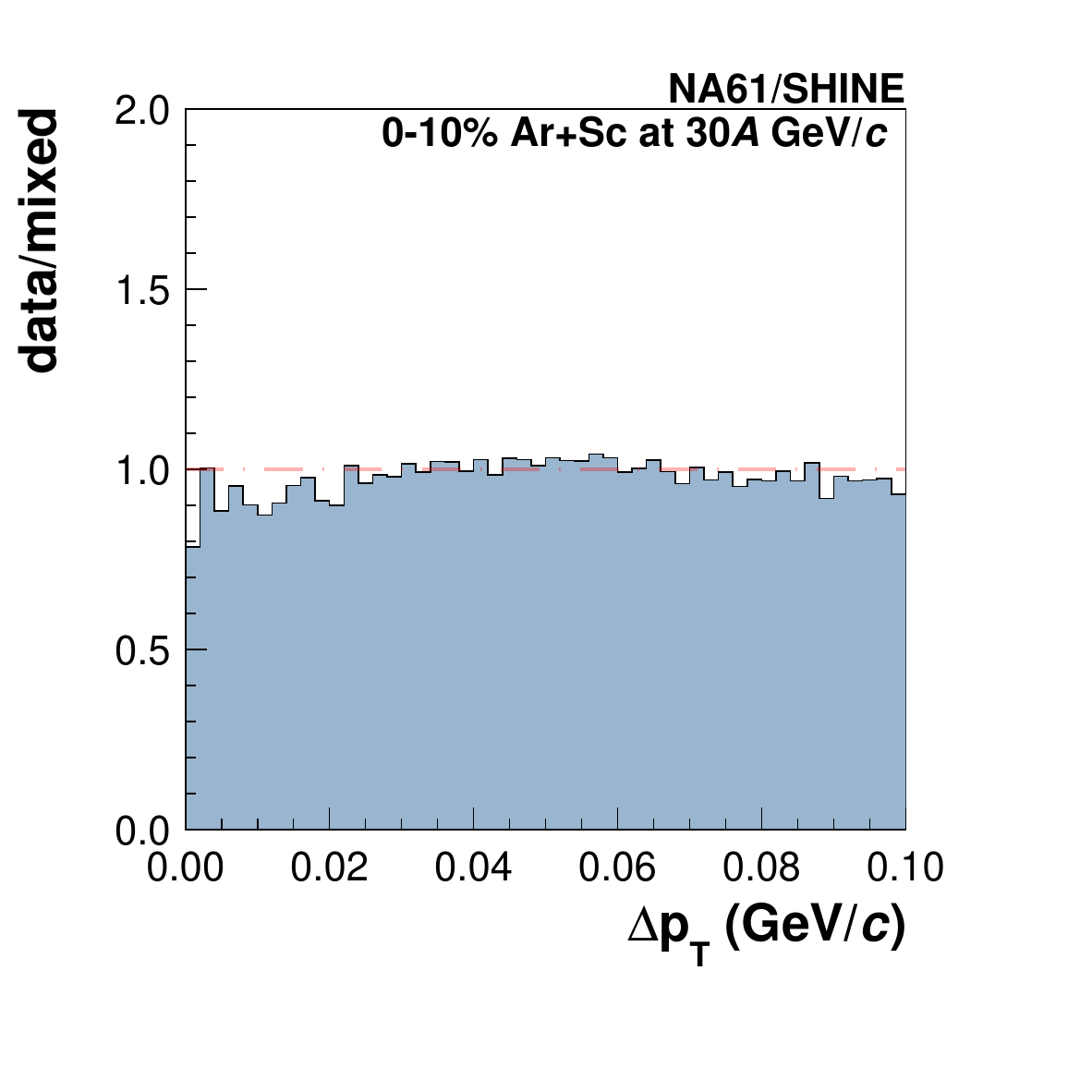}\\
\includegraphics[height=5cm, width=5cm]{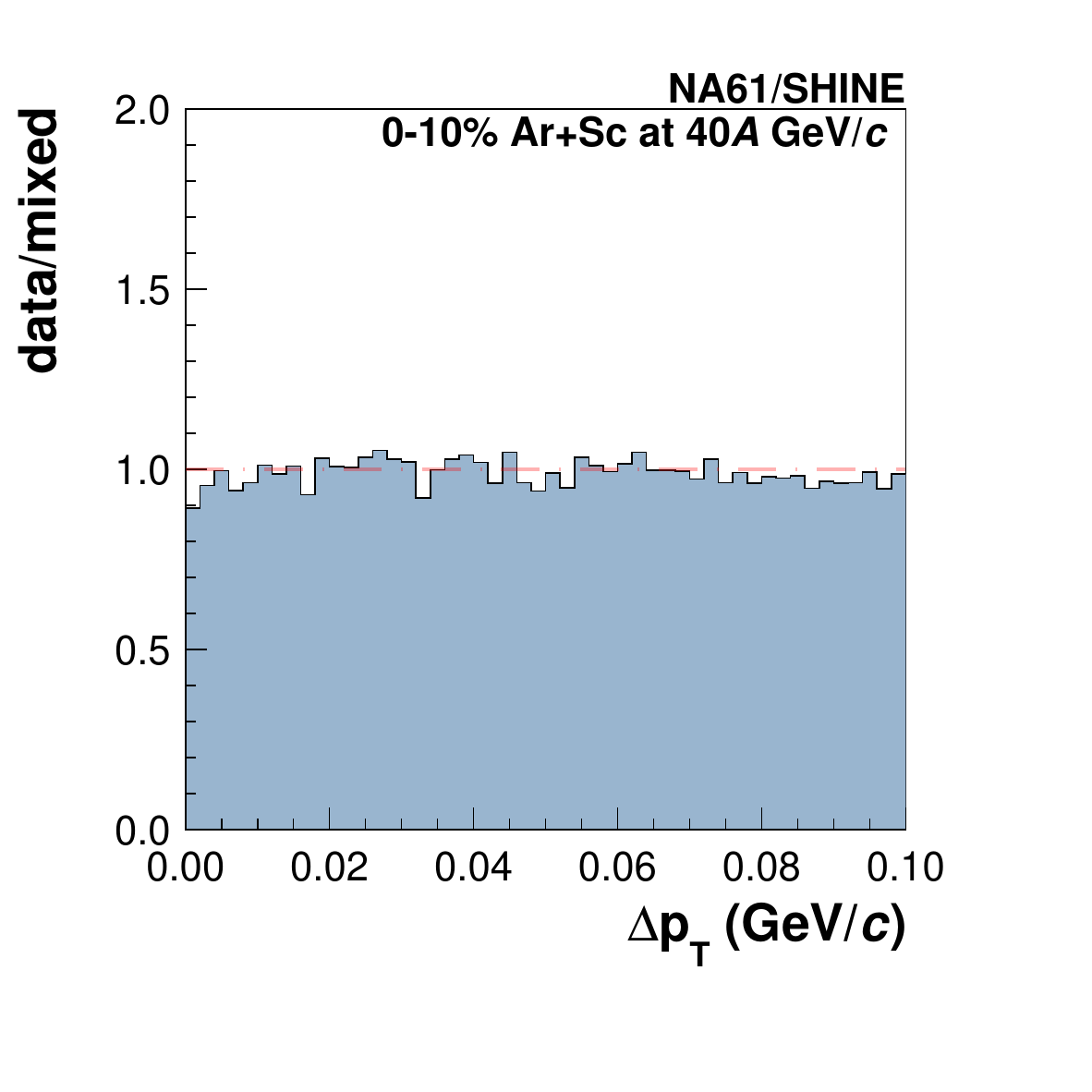}
\includegraphics[height=5cm, width=5cm]{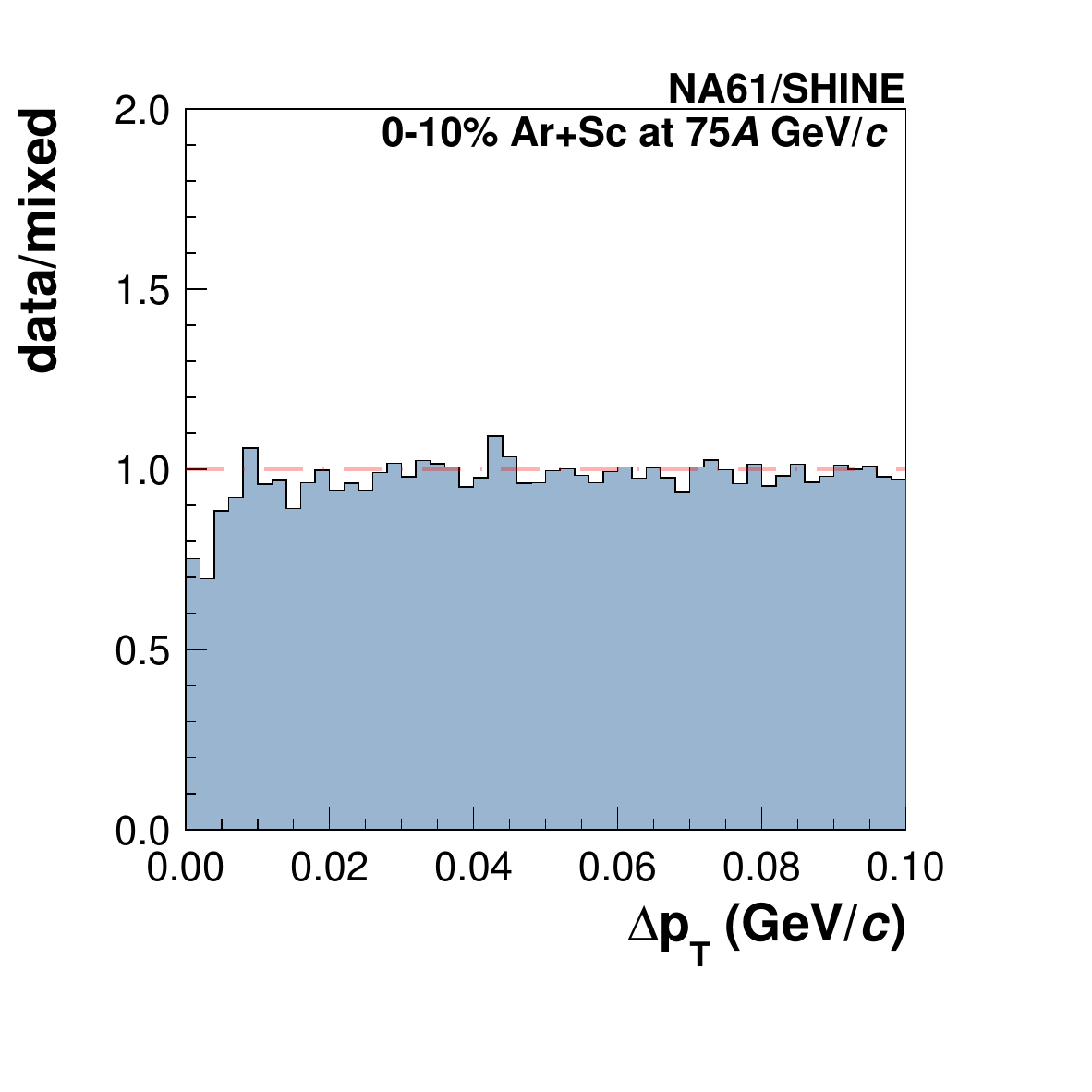}
\rule{42em}{0.5pt}
\caption{Two-particle correlation functions in $\Delta p_{T}$ for selected proton candidates within analysis acceptance for central \ArSc collisions at 13\A--75\AGeVc are shown. The data distribution includes the mTTD cut, whereas the mixed one does not. }
\label{fig:DeltapT-arsc}
\end{figure}

The two-particle correlation function, $\Delta p_{T}$, of selected proton candidates within the analysis acceptance for central \ArSc collisions at 13\A--75\AGeVc is shown in Fig.~\ref{fig:DeltapT-arsc}. The correlation function is the ratio of normalized $\Delta p_{T}$ distributions for data and mixed events. The data distribution includes the mTTD cut, whereas the mixed one does not. The decrease of the correlation function at $\Delta p_{T} \approx 0$ is due to anti-correlation introduced by the mTTD cut.
\subsection{Subdivisions in cumulative transverse momentum space}
\label{sec:results_cumulative}

\begin{figure}[!ht]
    \centering
    \includegraphics[width=.33\textwidth]{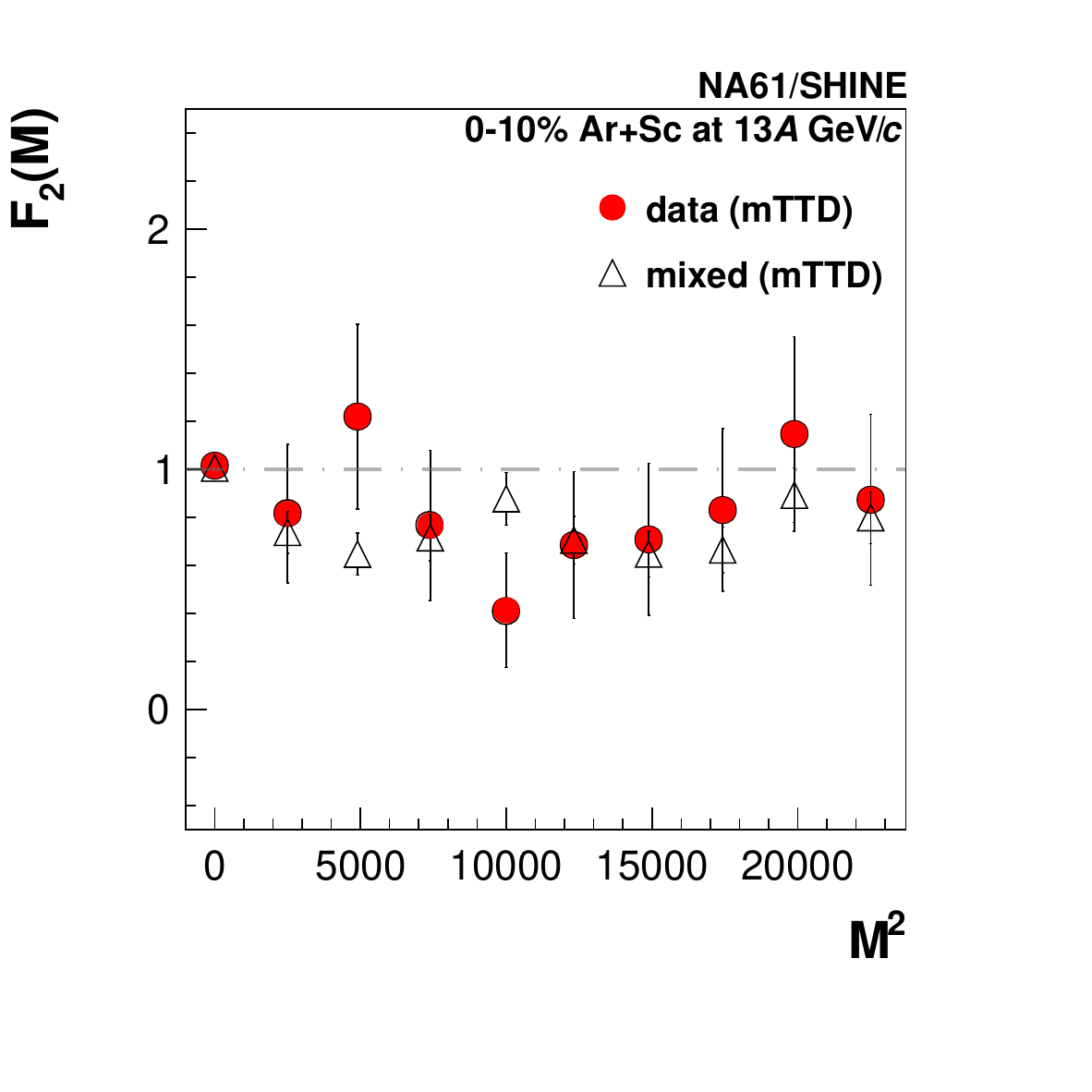}\hfill
    \includegraphics[width=.33\textwidth]{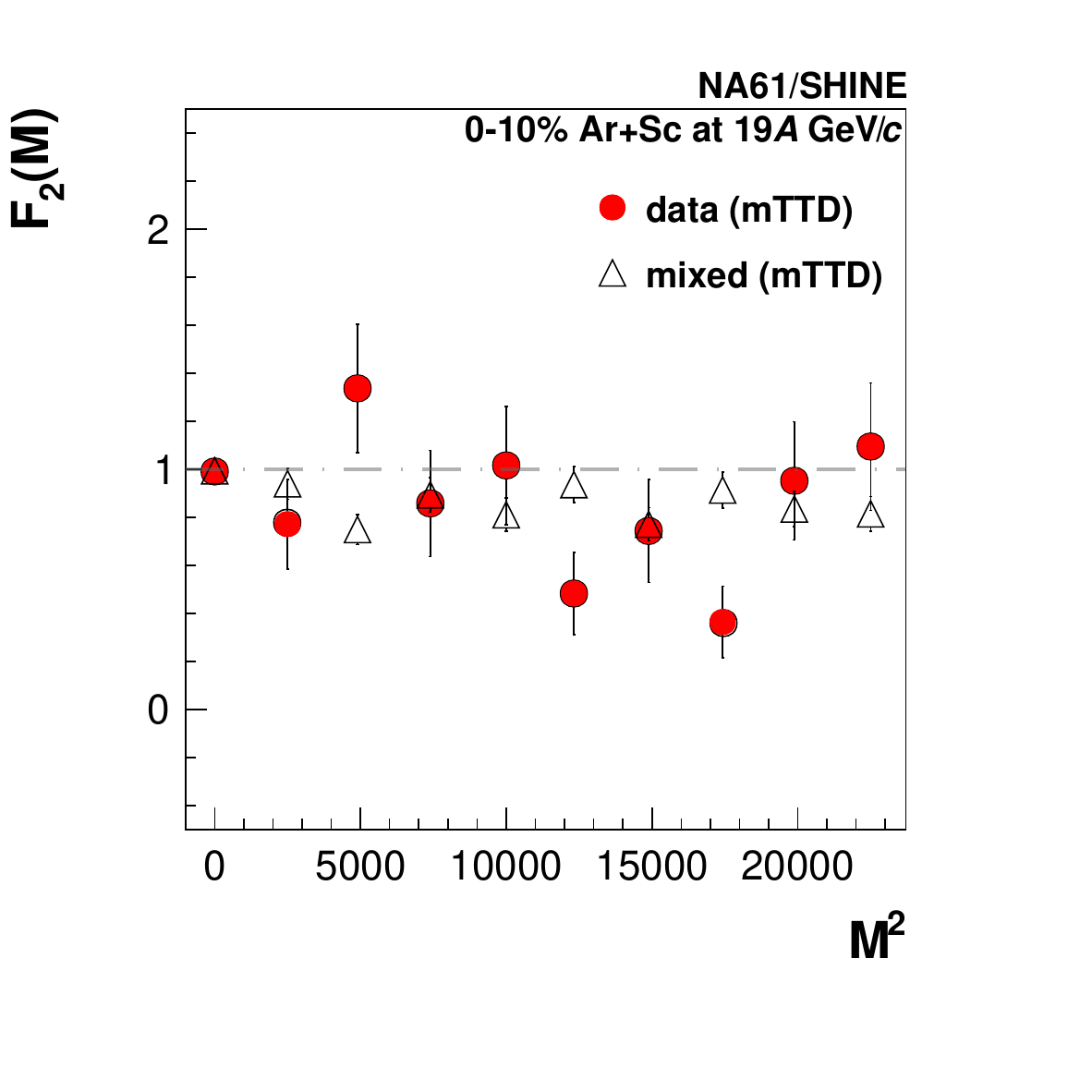}\hfill
    \includegraphics[width=.33\textwidth]{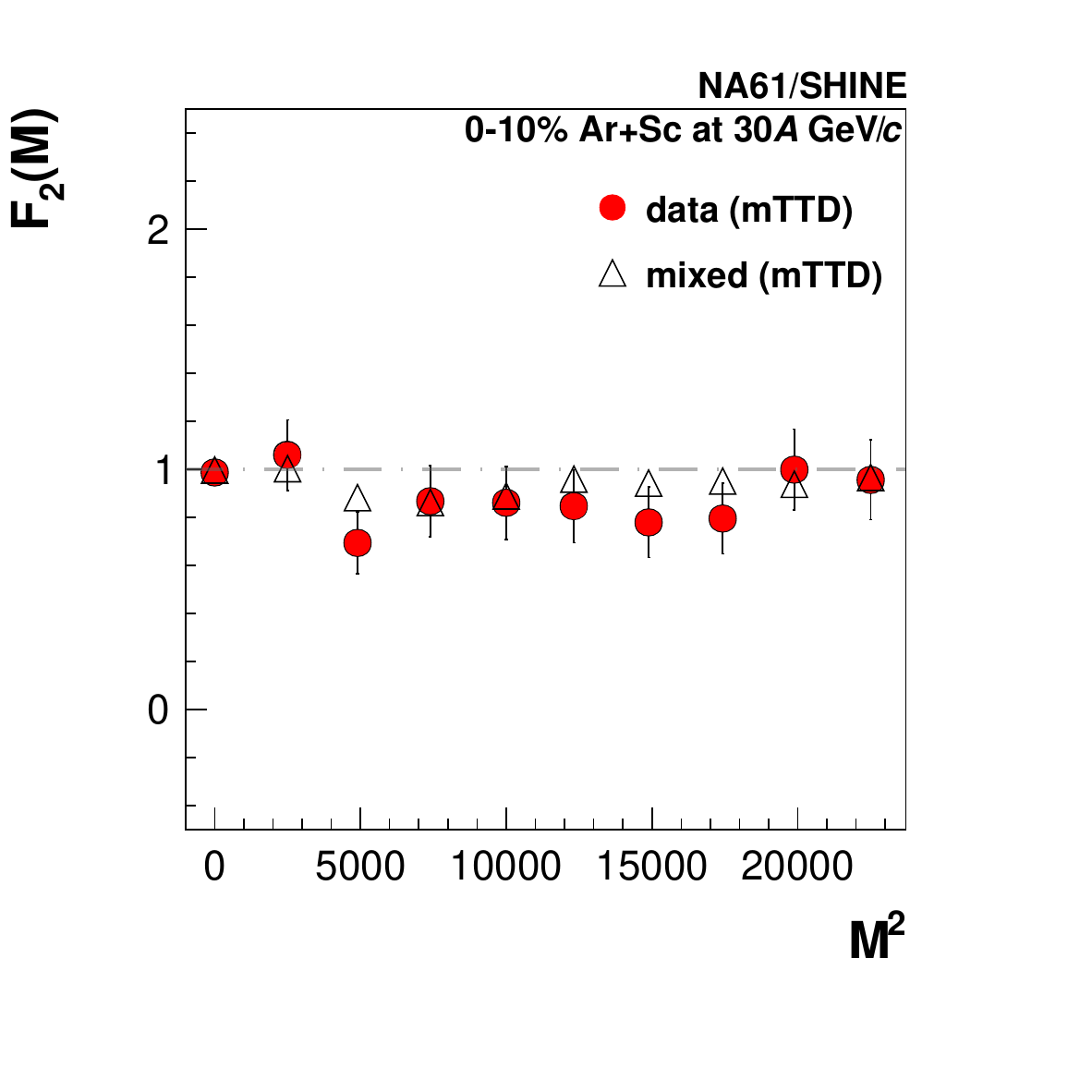}\\
    \includegraphics[width=.33\textwidth]{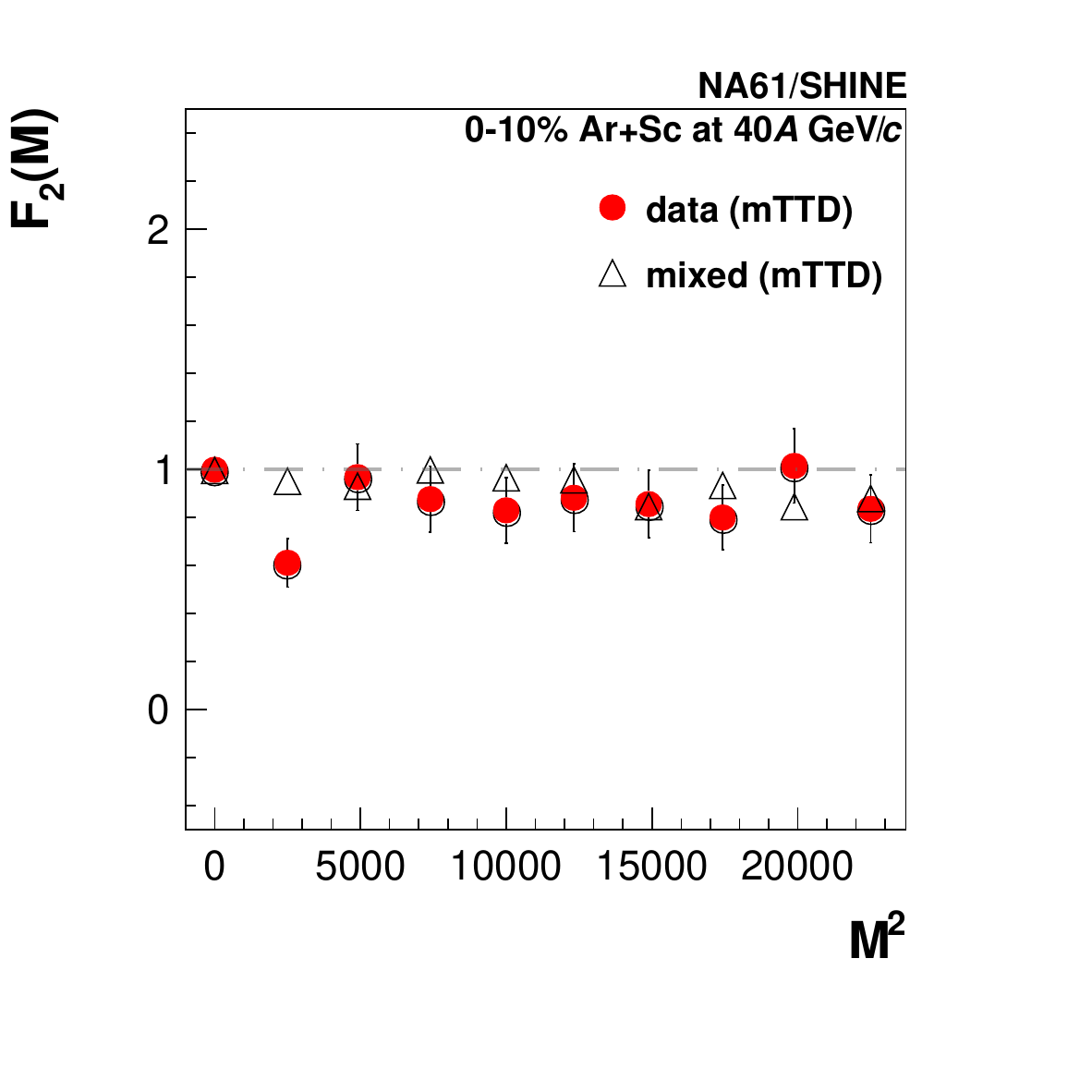}\qquad
    \includegraphics[width=.33\textwidth]{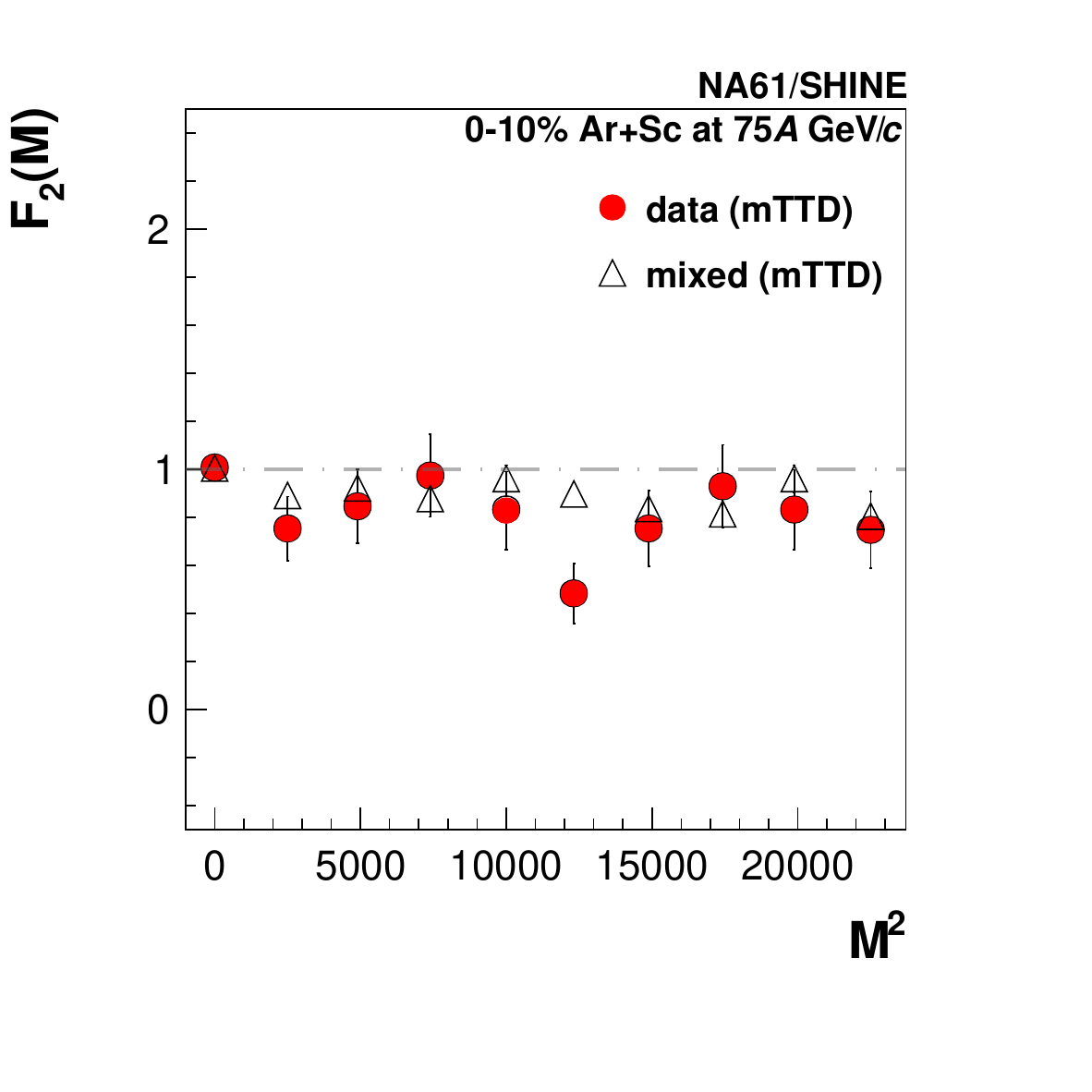}
    \rule{42em}{0.5pt}
    \caption{
    Results on the dependence of the scaled factorial moment of proton multiplicity distribution
    on the number of subdivisions in cumulative transverse momentum space $M^{2}$ for $1^{2} \leq M^{2} \leq 150^{2}$.
    Results are shown for 0--10\% central \ArSc collisions at 13\A--75\AGeVc. Closed red circles indicate the experimental data. Corresponding results for mixed events (open triangles) are also shown.  Both the data and mixed events include the mTTD cut. Only statistical uncertainties are indicated.
    }
    \label{fig:results-cum-full}
\end{figure}

\begin{figure}[!ht]
    \centering
    \includegraphics[width=.33\textwidth]{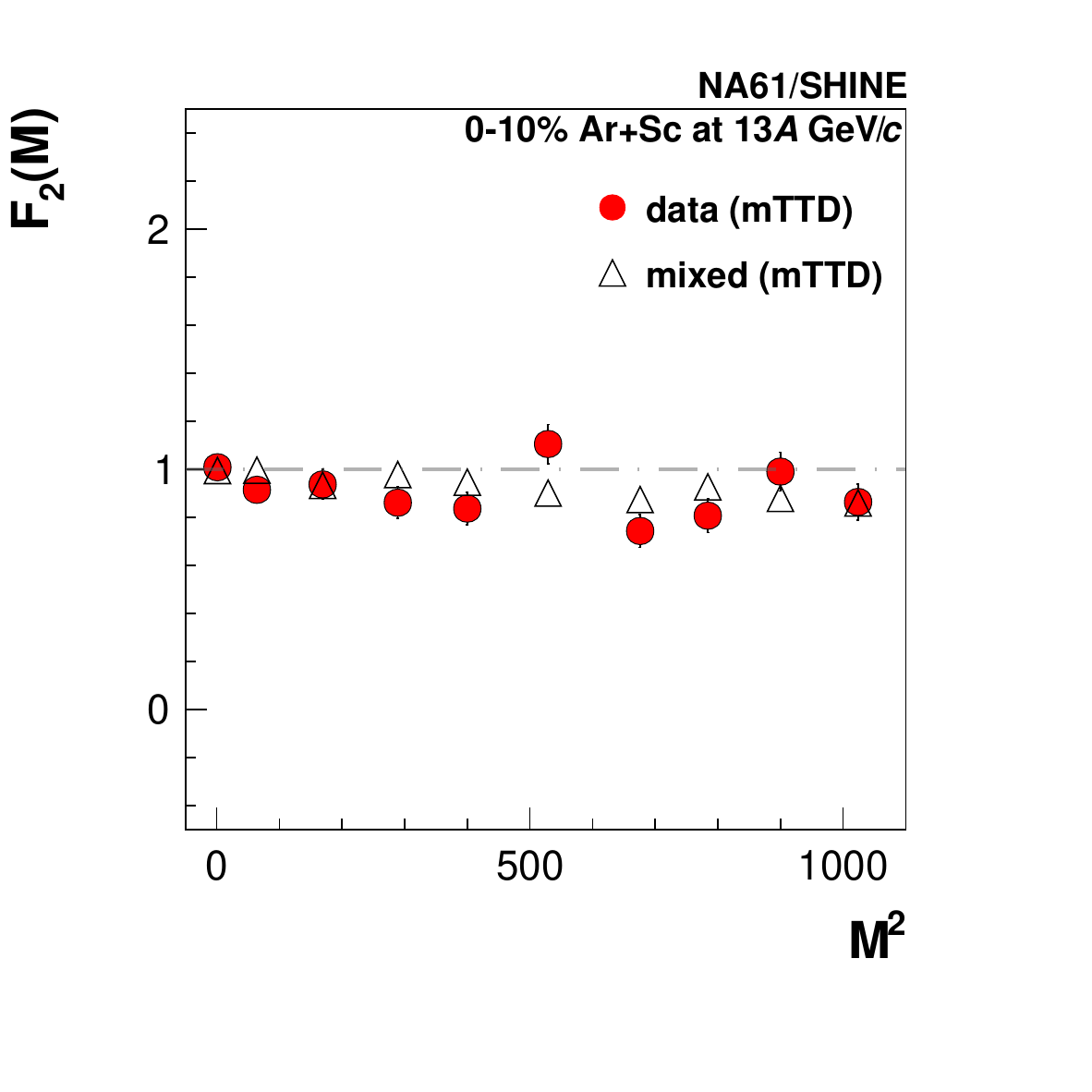}\hfill
    \includegraphics[width=.33\textwidth]{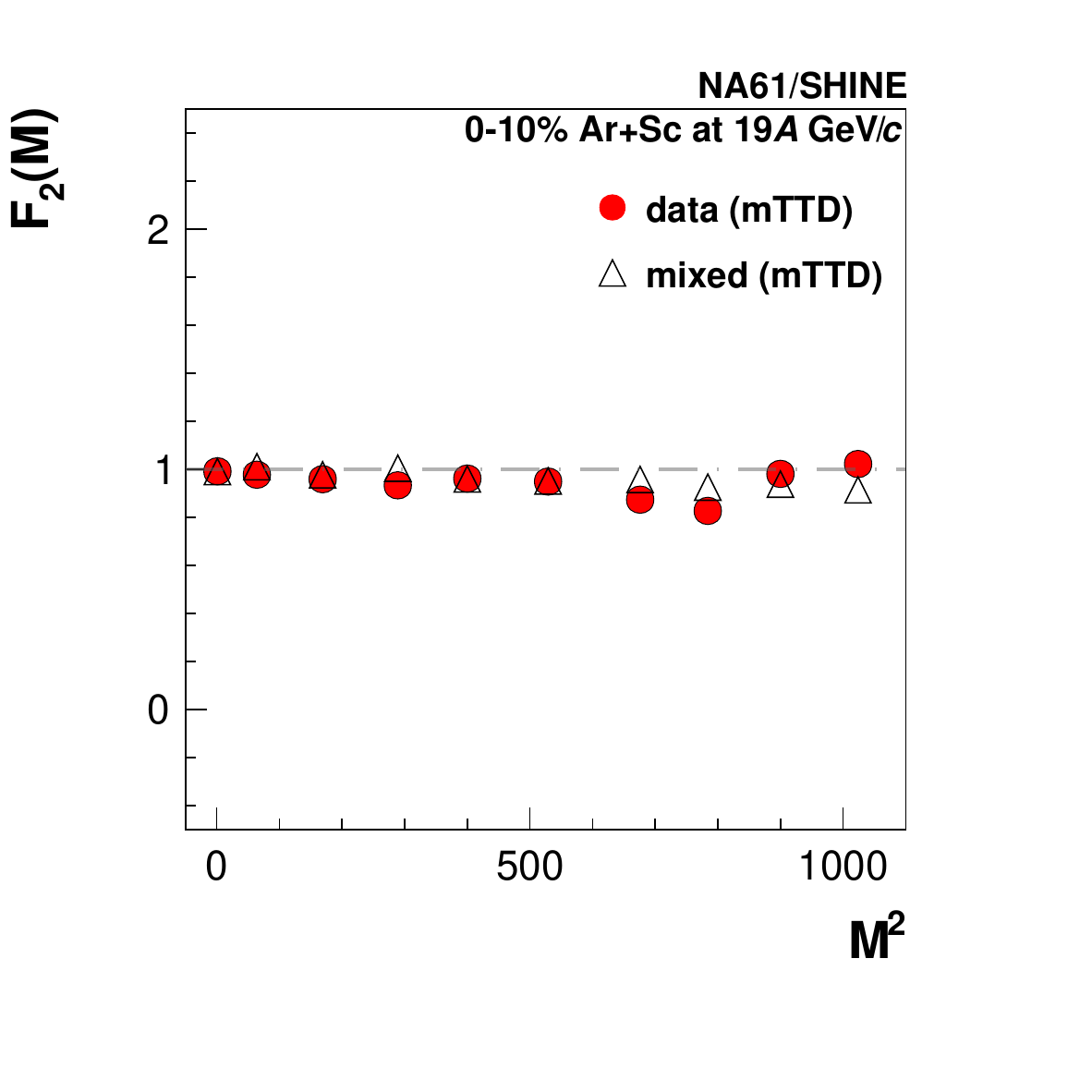}\hfill
    \includegraphics[width=.33\textwidth]{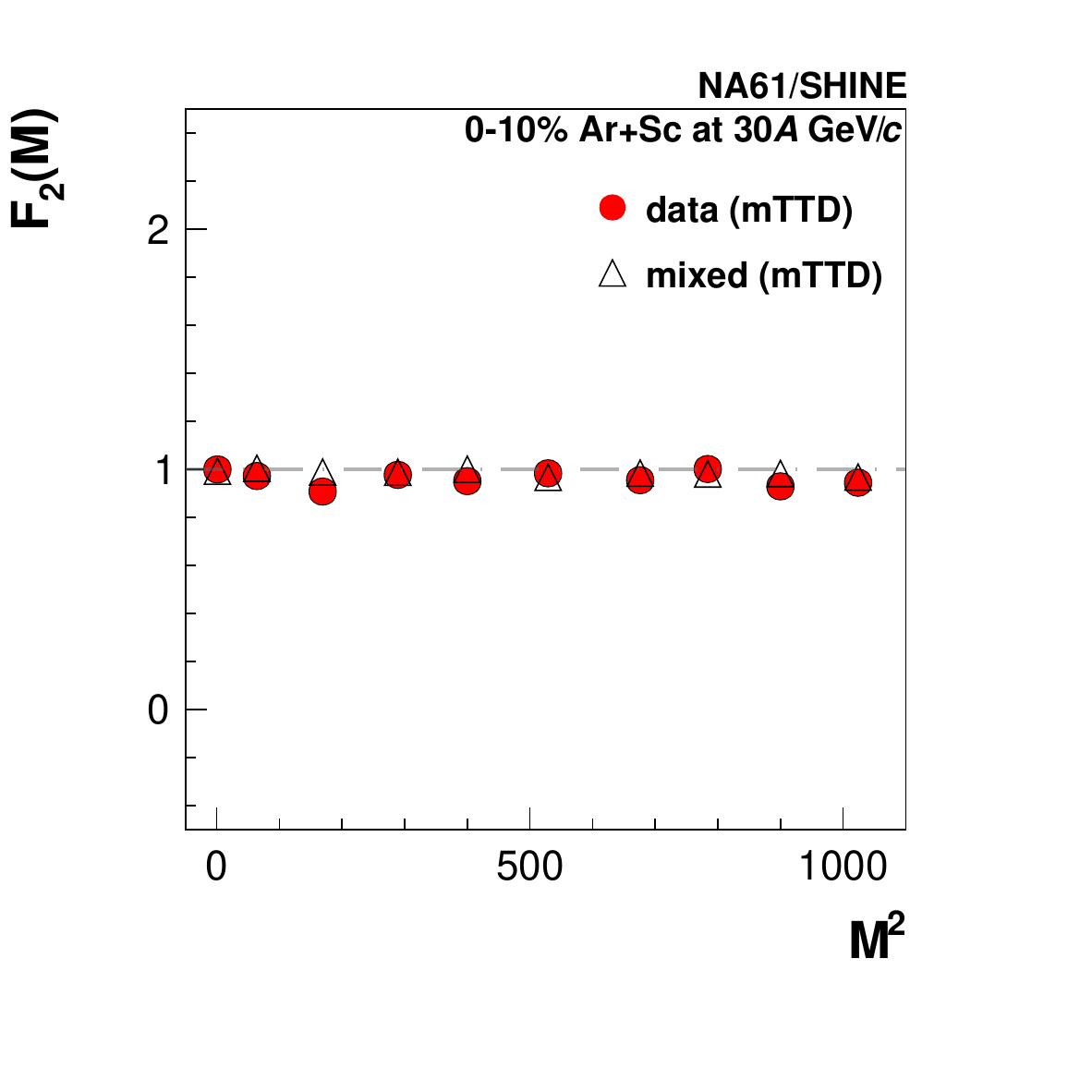}\\
    \includegraphics[width=.33\textwidth]{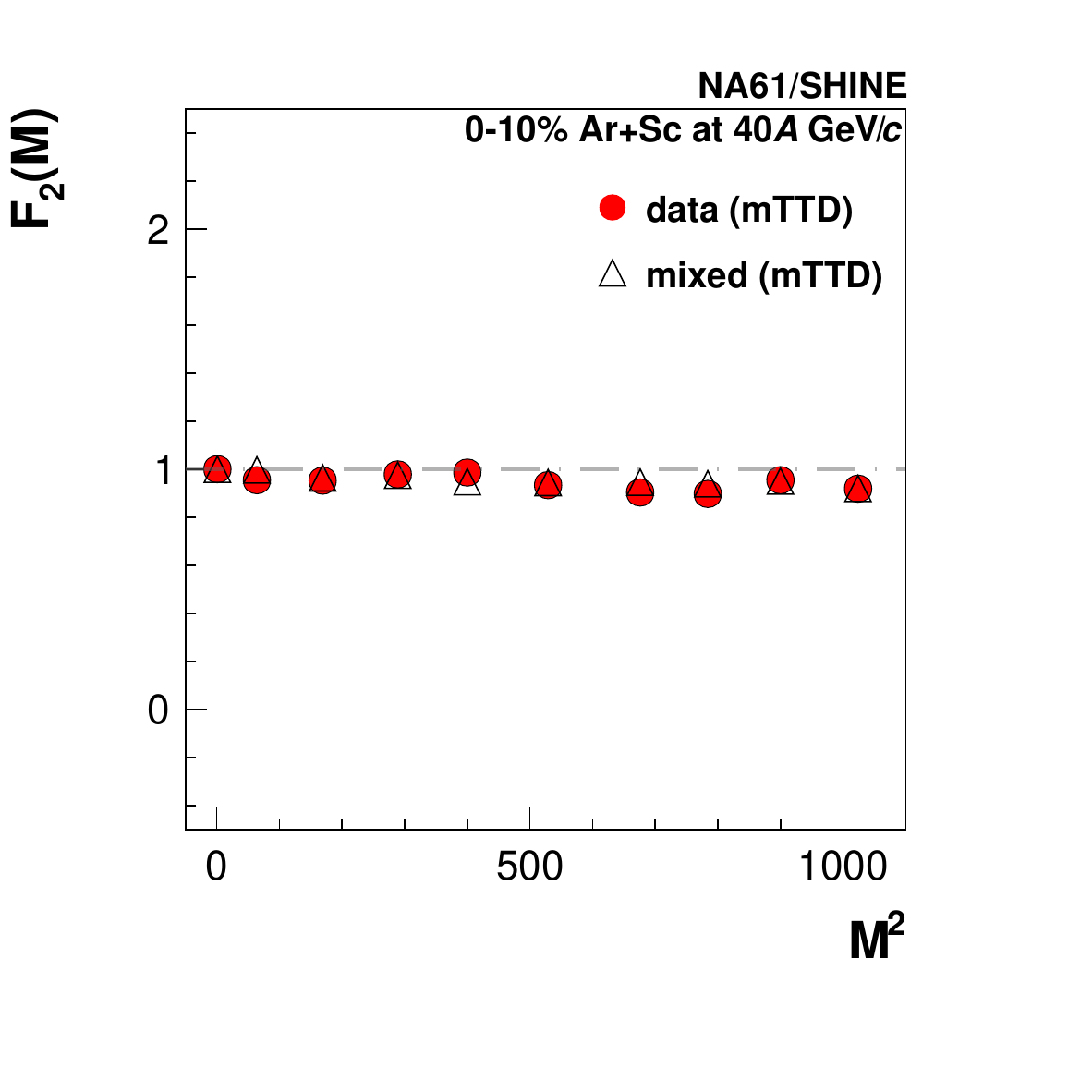}\qquad
    \includegraphics[width=.33\textwidth]{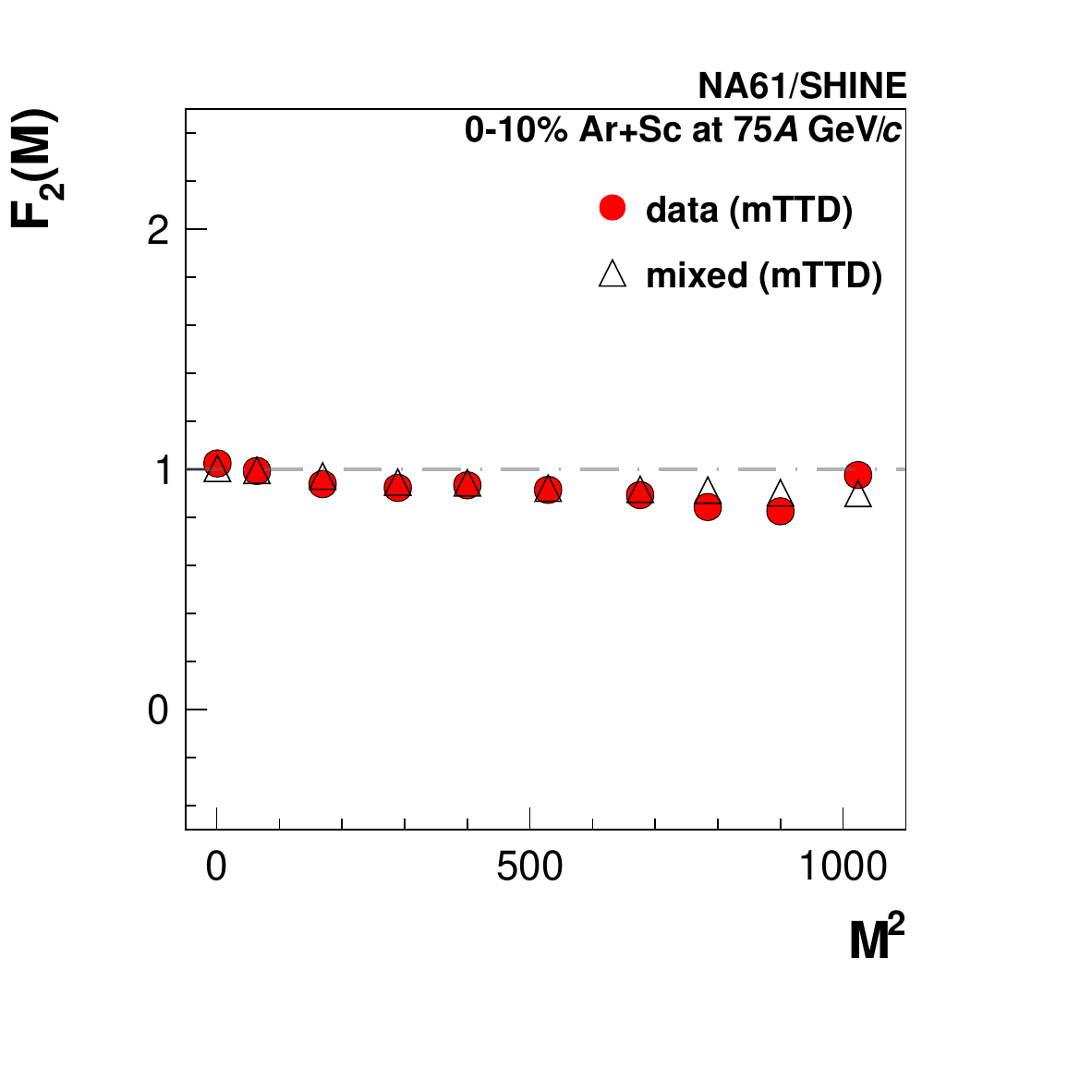}
    \rule{42em}{0.5pt}
    \caption{
      Results on the dependence of the scaled factorial moment of proton multiplicity distribution
       on the number of subdivisions in cumulative transverse momentum space $M^{2}$ for $1^{2} \leq M^{2} \leq 32^{2}$. Results are shown for 0--10\% central \ArSc collisions at 13\A--75\AGeVc. Closed red circles indicate the experimental data. Corresponding results for mixed events (open triangles) are also shown.  Both the data and mixed events include the mTTD cut. Only statistical uncertainties are indicated.
    }
    \label{fig:results-cum-small}
\end{figure}

Figures~\ref{fig:results-cum-full} and~\ref{fig:results-cum-small} present the dependence of the factorial moment on the number of subdivisions in cumulative-transverse momentum space for the maximum
subdivision number of $M = 150$ and $M = 32$, respectively. The latter, coarse subdivision, was introduced to limit the effect of experimental momentum resolution; see Ref.~\cite{Samanta:2021usk} and below for details.
The experimental results are shown for 0--10\% central \ArSc collisions at 13\A--75\AGeVc.
As a reference, the corresponding results for mixed events are also shown.

By construction, the multiplicity distribution of protons in mixed events for $M = 1$ equals the corresponding distribution for the data. In mixed events, the only correlation of particles in the transverse momentum space is due to the mTTD cut. Both the data and mixed events include the mTTD cut. The mTTD cut is necessary to properly account for the detector resolution -- losses of close-in-space tracks.

The results for subdivisions in cumulative transverse momentum space, $F_{2}(M)$ for $M > 1$ are systematically below $F_{2}(M=1)$. It is likely due to the anti-correlation generated by the mTTD cut to the \mbox{data (see Sec.~\ref{sec:track-pair-selection} for details).} 

The experimental results show no increase of $F_{2}(M)$ with $M^{2}$.
There is no indication of the critical fluctuations for selected proton candidates.

\clearpage
\subsection{Subdivisions in transverse momentum space}

\begin{figure}[!ht]
    \centering
    \includegraphics[width=.33\textwidth]{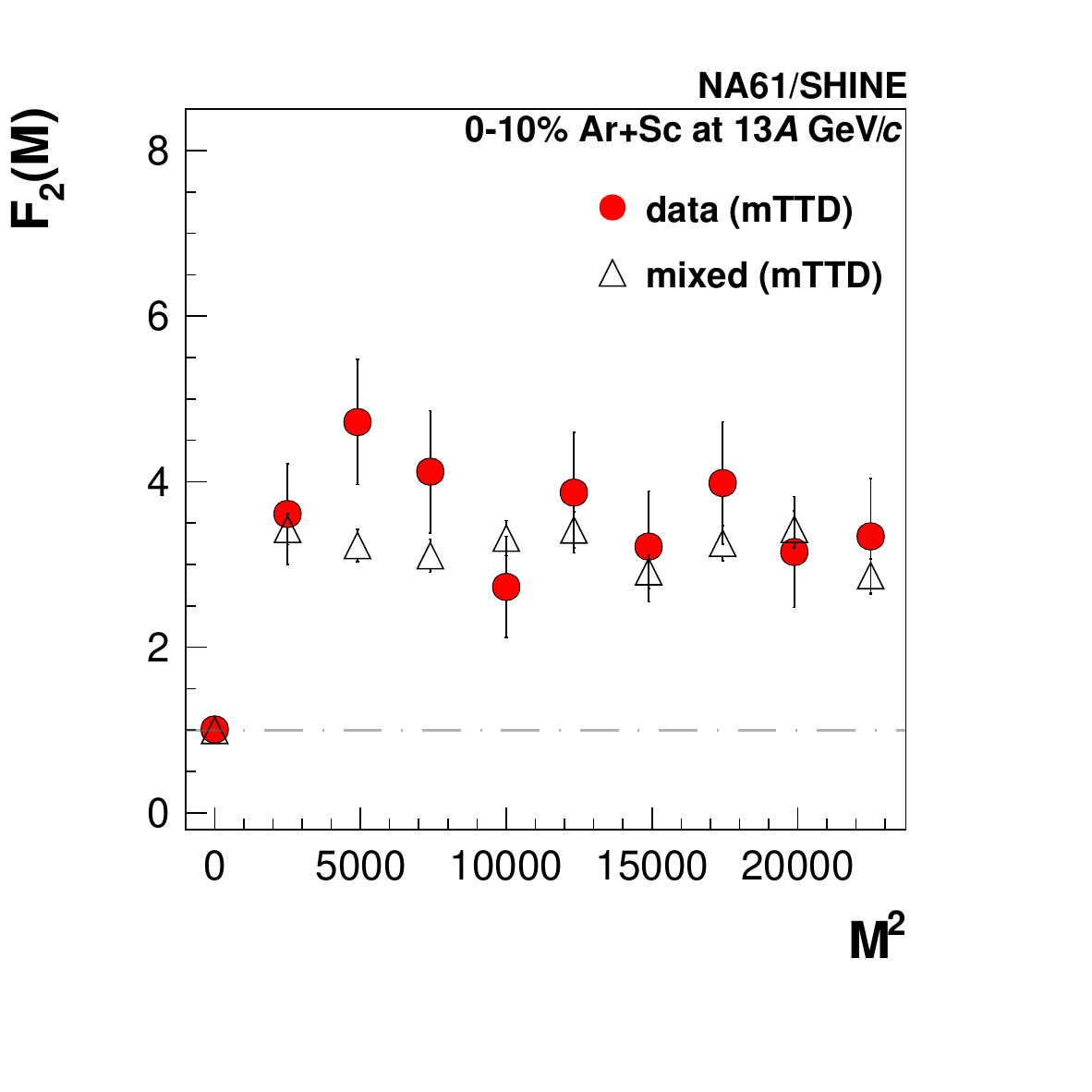}\hfill
    \includegraphics[width=.33\textwidth]{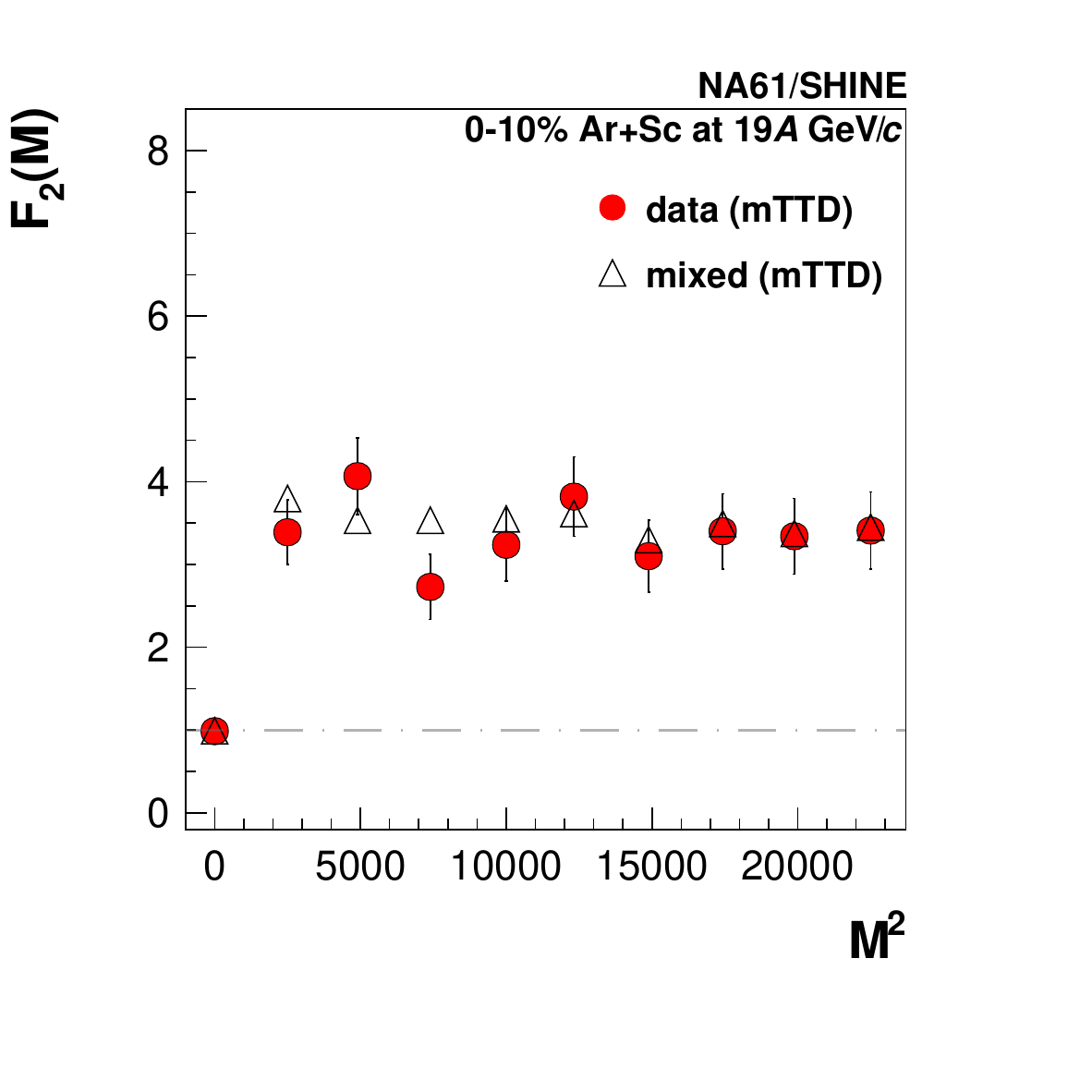}\hfill
    \includegraphics[width=.33\textwidth]{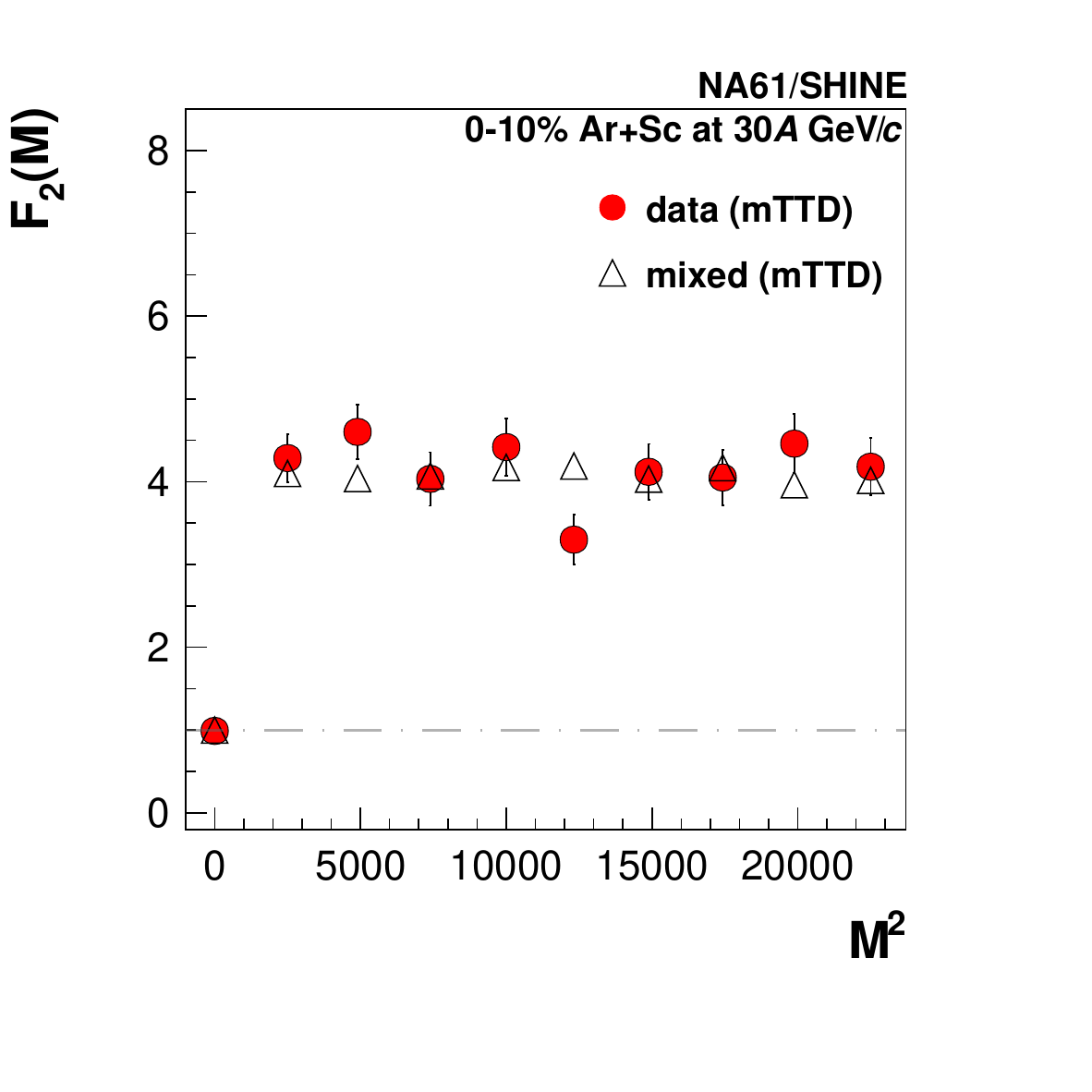}\\
    \includegraphics[width=.33\textwidth]{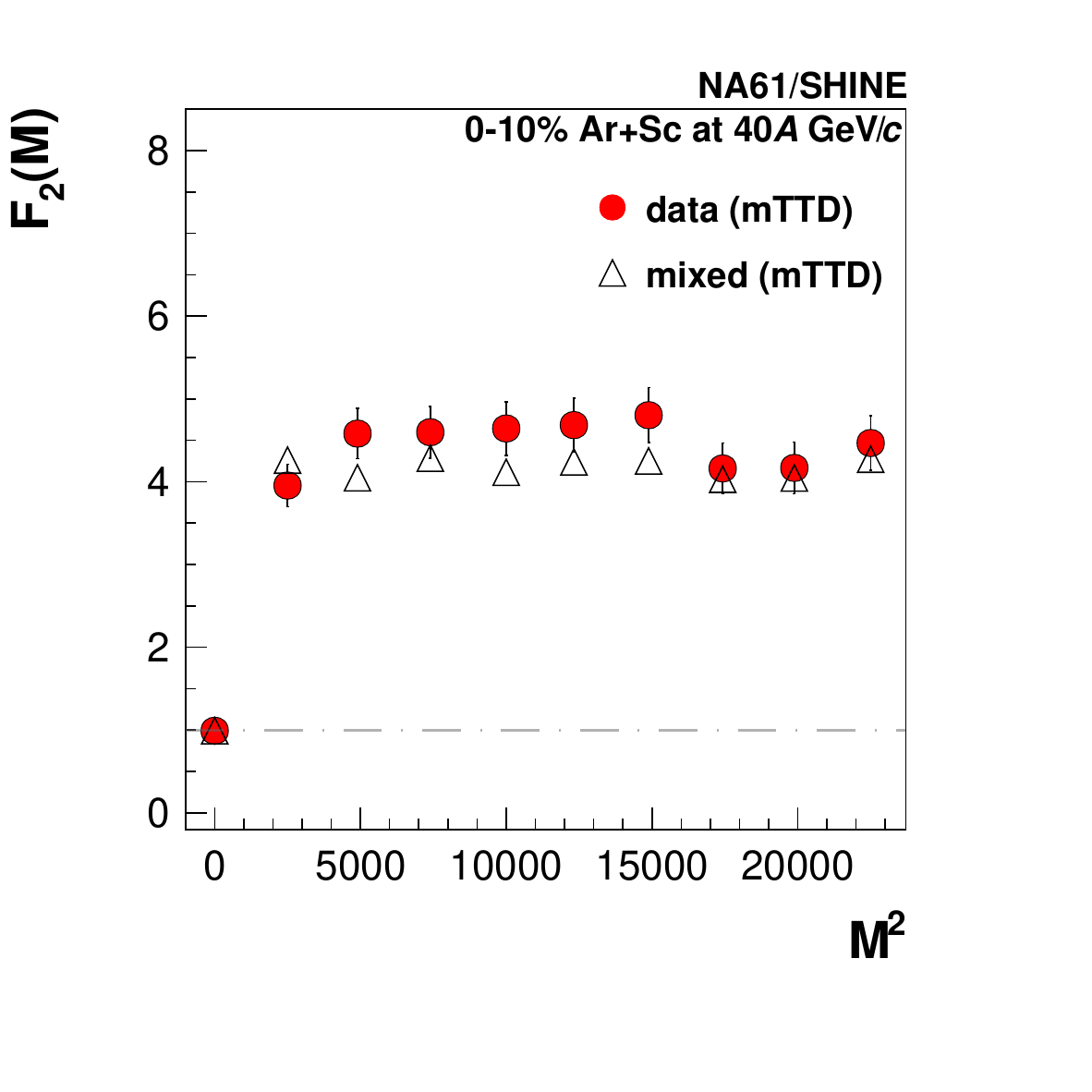}\qquad
    \includegraphics[width=.33\textwidth]{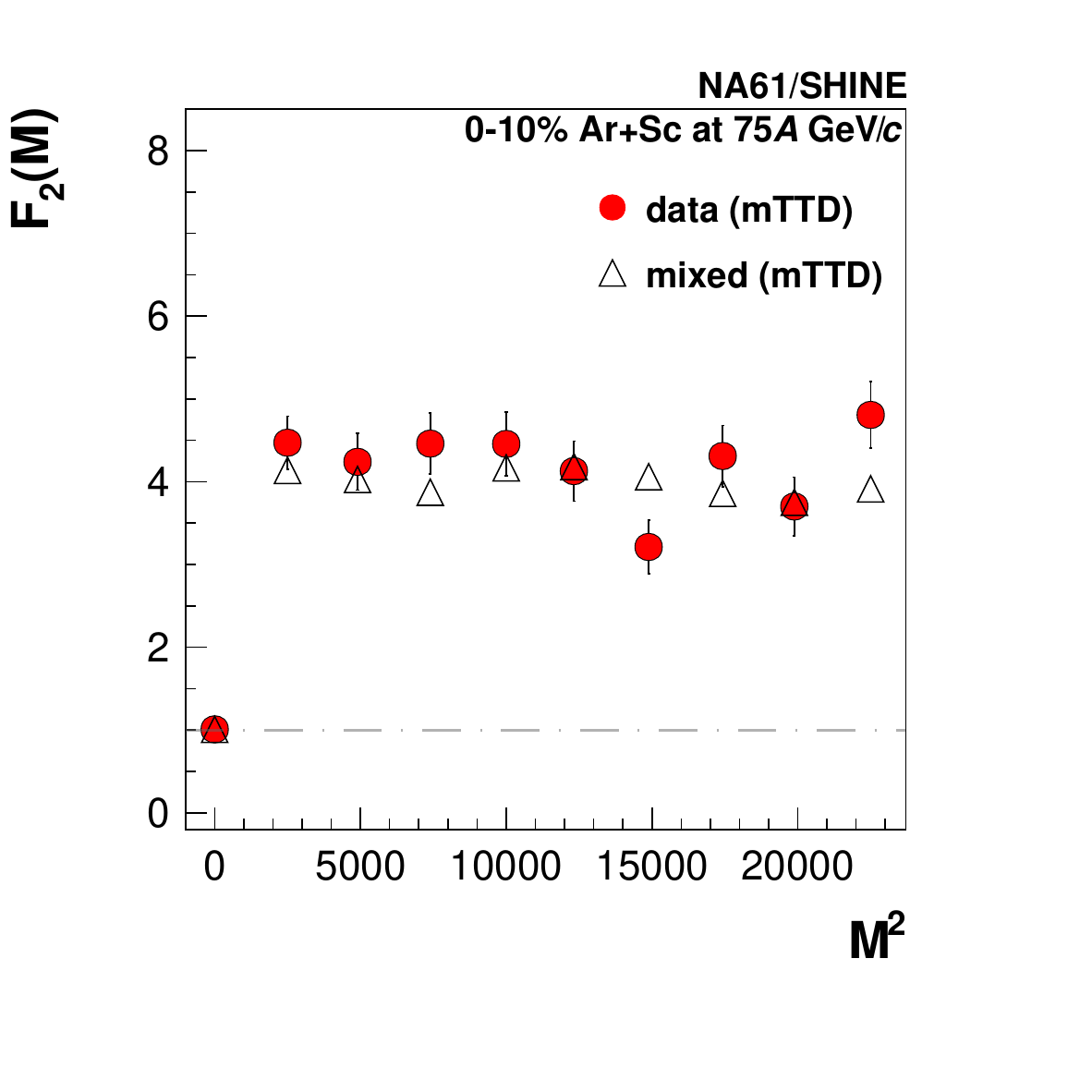}
    \rule{42em}{0.5pt}
    \caption{
       Results on the dependence of the scaled factorial moment of proton multiplicity distribution on the number of subdivisions in transverse momentum space $M^{2}$ for $1^{2} \leq M^{2} \leq 150^{2}$. Results are shown for 0--10\%  central \ArSc collisions at 13\A--75\AGeVc.  Closed red circles indicate the experimental data. Corresponding results for mixed events (open triangles) are also shown.  Both the data and mixed events include the mTTD cut. Only statistical uncertainties are indicated.
    }
    \label{fig:results-noncum-full}
\end{figure}

\begin{figure}[!hbt]
    \centering
    \includegraphics[width=.33\textwidth]{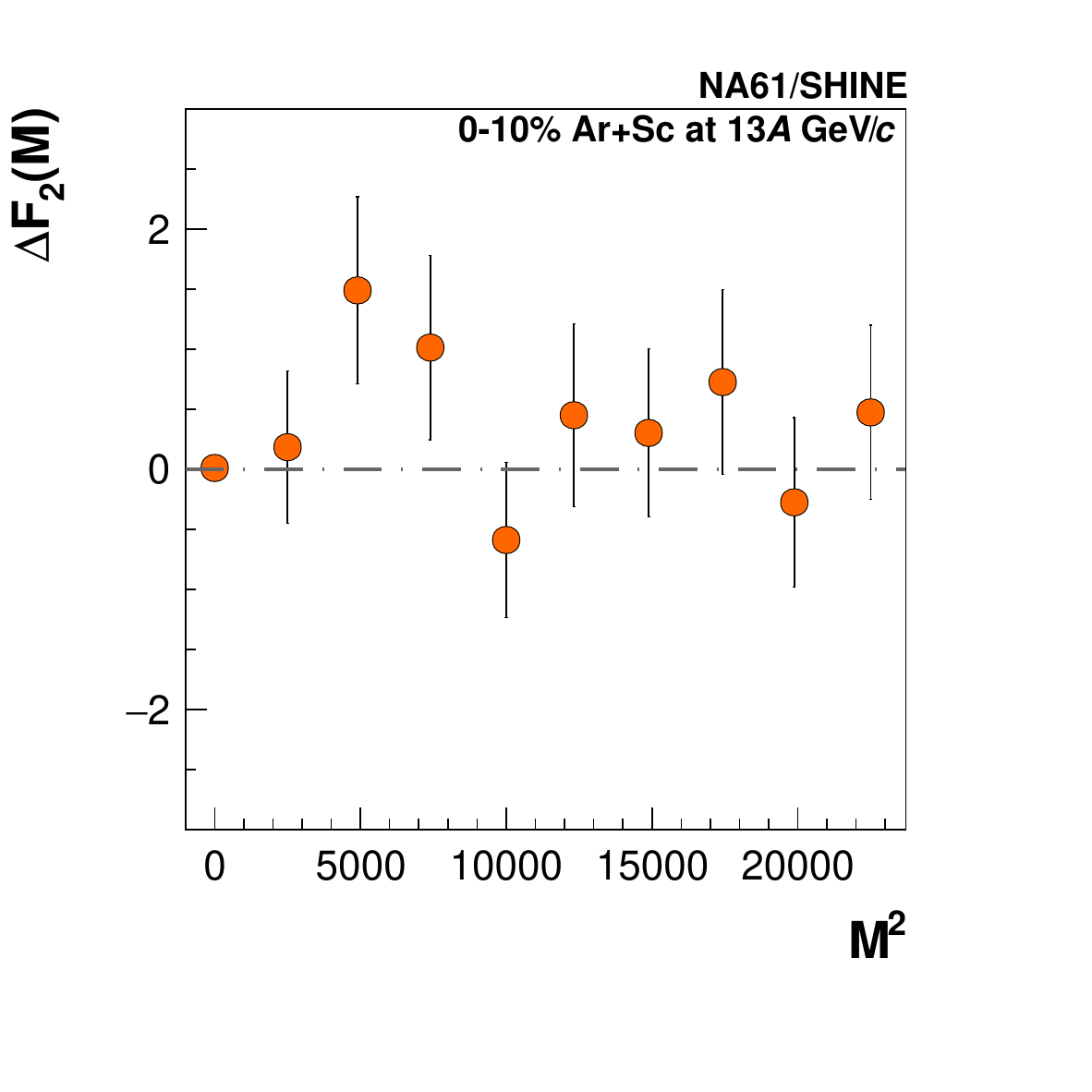}\hfill
    \includegraphics[width=.33\textwidth]{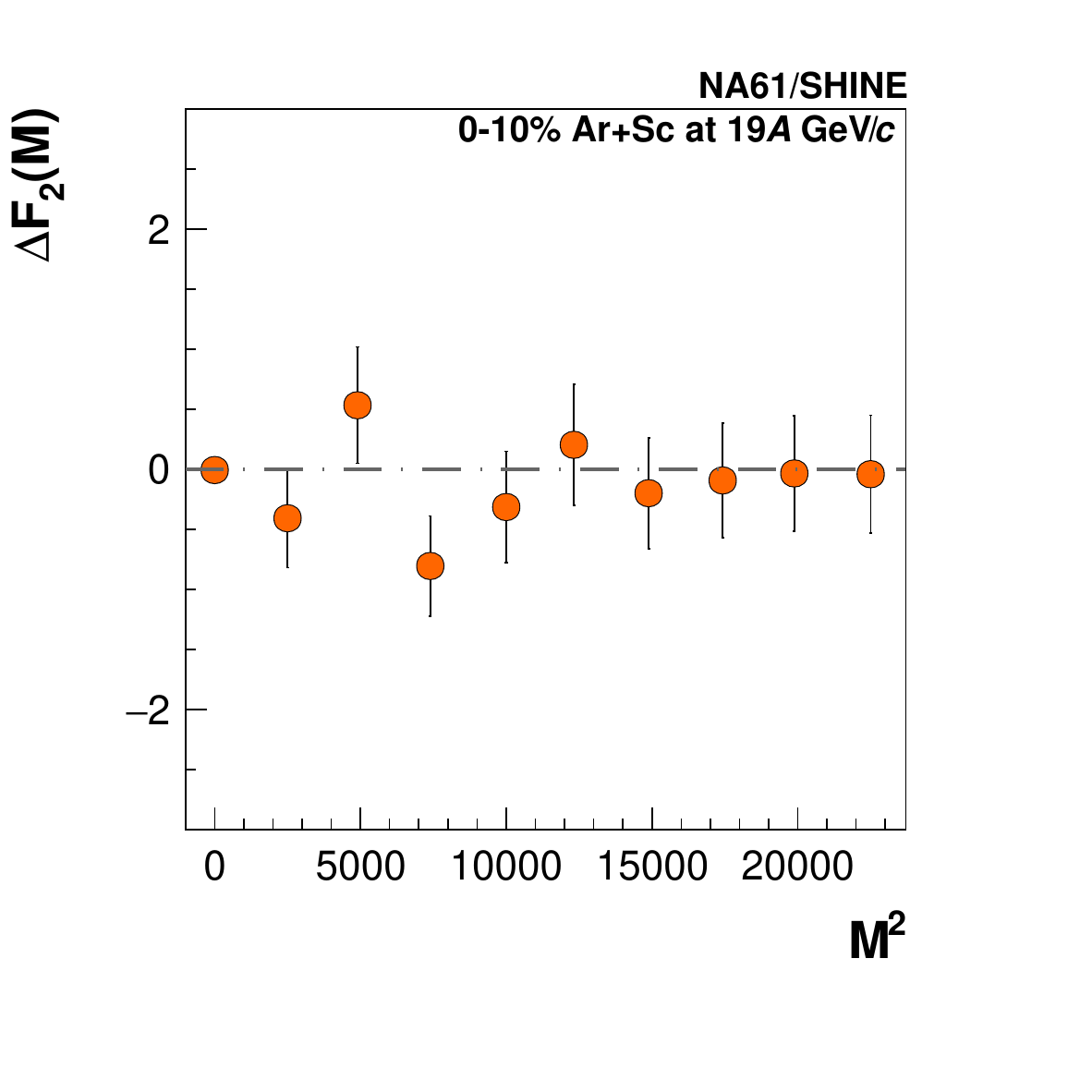}\hfill
    \includegraphics[width=.33\textwidth]{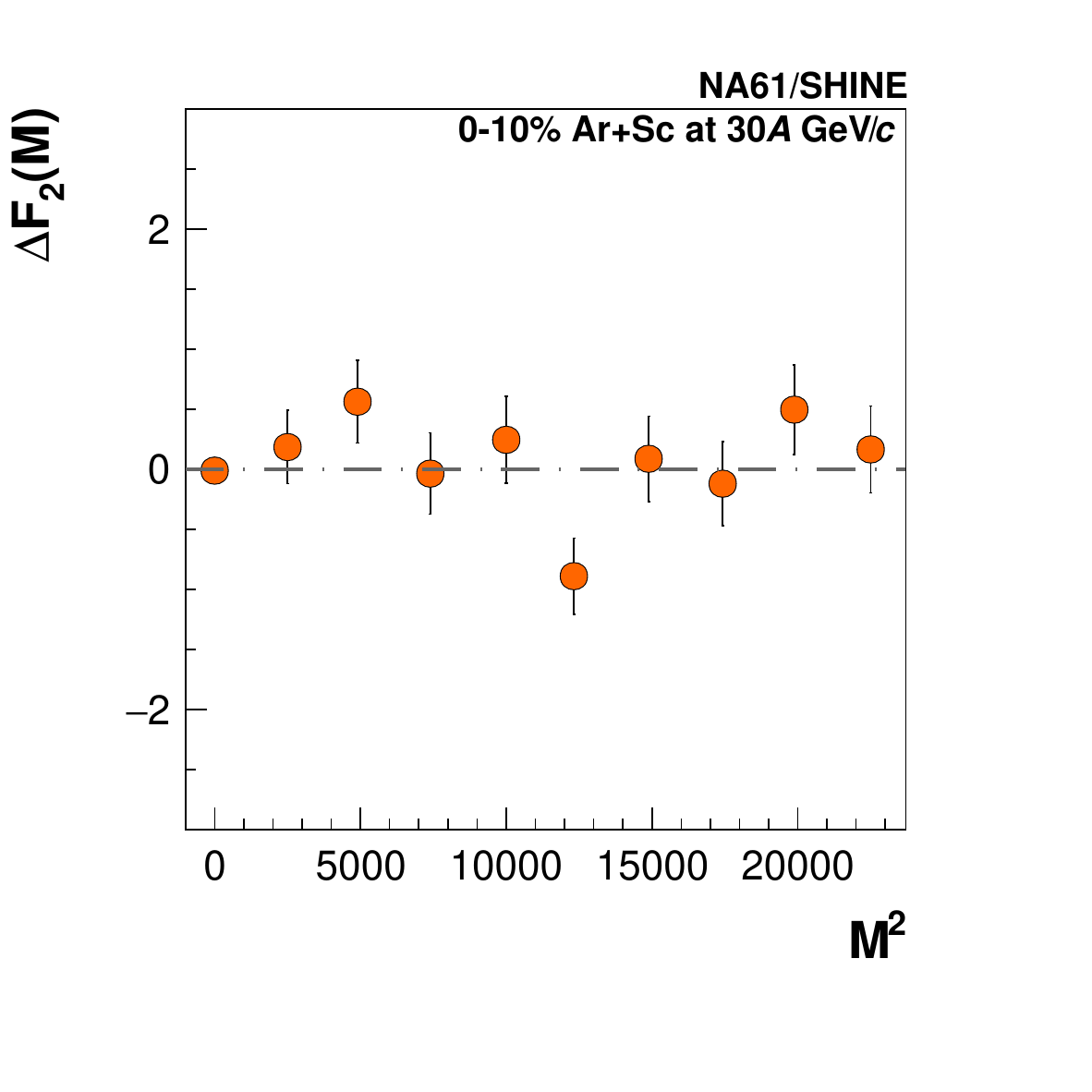}\\
    \includegraphics[width=.33\textwidth]{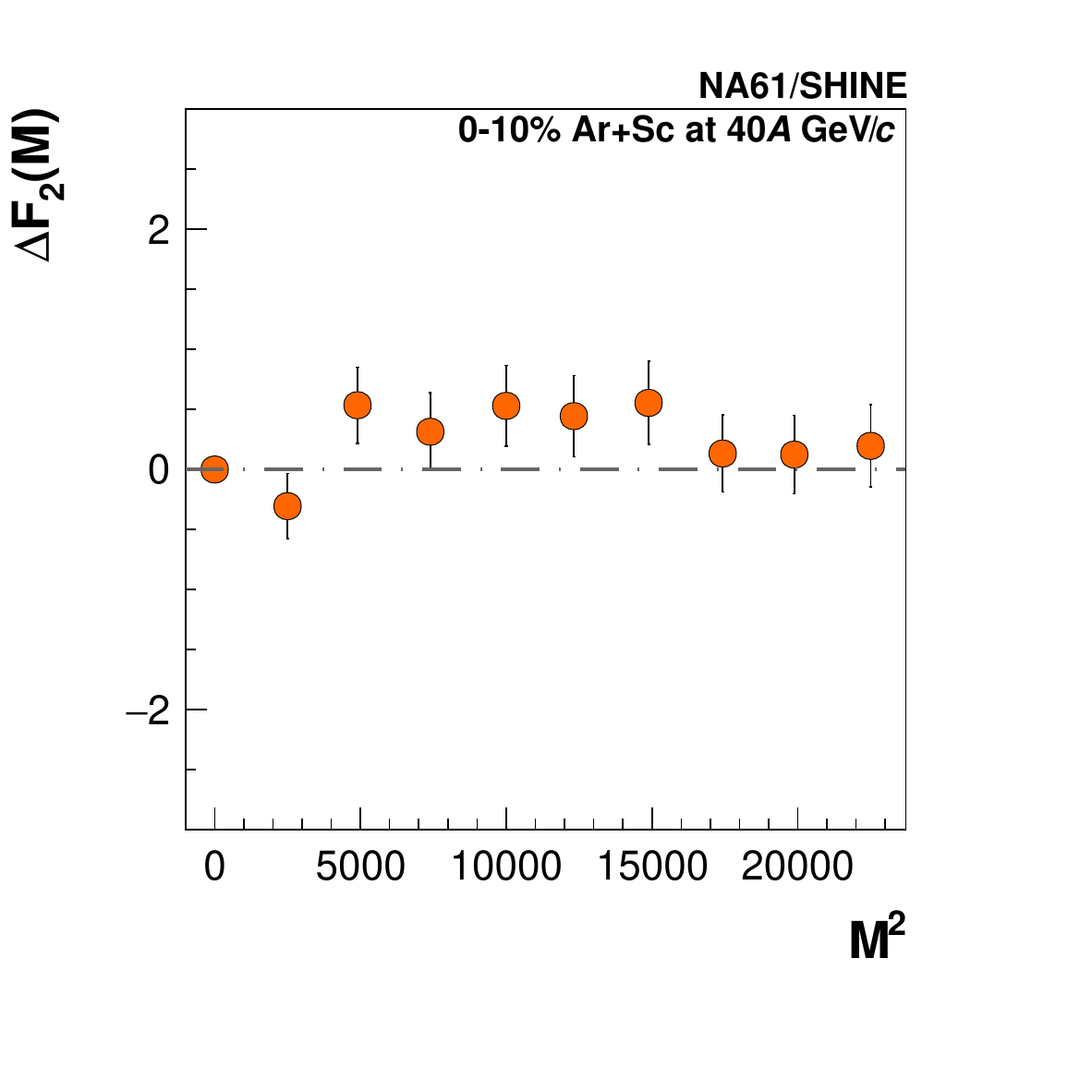}\qquad
    \includegraphics[width=.33\textwidth]{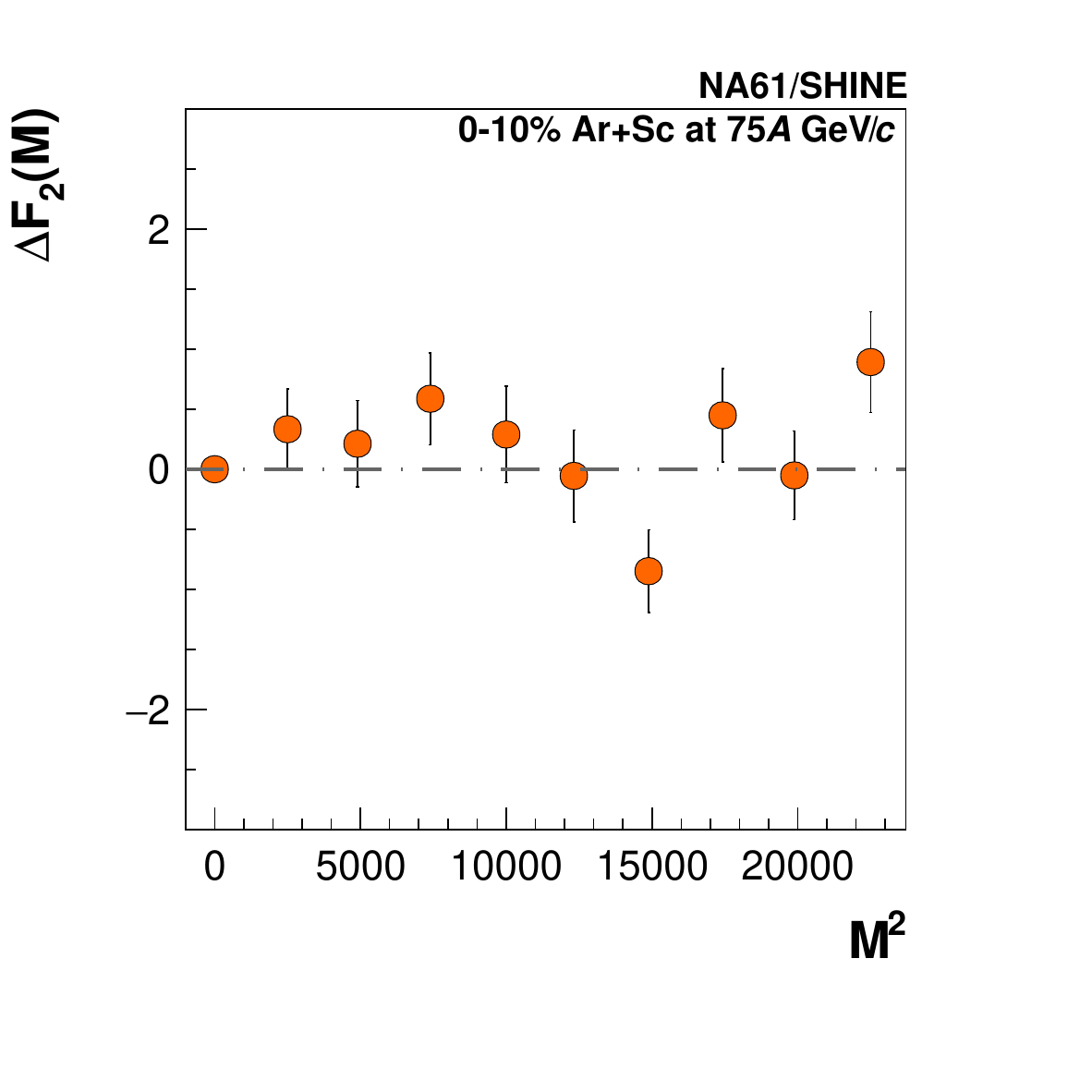}
     \rule{42em}{0.5pt}
    \caption{Results on the dependence of $\Delta F_{2}(M)$ of proton multiplicity distribution
    on the number of subdivisions in transverse momentum space $M^{2}$ for $1^{2} \leq M^{2} \leq 150^{2}$. Results for 0--10\% central \ArSc collisions at \mbox{13\A--75\AGeVc} are shown. Both the data and mixed events include the mTTD cut. Only statistical uncertainties are indicated.}
\label{fig:results-deltaF2-full-arsc}
\end{figure}

\begin{figure}[!ht]
    \centering
    \includegraphics[width=.33\textwidth]{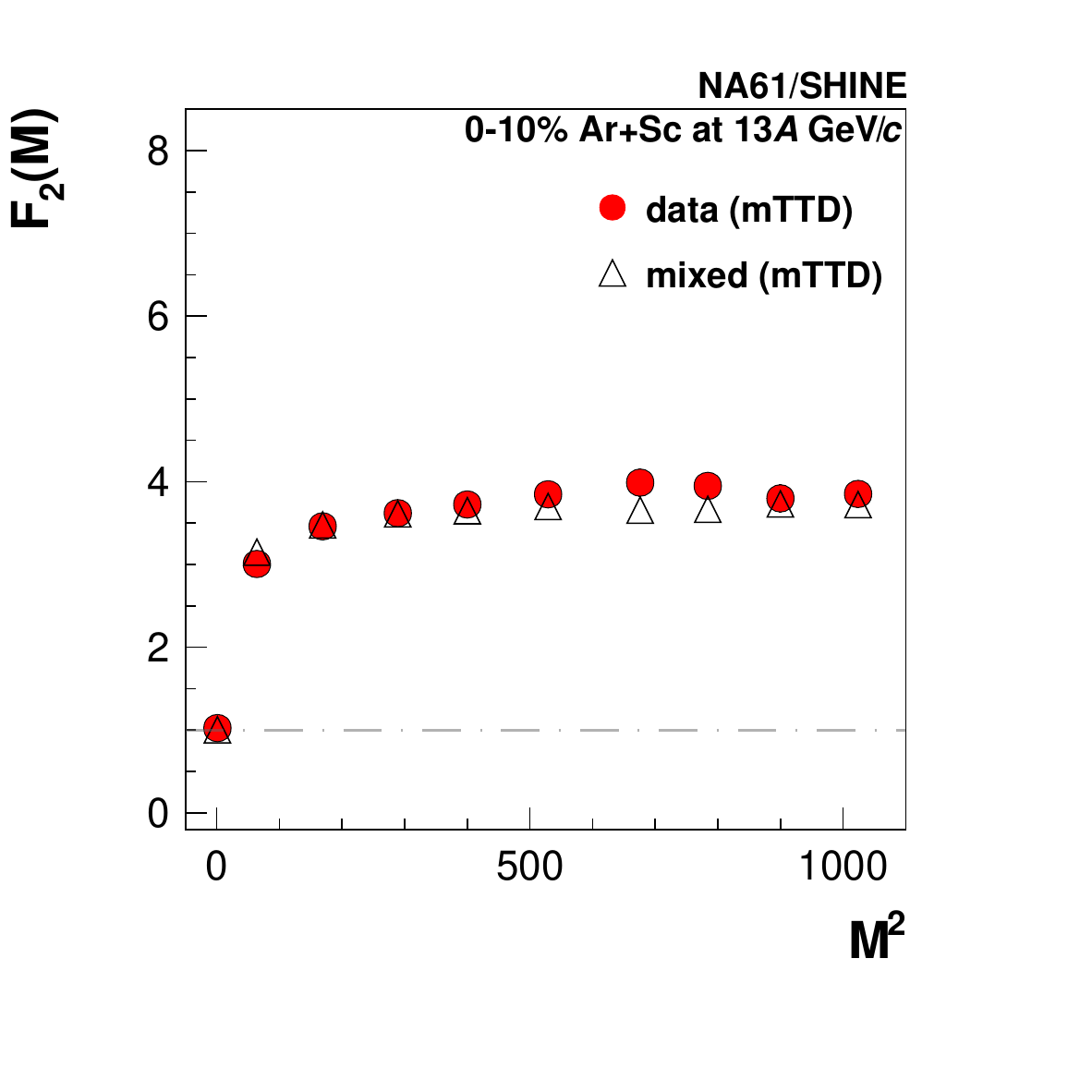}\hfill
    \includegraphics[width=.33\textwidth]{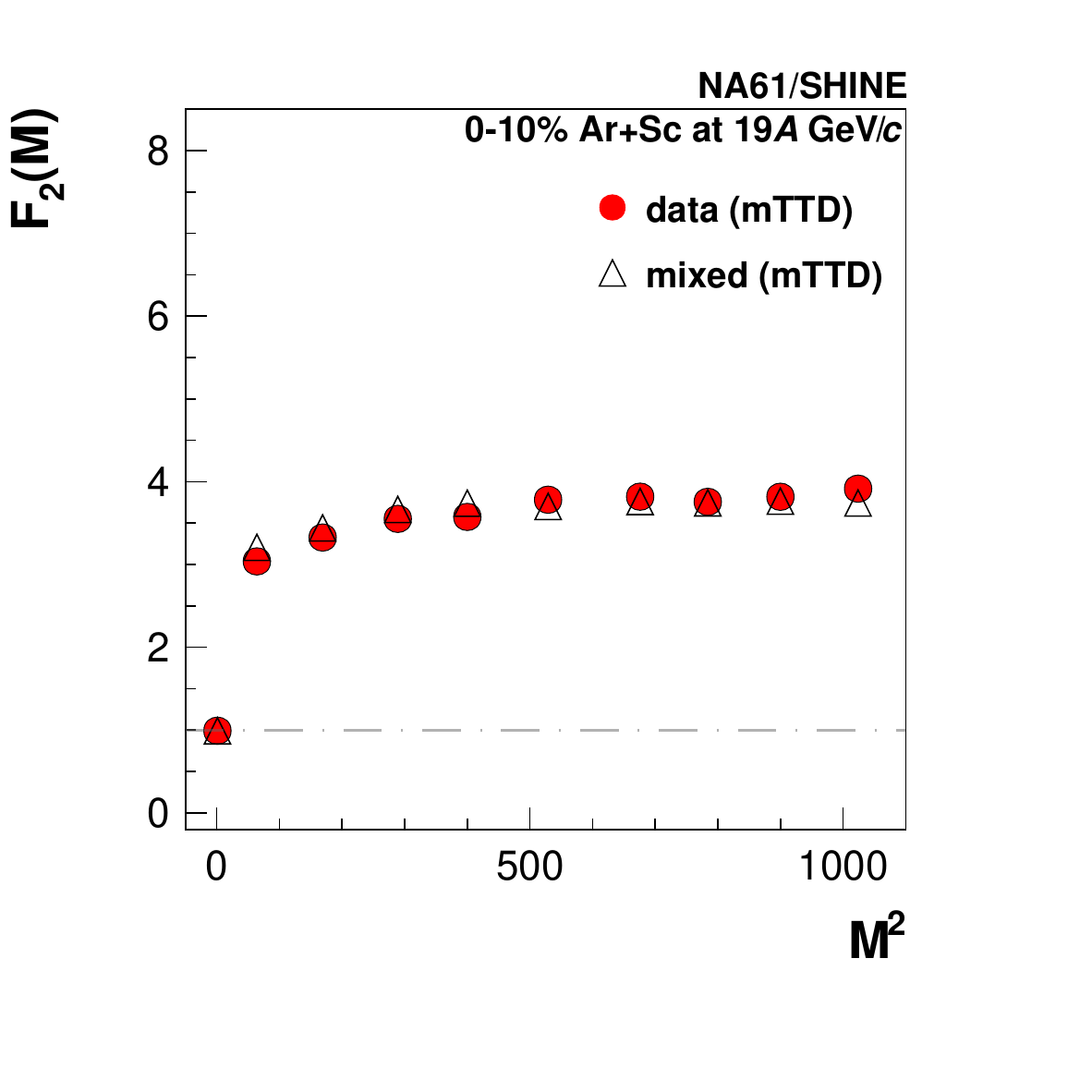}\hfill
    \includegraphics[width=.33\textwidth]{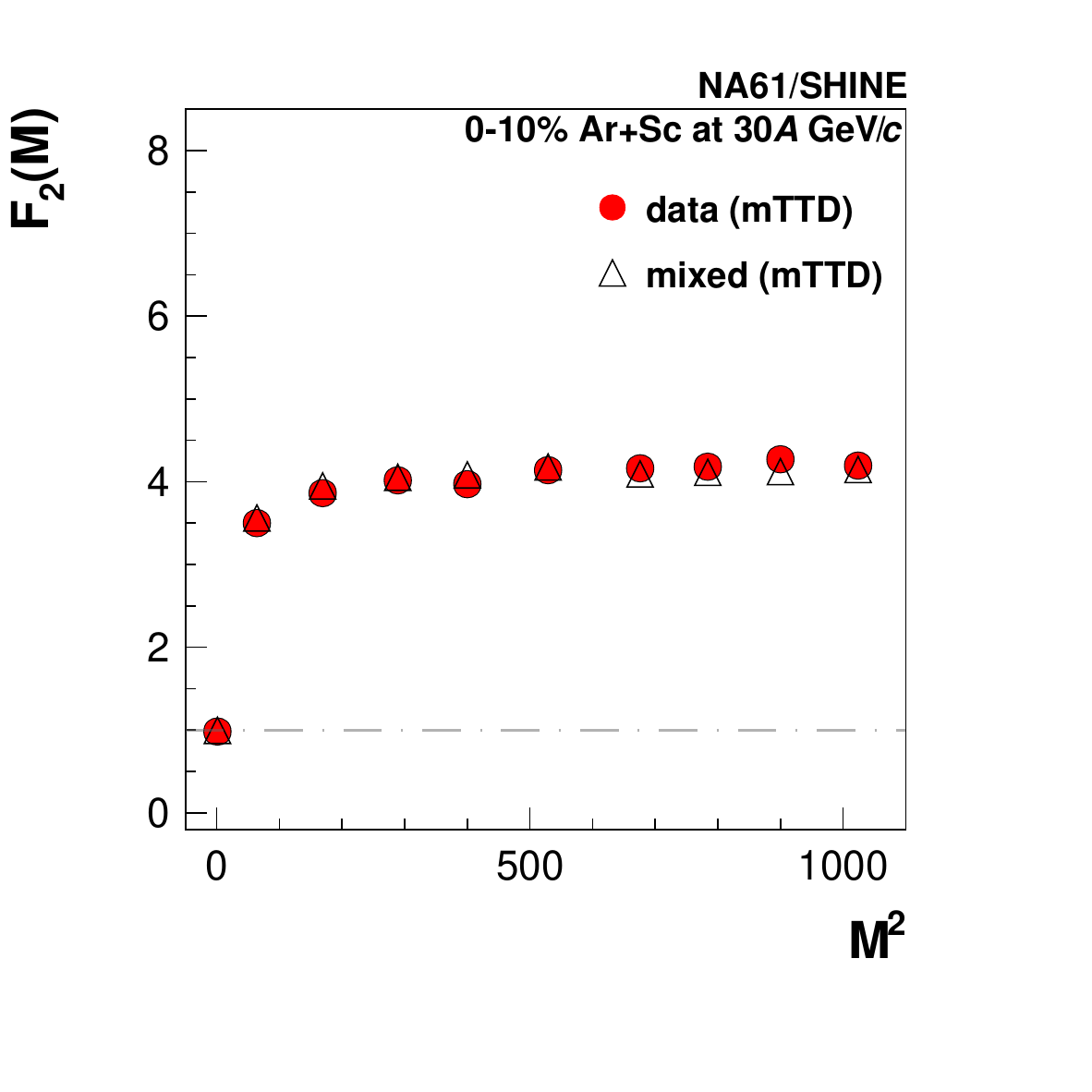}\\
    \includegraphics[width=.33\textwidth]{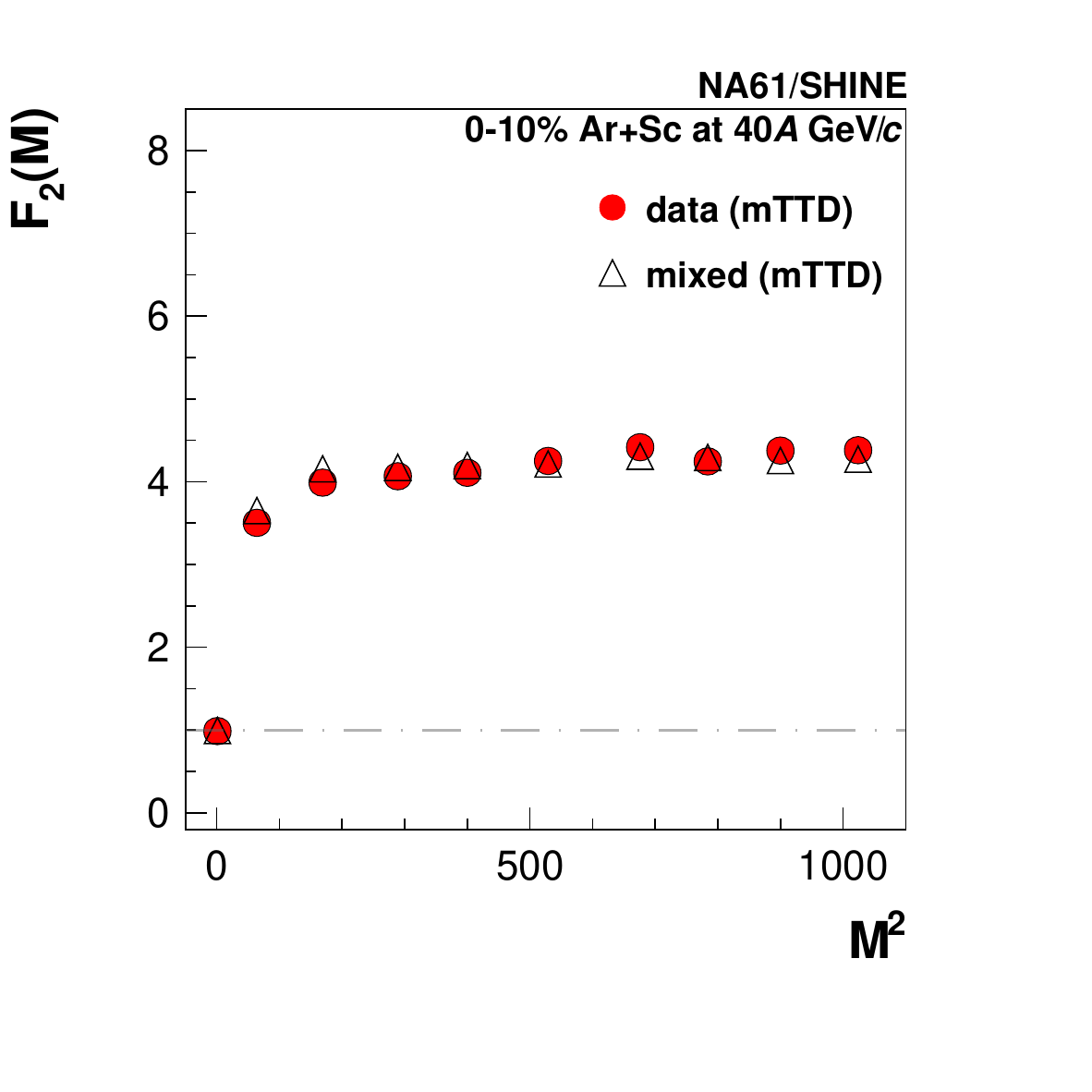}\qquad
    \includegraphics[width=.33\textwidth]{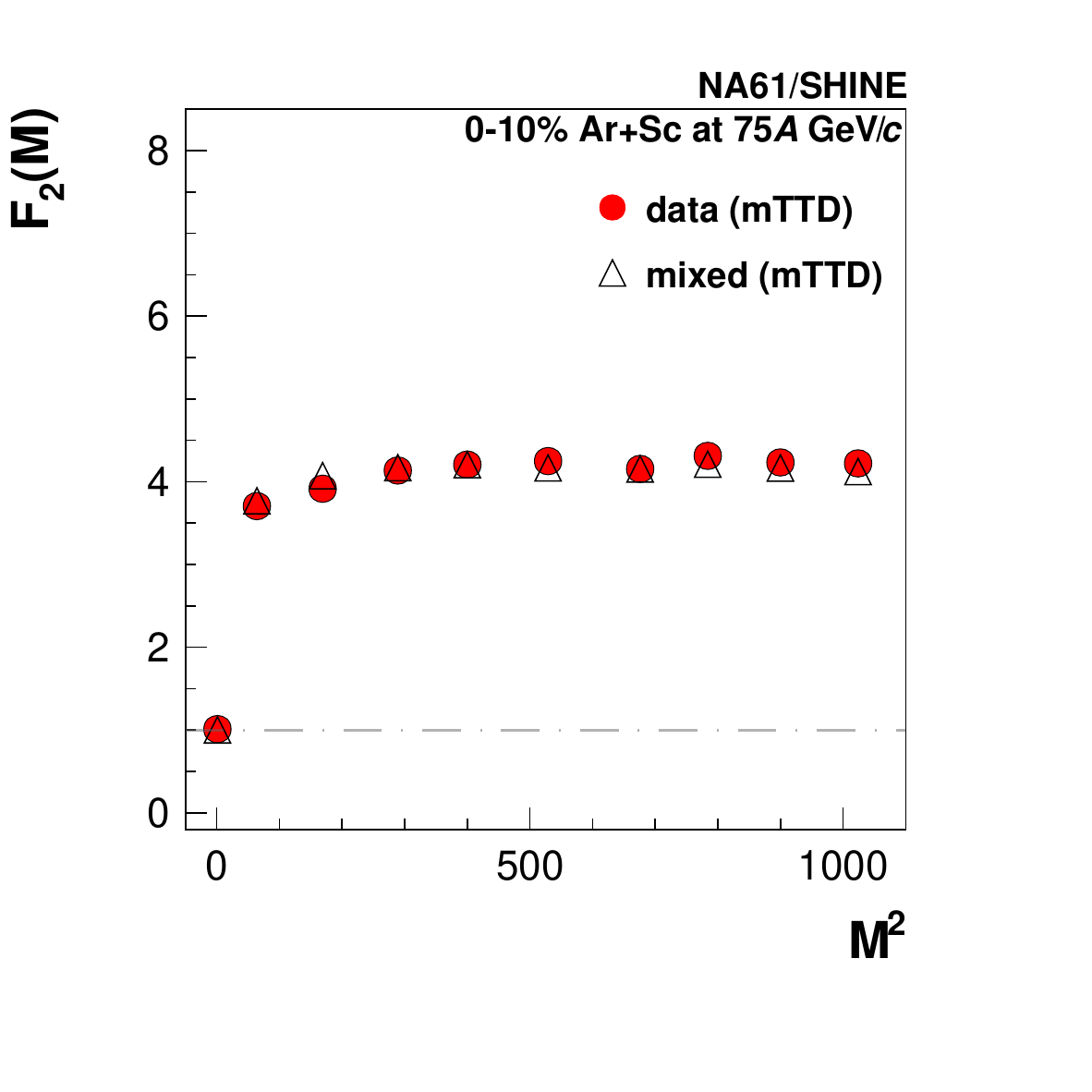}
    \rule{42em}{0.5pt}
    \caption{
       Results on the dependence of the scaled factorial moment of proton multiplicity distribution on the number of subdivisions in transverse momentum space $M^{2}$ for $1^{2} \leq M^{2} \leq 32^{2}$. Results are shown for 0--10\% central \ArSc collisions at 13\A--75\AGeVc. Closed red circles indicate the experimental data. Corresponding results for mixed events (open triangles) are also shown.  Both the data and mixed events include the mTTD cut. Only statistical uncertainties are indicated.
    }
    \label{fig:results-noncum-small}
\end{figure}

\begin{figure}[!hbt]
    \centering
    \includegraphics[width=.33\textwidth]{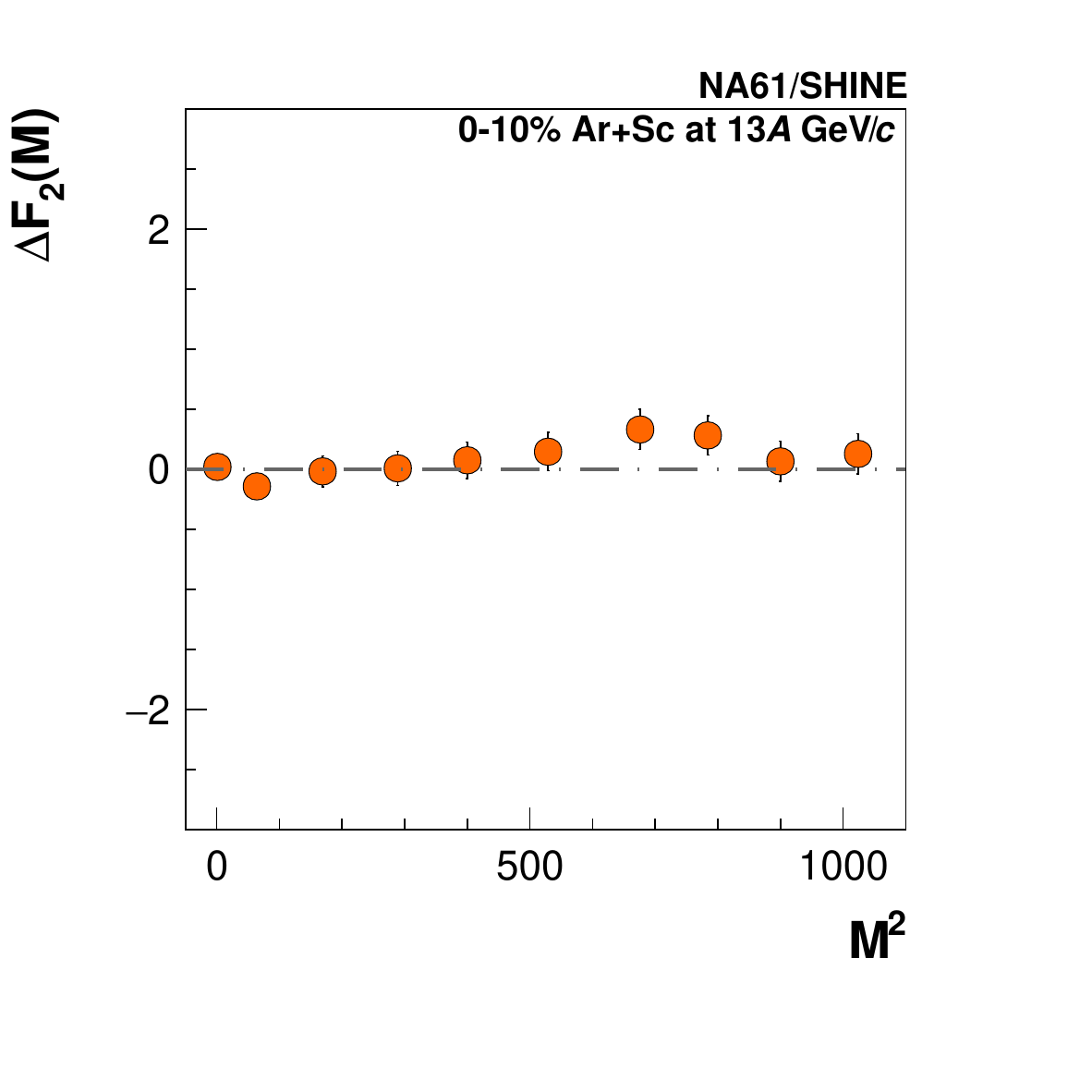}\hfill
    \includegraphics[width=.33\textwidth]{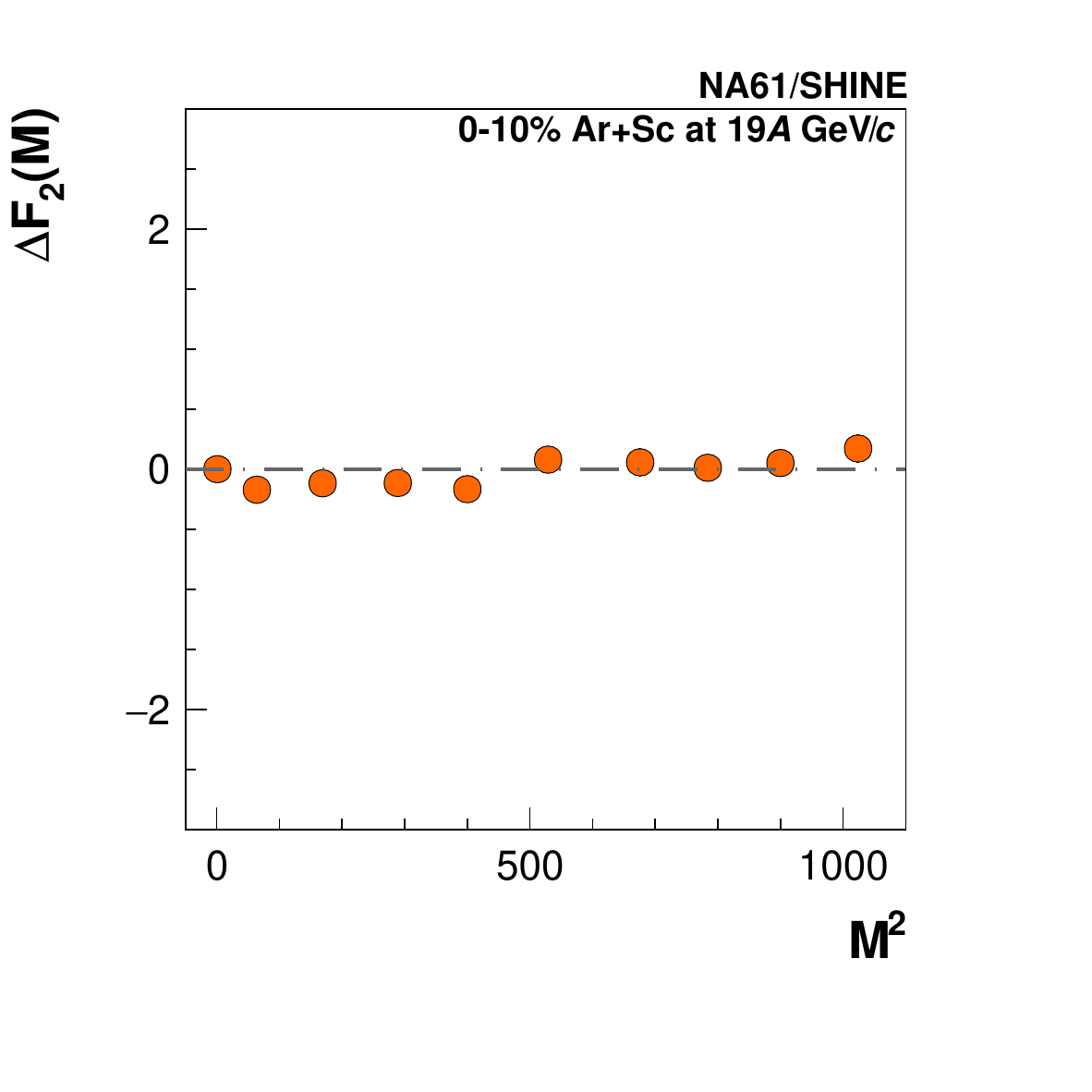}\hfill
    \includegraphics[width=.33\textwidth]{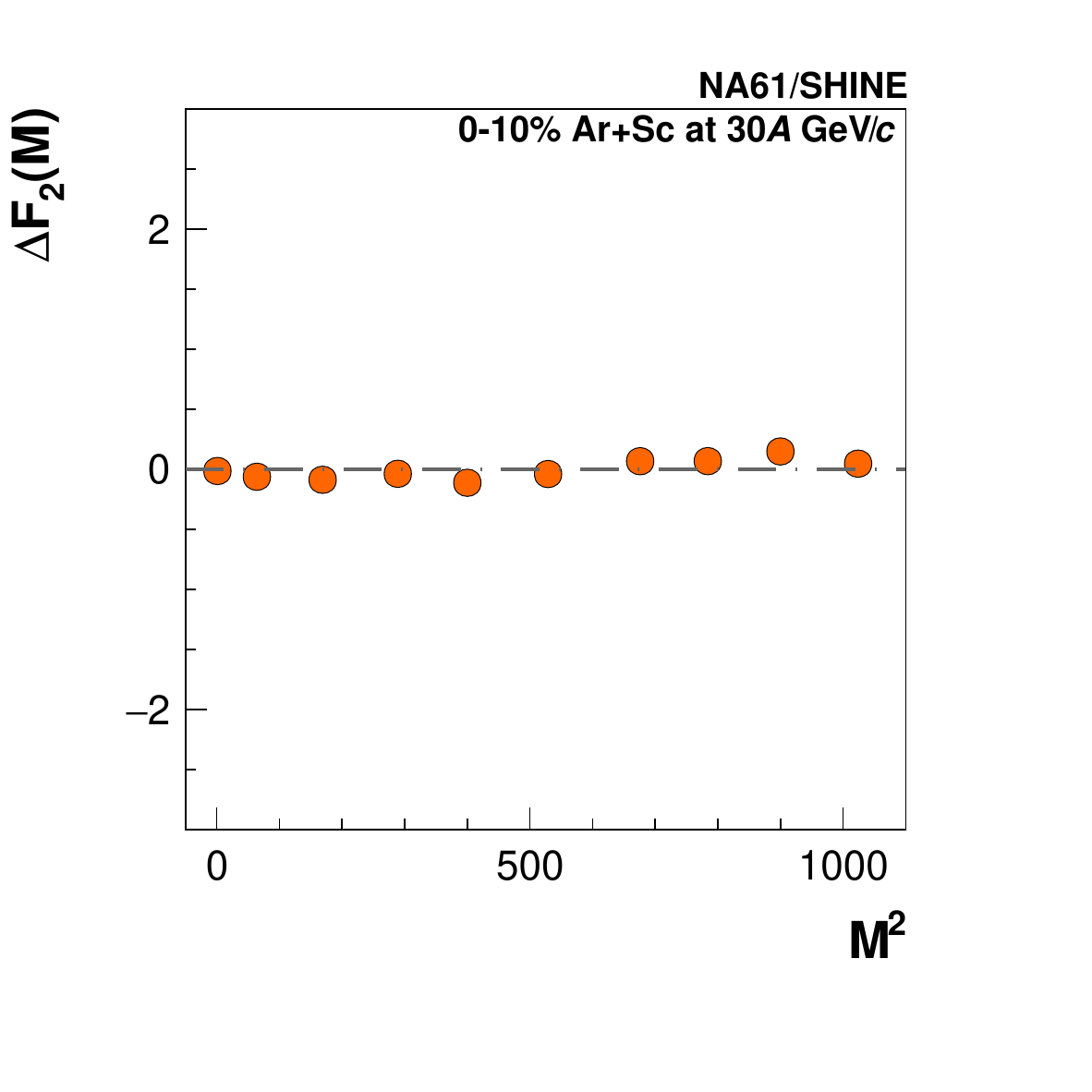}\\
    \includegraphics[width=.33\textwidth]{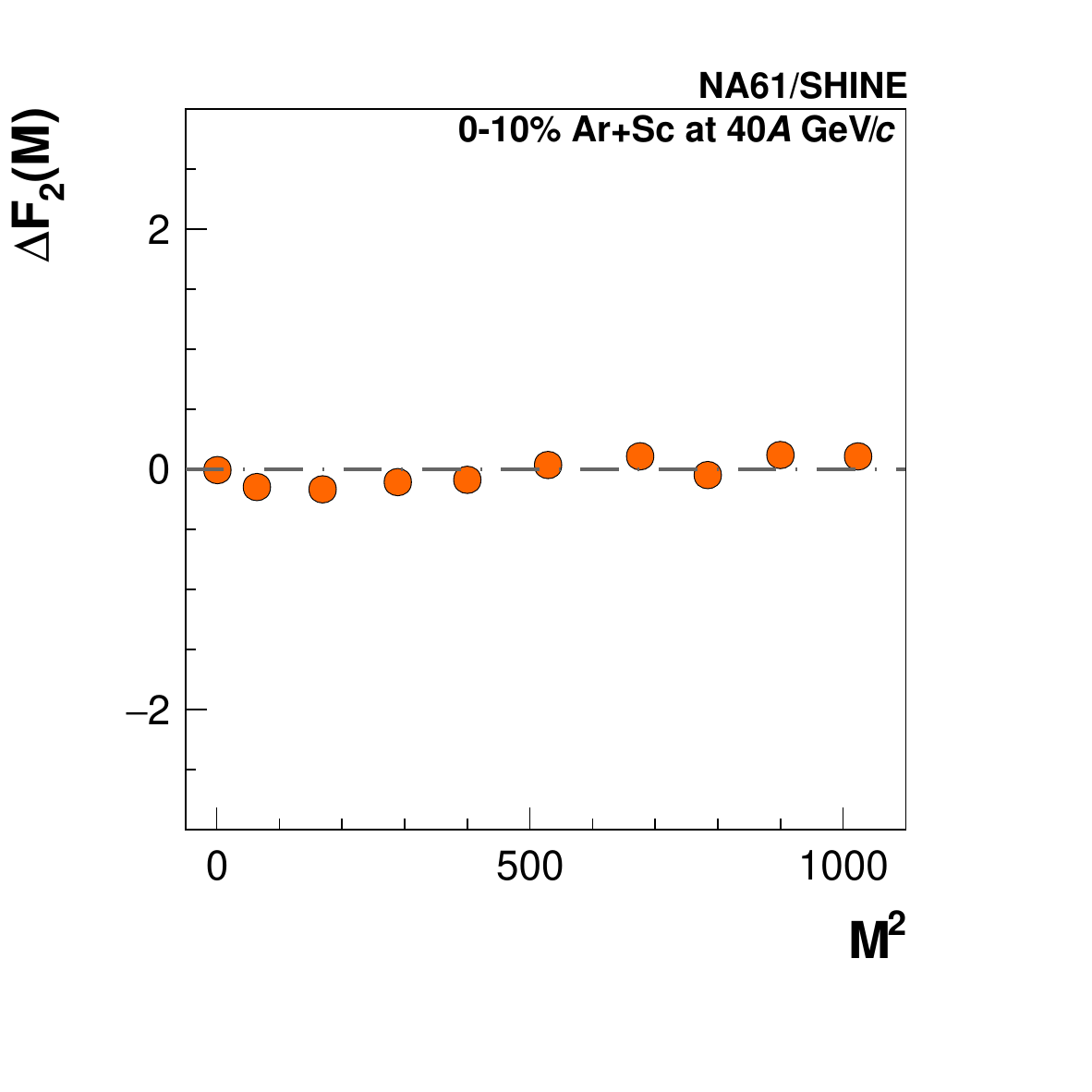}\qquad
    \includegraphics[width=.33\textwidth]{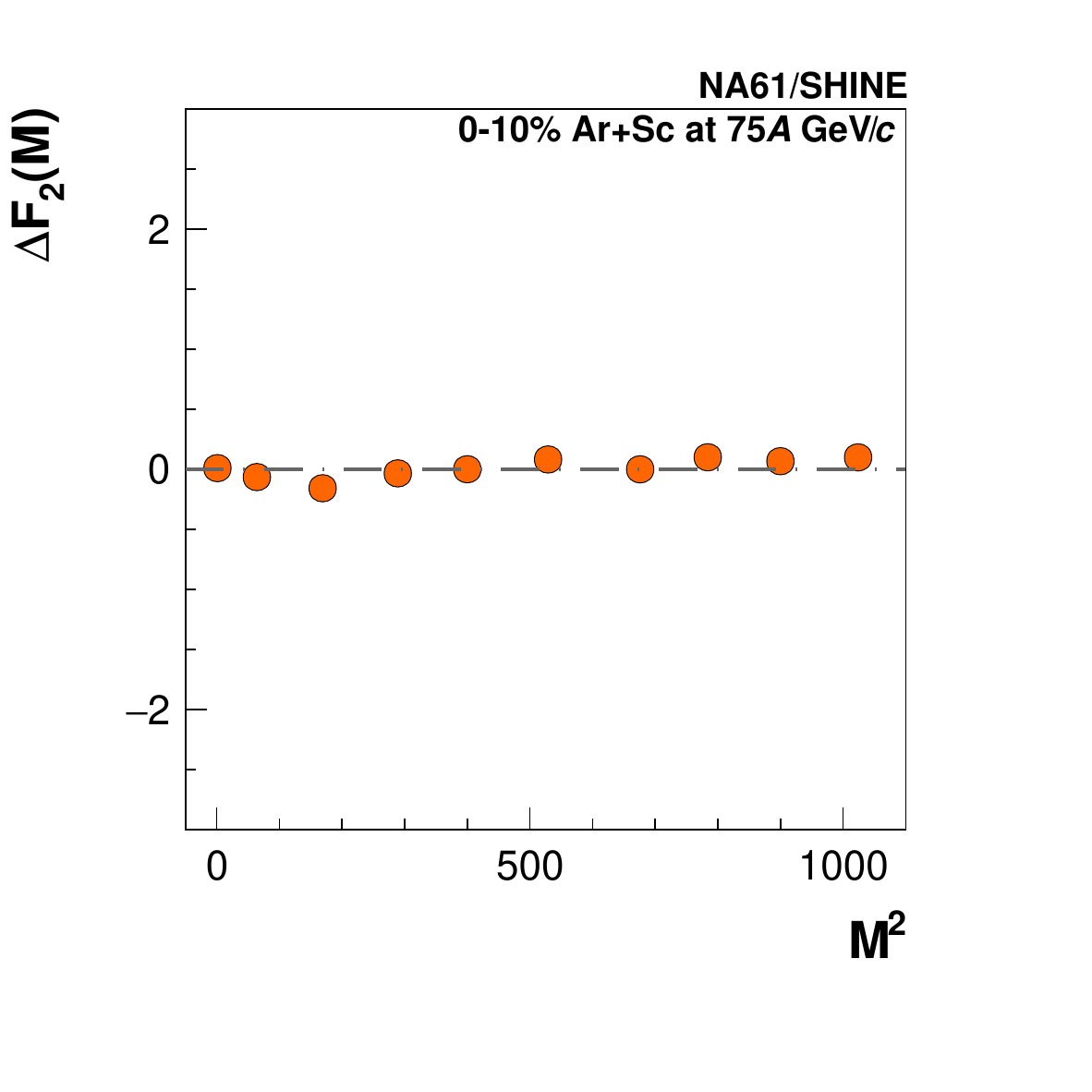}
    \rule{42em}{0.5pt}
    \caption{Results on the dependence of $\Delta F_{2}(M)$ of proton multiplicity distribution
    on the number of subdivisions in transverse momentum space $M^{2}$ for $1^{2} \leq M^{2} \leq 32^{2}$. Results for 0--10\% central \ArSc collisions at \mbox{13\A--75\AGeVc} are shown. Both the data and mixed events include the mTTD cut. Only statistical uncertainties are indicated.}
    \label{fig:results-deltaF2-Small-arsc}
\end{figure}

Figure~\ref{fig:results-noncum-full} and~\ref{fig:results-noncum-small} present the results which correspond to the results shown in Figs.~\ref{fig:results-cum-full}, and~\ref{fig:results-cum-small} but subdivisions are done in the transverse momentum space.
By construction, $F_2(1)$ values are equal for subdivisions in cumulative 
transverse-momentum space and transverse-momentum space. But for the latter, $F_2(M)$ rapidly increases from the value $F_{2}(M=1)$ to approximate plateau at $M\approx 20$. This dependence is primarily due to the non-uniform shape of the single-particle transverse momentum distributions, see Sec.~\ref{sec:sfm_cumulative}. It can be accounted for by comparing the experimental data results with those obtained for the mixed events using $\Delta F_{2}(M)$. The dependence of $\Delta F_{2}(M)$ on the number of sub-divisions, $M^{2}$ are shown in Figs.~\ref{fig:results-deltaF2-full-arsc} and~\ref{fig:results-deltaF2-Small-arsc} for fine and coarse binning.

The experimental results presented in Figs.~\ref{fig:results-noncum-full}--\ref{fig:results-deltaF2-Small-arsc} do not show any significant difference to the results for mixed events with the mTTD cut on $M^{2}$ ($\Delta F_{2}(M)\approx0$). There is no indication of the critical fluctuations for selected protons.

\clearpage

\section{Comparison with models}
\label{sec:models}

This section presents a comparison of the experimental results with
two models. The first one, \EposLong~\cite{Werner:2008zza}, takes into account numerous sources of particle correlations, in particular, conservation laws and resonance decays, but without critical fluctuations. The second one, the Power-law Model~\cite{Czopowicz:2022}, produces pairs of particles correlated by the
power law inside the pair and fully uncorrelated particles.

\subsection{\Epos}
\label{sec:models-epos}

Minimum bias \ArSc events have been generated with \EposLong for comparison. Signals from the \NASixtyOne detector were simulated with \GeantThree software, and the generated events were reconstructed using the standard \NASixtyOne procedure.

 The number of forward spectators is used to select central \EposLong events of \ArSc collisions at 13\A--75\AGeVc. To calculate model predictions (pure \Epos), 0--10\% most central collisions
were analyzed. Protons and proton pairs within the single-particle and
two-particle acceptance maps were selected. Moreover, 60\% of accepted protons were randomly selected for the analysis to account for the limited proton identification in the experiment (see Fig.~\ref{fig:dEdx}).

\begin{figure}[!ht]
    \centering
    \includegraphics[width=.33\textwidth]{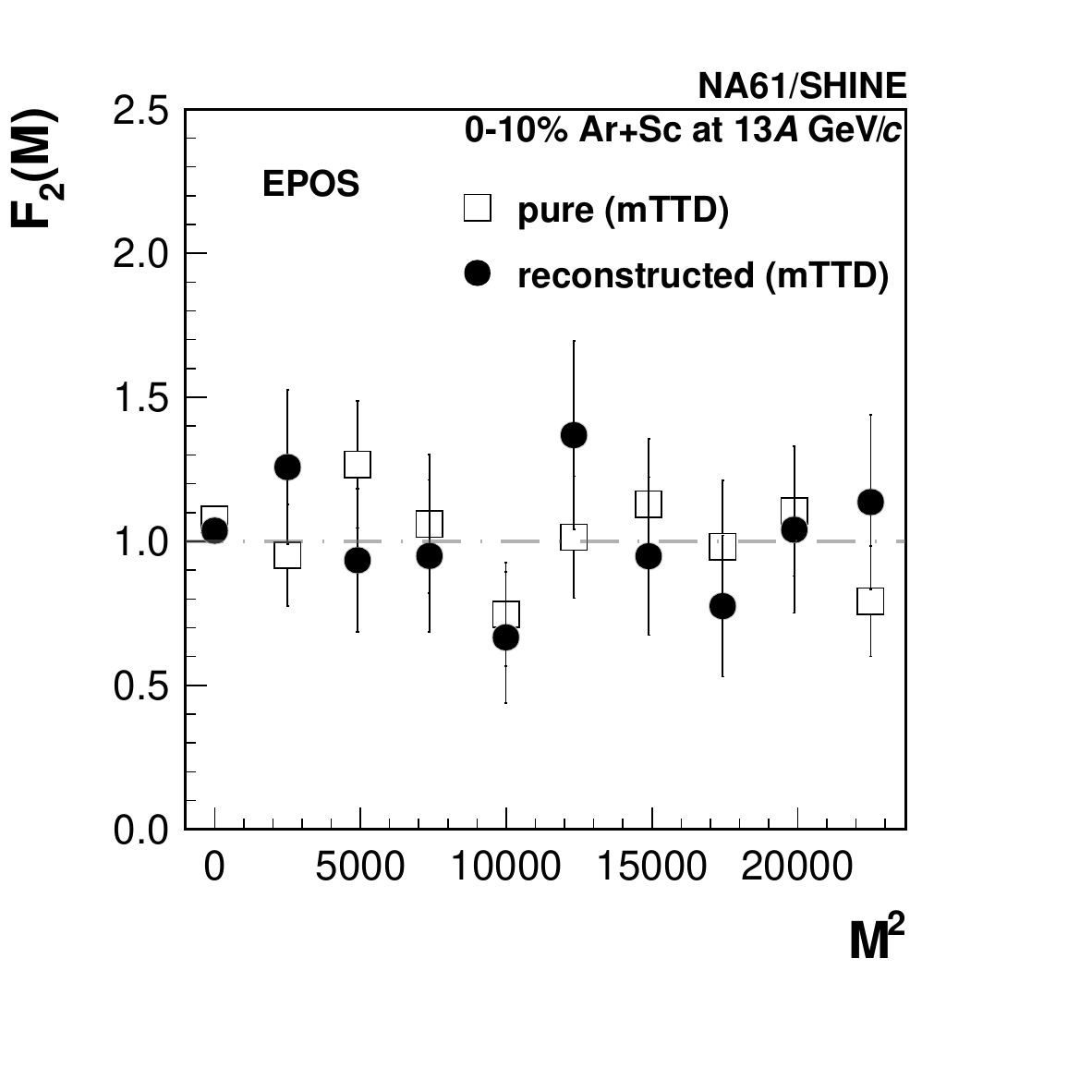}\hfill
    \includegraphics[width=.33\textwidth]{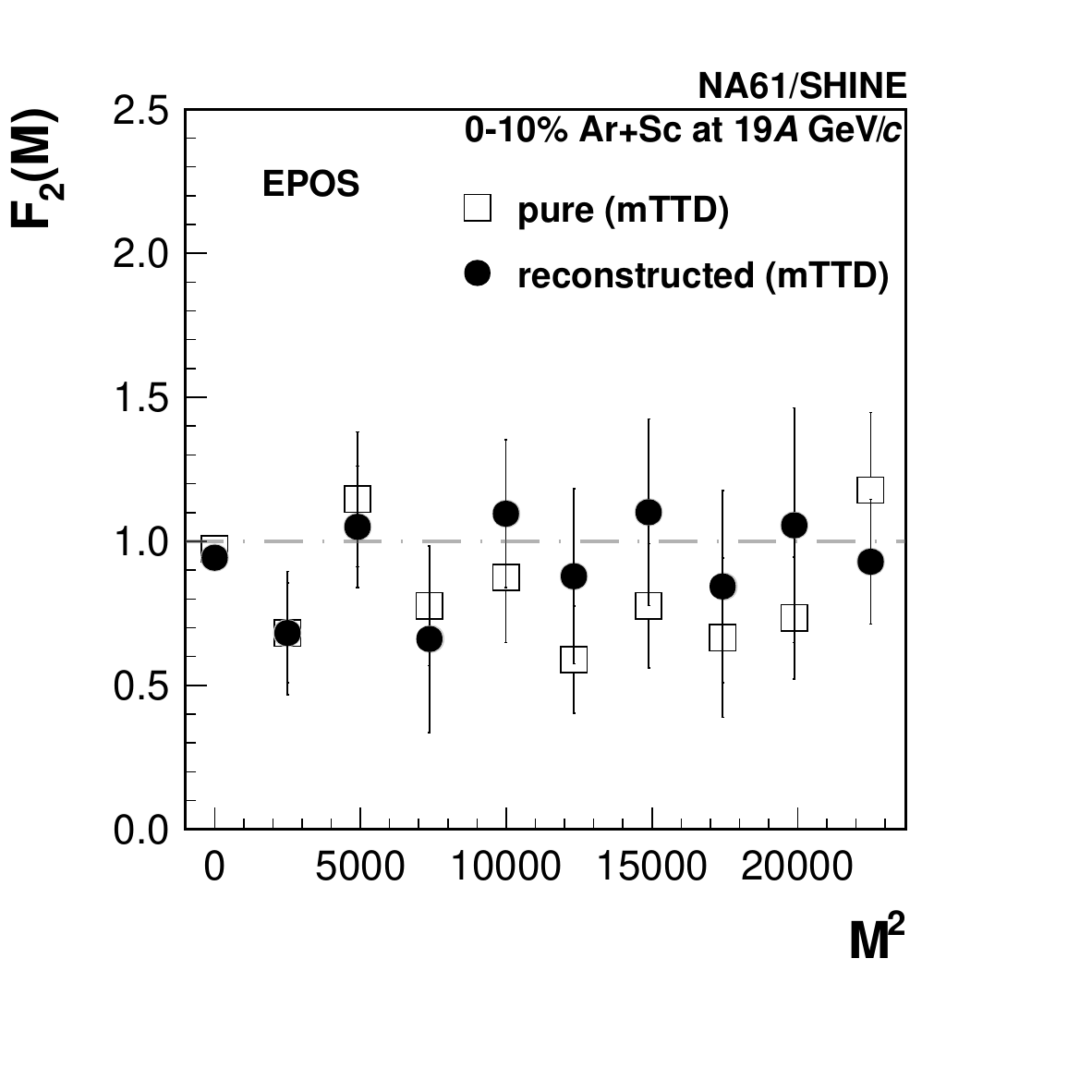}\hfill
    \includegraphics[width=.33\textwidth]{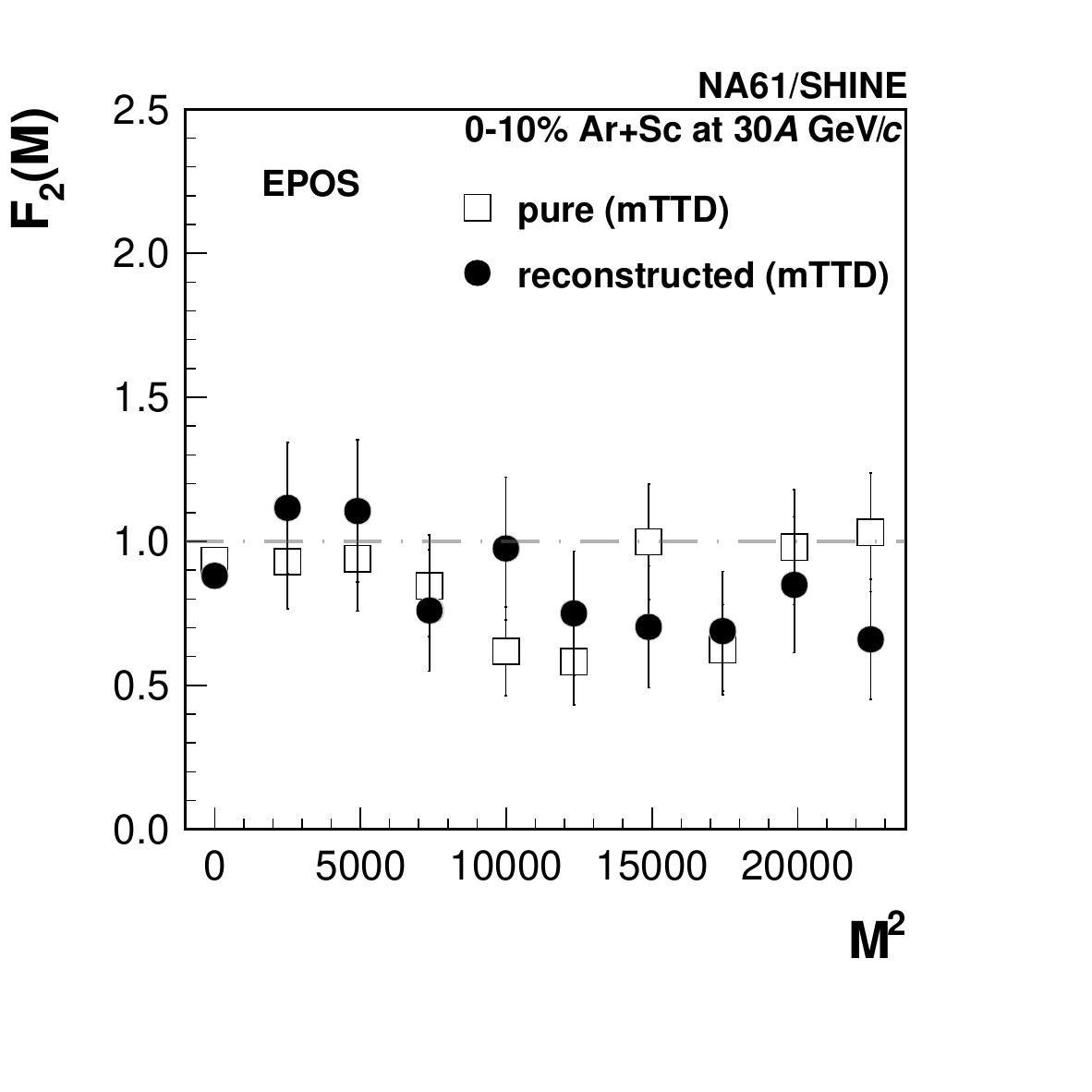}\\
    \includegraphics[width=.33\textwidth]{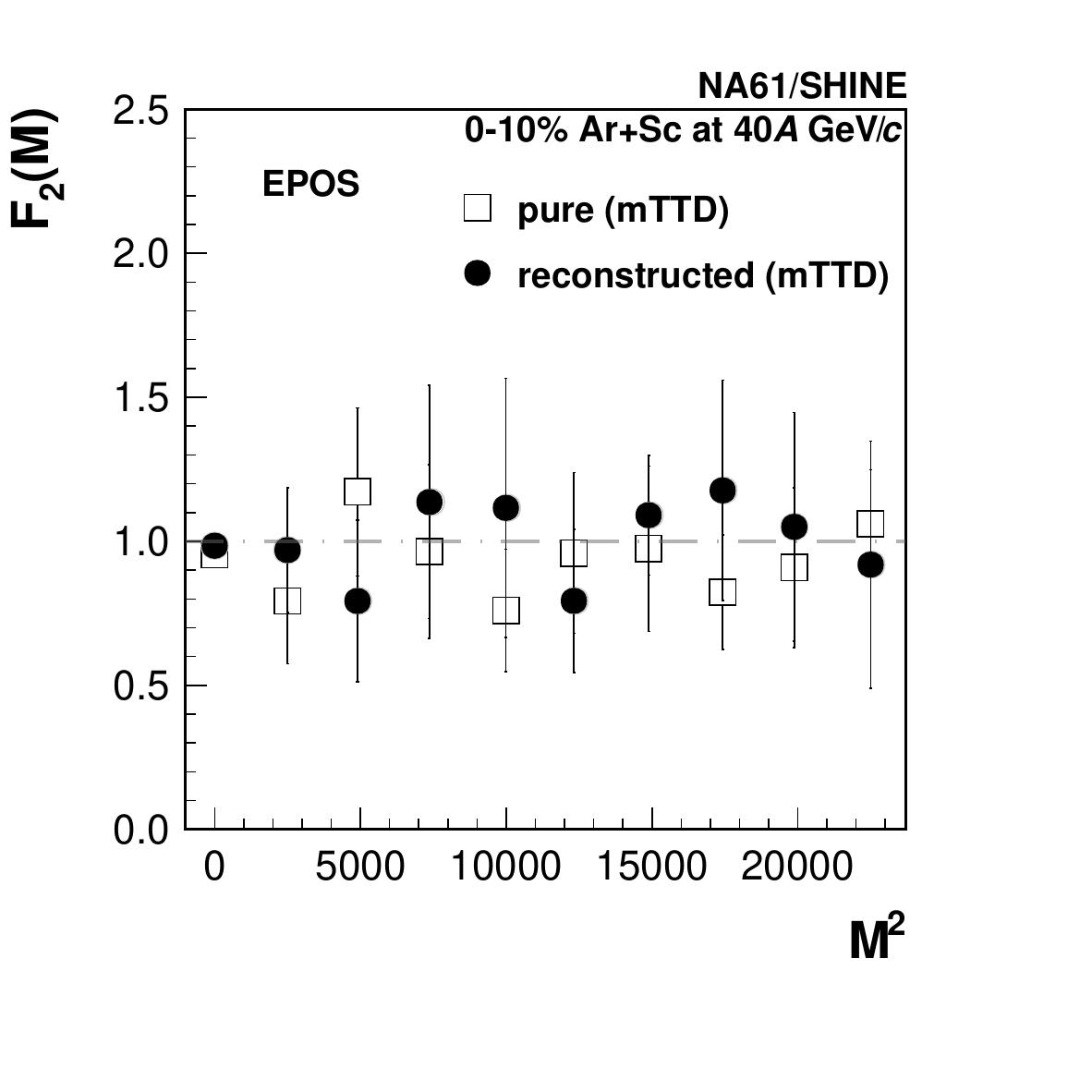}\qquad
    \includegraphics[width=.33\textwidth]{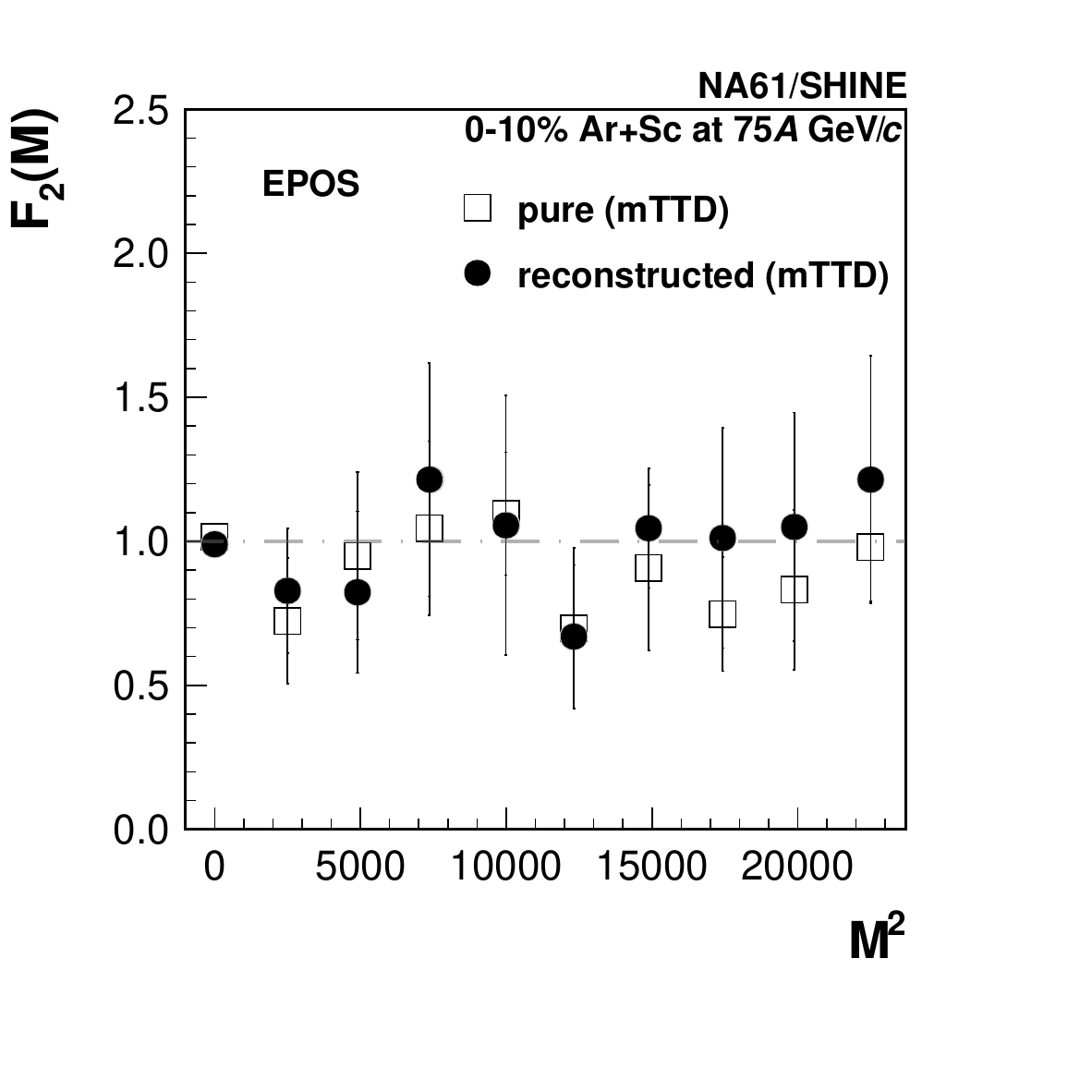}
    \rule{42em}{0.5pt}
    \caption{
        Results on the dependence of the scaled factorial moment of proton multiplicity distribution on the number of subdivisions in cumulative transverse momentum space, $M^{2}$ for $1^{2} \leq M^{2} \leq 150^{2}$,
        for events generated with \EposLong. Results are shown for 0--10\% central \ArSc collisions at 13\A--75\AGeVc. Closed black circles represent reconstructed \Epos, and open rectangles indicate pure (smeared) \Epos. Both pure and reconstructed \Epos include the mTTD cut. Only statistical uncertainties are indicated.
       }
    \label{fig:epos}
\end{figure}

Results for the reconstructed \Epos events were obtained as follows.
The model events were required to have the reconstructed primary vertex. Selected protons and proton pairs (matching the generated particles used for identification) were subjected to the same cuts as used for the experimental data analysis (see Sec.~\ref{sec:analysis}). The results for the pure and the reconstructed \Epos events are compared in Fig.~\ref{fig:epos} for central \ArSc collisions at 13\A--75\AGeVc. One concludes that for the \Epos-like physics, the experimental biases are smaller than the statistical uncertainties of the data.

\begin{figure}[!ht]
    \centering
    \includegraphics[width=.33\textwidth]{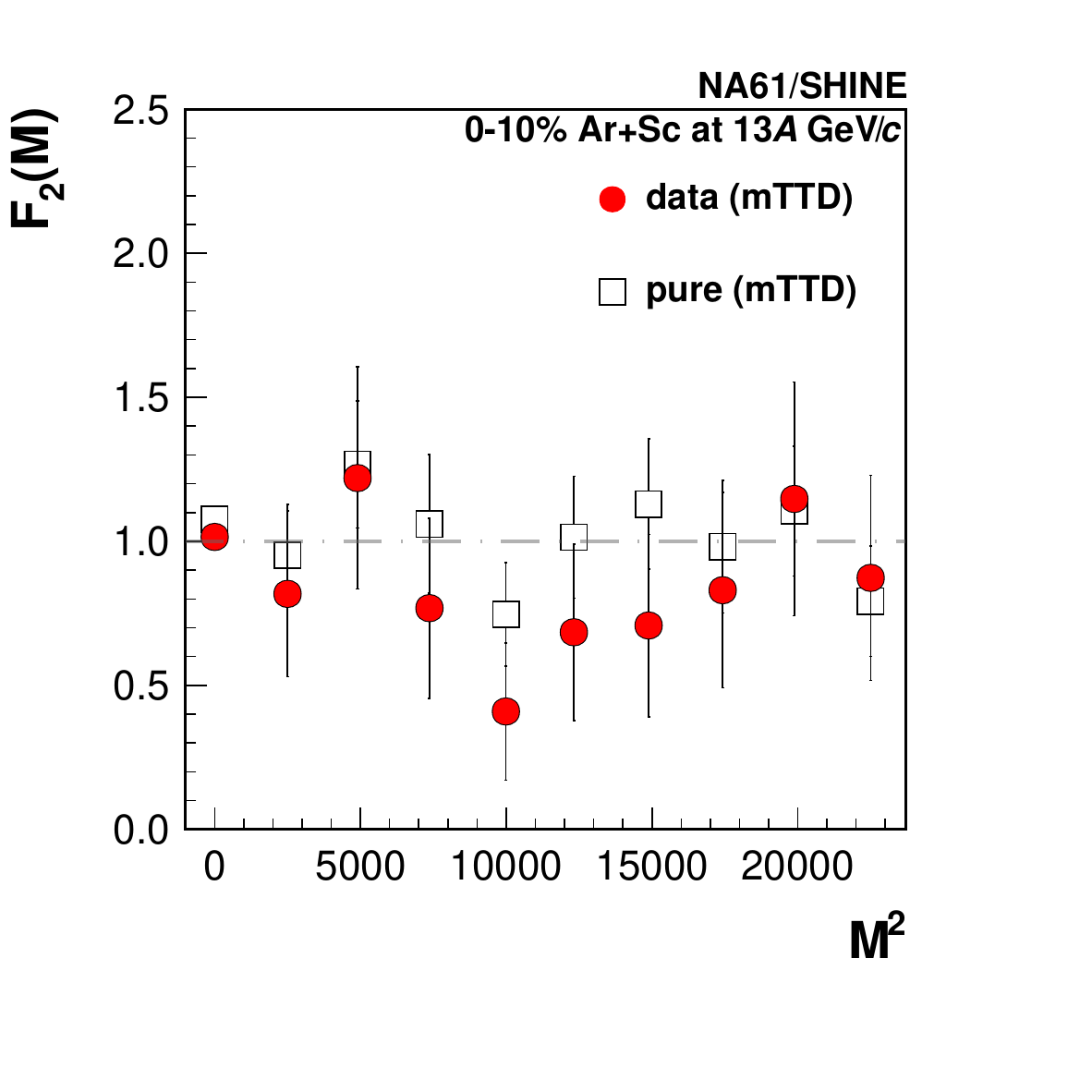}\hfill
    \includegraphics[width=.33\textwidth]{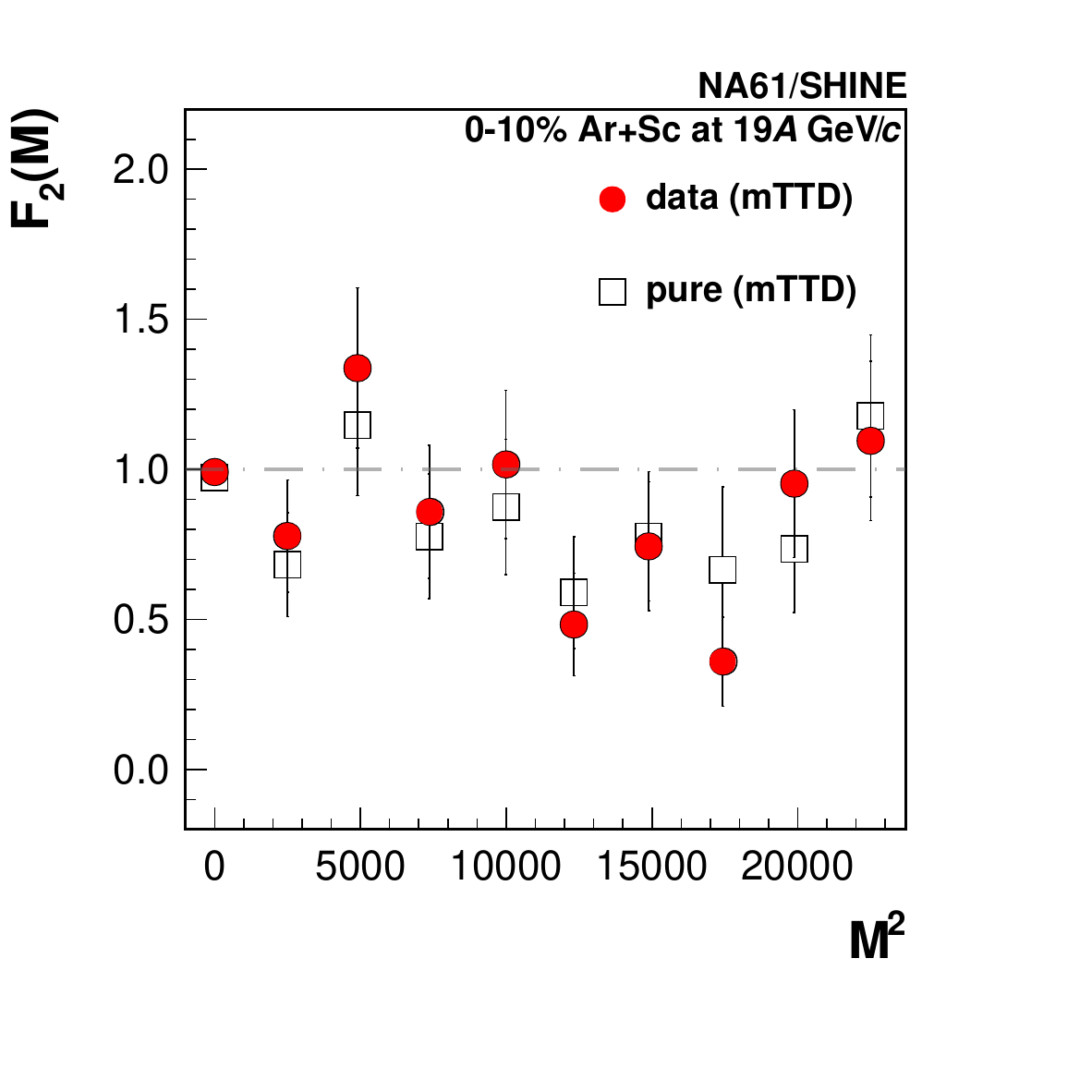}\hfill
    \includegraphics[width=.33\textwidth]{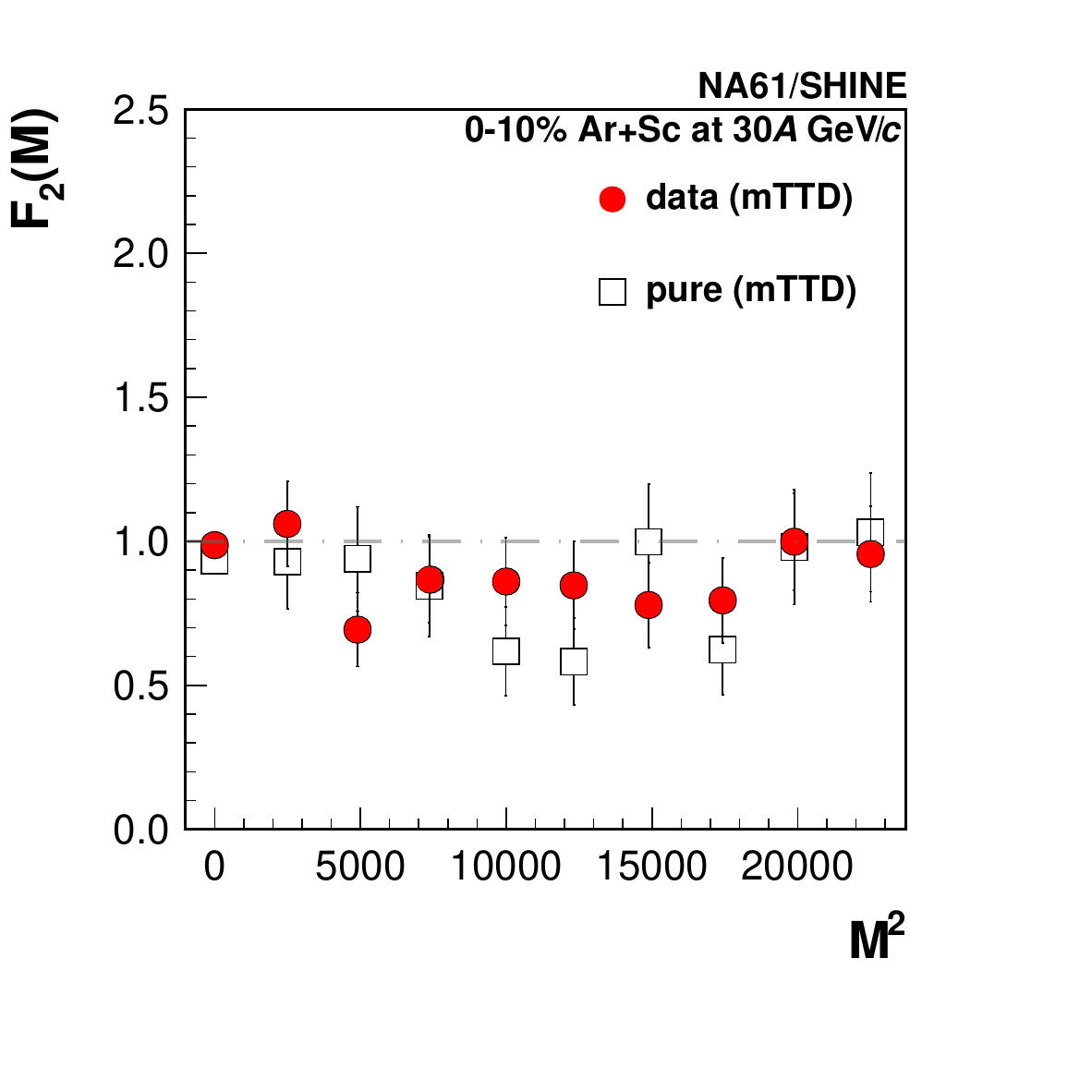}\\
    \includegraphics[width=.33\textwidth]{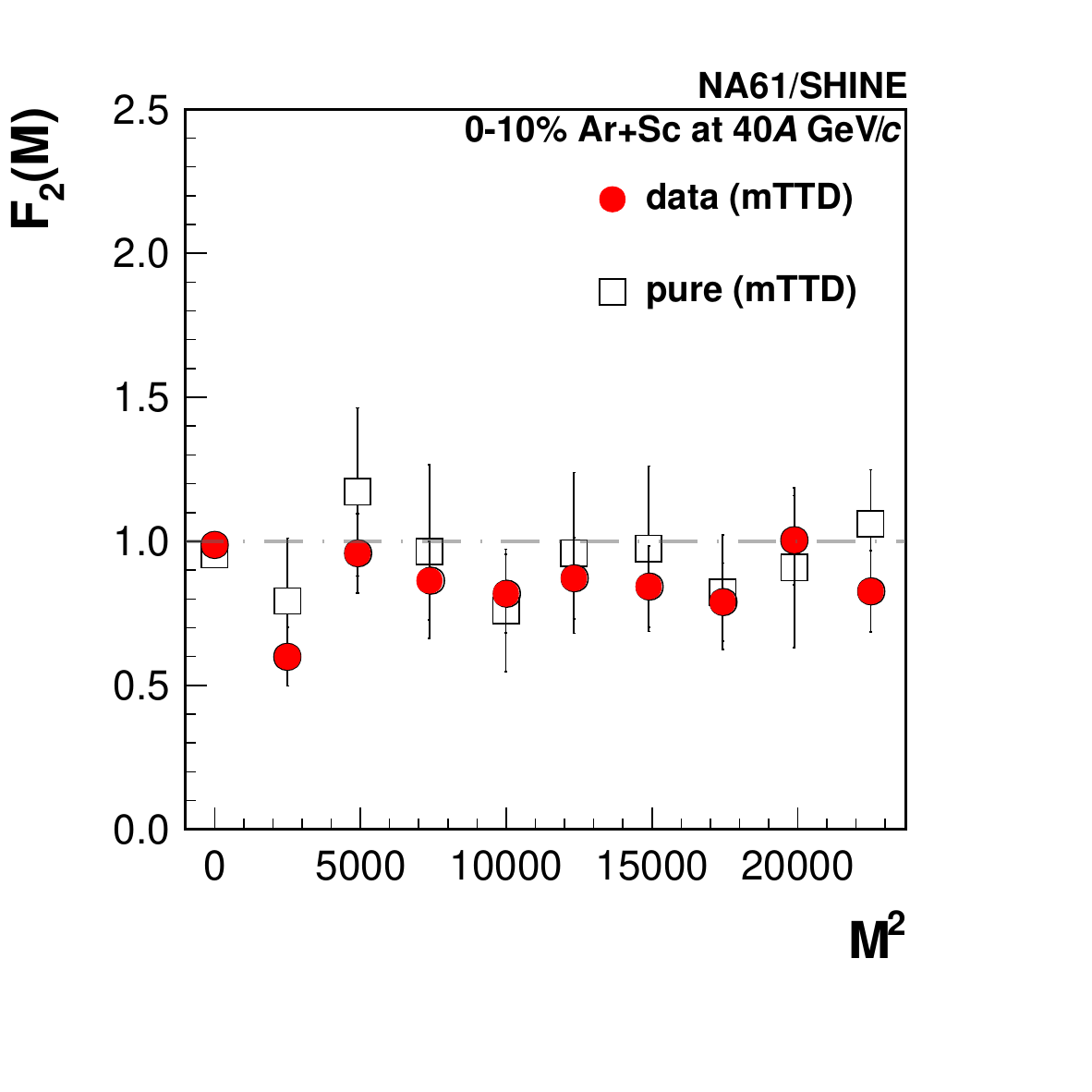}\qquad
    \includegraphics[width=.33\textwidth]{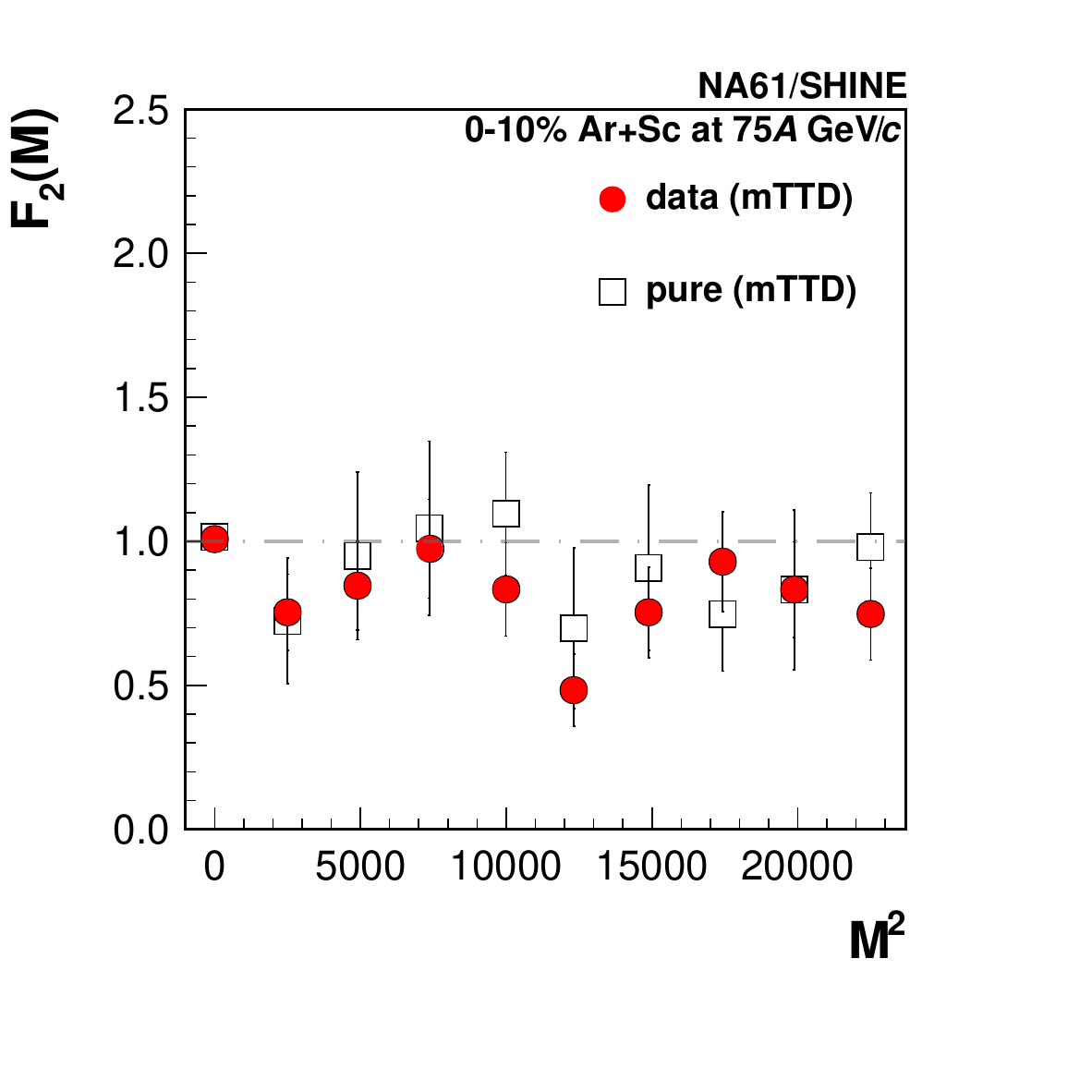}
    \rule{42em}{0.5pt}
    \caption{
        Results on the dependence of the scaled factorial moment of proton multiplicity distribution on the number of subdivisions in cumulative transverse momentum space $M^{2}$ for $1^{2} \leq M^{2} \leq 150^{2}$.
        Closed red circles indicate the experimental data. Corresponding results for the \EposLong model (open squares) were also shown for comparison. Results are shown for 0--10\% central \ArSc collisions at 13\A--75\AGeVc. Results for data and \Epos were obtained using the mTTD cut. Only statistical uncertainties are indicated.
       }
    \label{fig:epos-comparision}
\end{figure}

Finally, the experimental results are compared with the pure \Epos predictions for central \ArSc collisions at 13\A--75\AGeVc and are shown in Fig.~\ref{fig:epos-comparision}. No significant differences are found.

\clearpage
\subsection{Power-law Model}
\label{sec:models-power}

Inspired by expectations of the power-law correlations between particles near
the critical point,
the Power-law Model~\cite{Czopowicz:2023xcu} was developed to compare with the experimental result.
It generates momenta of uncorrelated and correlated protons with a given single-particle transverse
momentum distribution in events with a given multiplicity distribution. The correlated protons are produced in pairs, and their correlation inside the pair is given the power law as described in Ref.~\cite{Czopowicz:2023xcu}. The model has two controllable
parameters:
\begin{enumerate}[(i)]
    \item fraction of correlated particles, \textit{r},
    \item strength of the correlation (the power-law exponent), $\phi$.
\end{enumerate}

The transverse momentum of particles is drawn from the input transverse momentum distribution.
Correlated-particle pairs' transverse momentum difference follows a power-law distribution:
\begin{equation}
    \rho(|\Delta\overrightarrow{p_{T}}|) \sim |\Delta\overrightarrow{p_{T}}|^{-\phi},
\end{equation}
where the exponent $0 \leq \phi < 1$. For $r=0$, the Power-law Model results correspond to the mixed event results.
The exponent $\phi$ is related to the intermittency index $\varphi_{2}$ as:
\begin{equation}
  \varphi_{2} = \frac{\phi + 1}{2}.
\end{equation}

Azimuthal angle distribution is assumed to be uniform.
The momentum component along the beamline, $p_{z}$, is calculated assuming a uniform rapidity
distribution from $0$ to $0.75$ and proton mass.

\begin{figure}
    \centering
    \includegraphics[width=.45\textwidth]{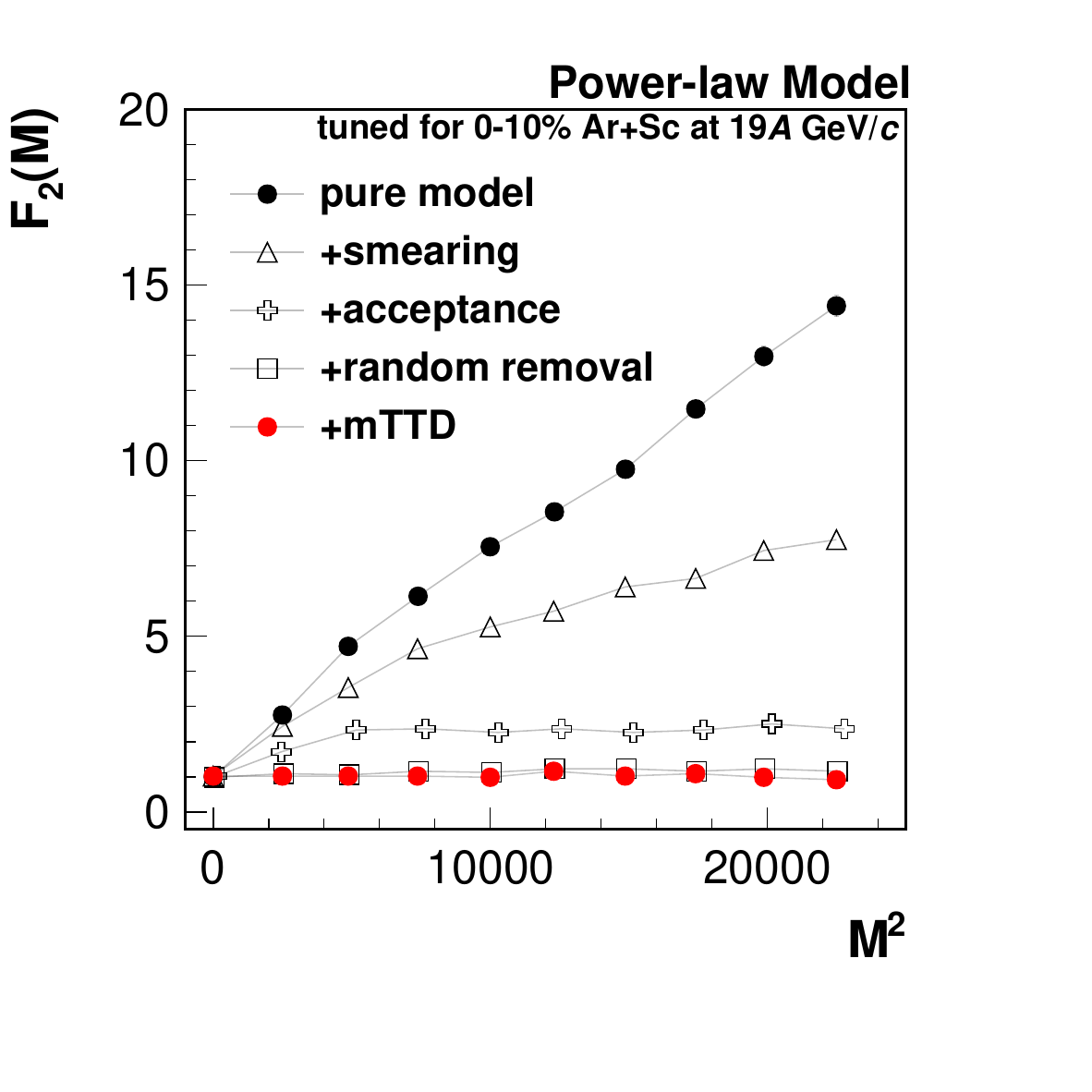}
    \includegraphics[width=.45\textwidth]{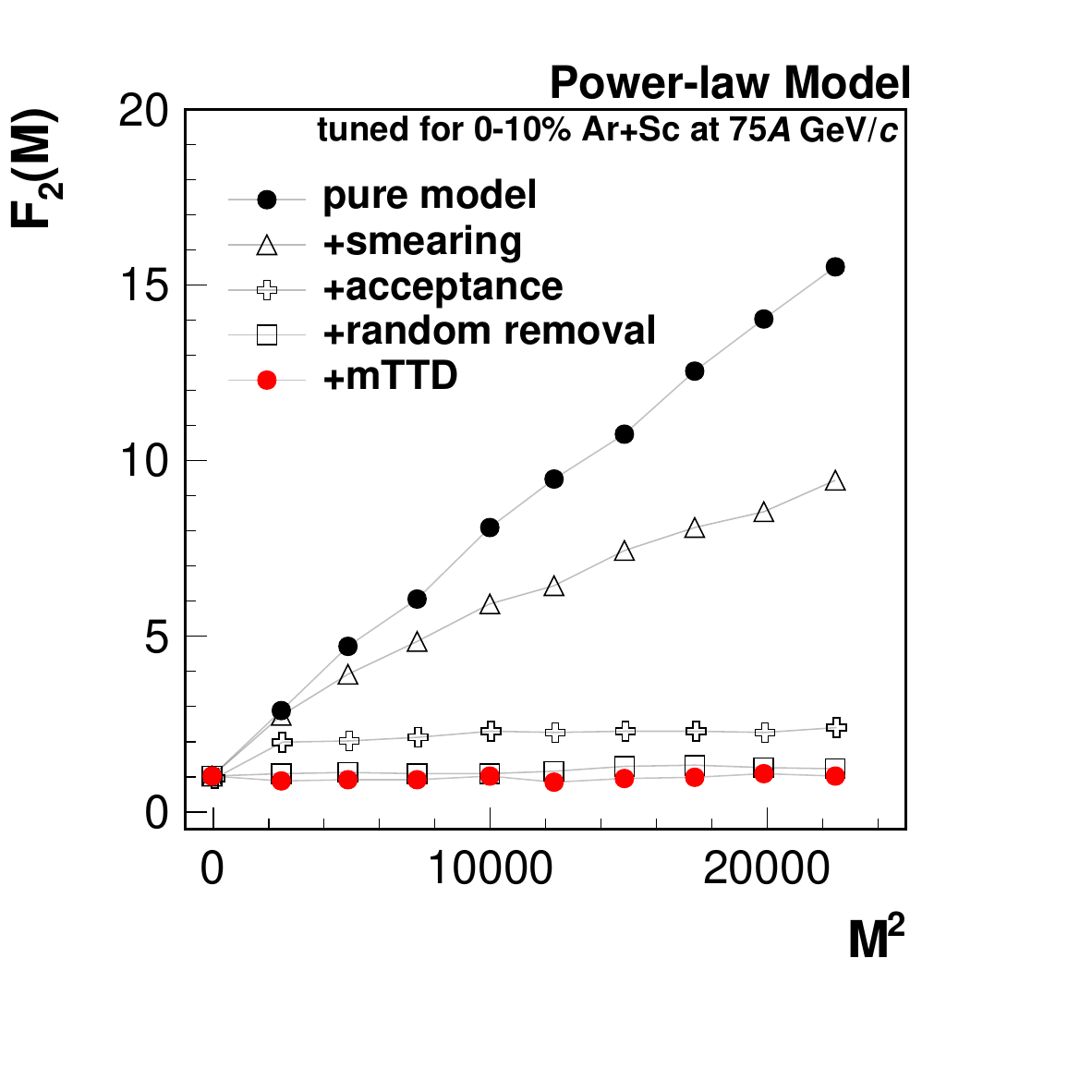}
    \rule{42em}{0.5pt}
    \caption{Dependence of the scaled factorial moment on the number of subdivisions in the
    cumulative transverse momentum for the Power-law Model with the power-law exponent set to 0.80
    and fraction of correlated particles to 2\%.
    Each line presents a result with a different effect included separately, and the red
    circles all of them together for central \ArSc collisions at 19\AGeVc (\textit{left}) and 75\AGeVc (\textit{right}).}
    \label{fig:model-biases}
\end{figure}

Many high-statistics data sets have been produced using the Power-law Model with multiplicity and inclusive transverse momentum distributions taken from experimental data. Each data set has a different fraction of correlated particles (varying from 0 to 2\%) and a
different power-law exponent (varying from 0.00 to 0.95).
The following effects have been included:
\begin{enumerate}[(i)]

    \item Gaussian smearing of momentum components to mimic reconstruction resolution of the momentum
    (see Eqs.~\ref{eq:smearing}),
    \item single-particle acceptance map (see Sec.~\ref{sec:maps}),
    \item two-particle acceptance map (see Sec.~\ref{sec:track-pair-selection}),
    \item random exchange of 40\% of correlated particles with uncorrelated ones to simulate 60\%
    acceptance of protons (preserves the desired multiplicity distribution but requires generating more
    correlated pairs at the beginning) (see Sec.~\ref{sec:protonselection}).

\end{enumerate}
The influence of each of the above effects separately and all of them applied together on
$F_{2}(M)$ is shown in Fig.~\ref{fig:model-biases} for \textit{r} = 0.02 and $\phi$ = 0.8, and fine subdivisions.

\begin{figure}[!ht]
    \centering
    \includegraphics[width=.45\textwidth]{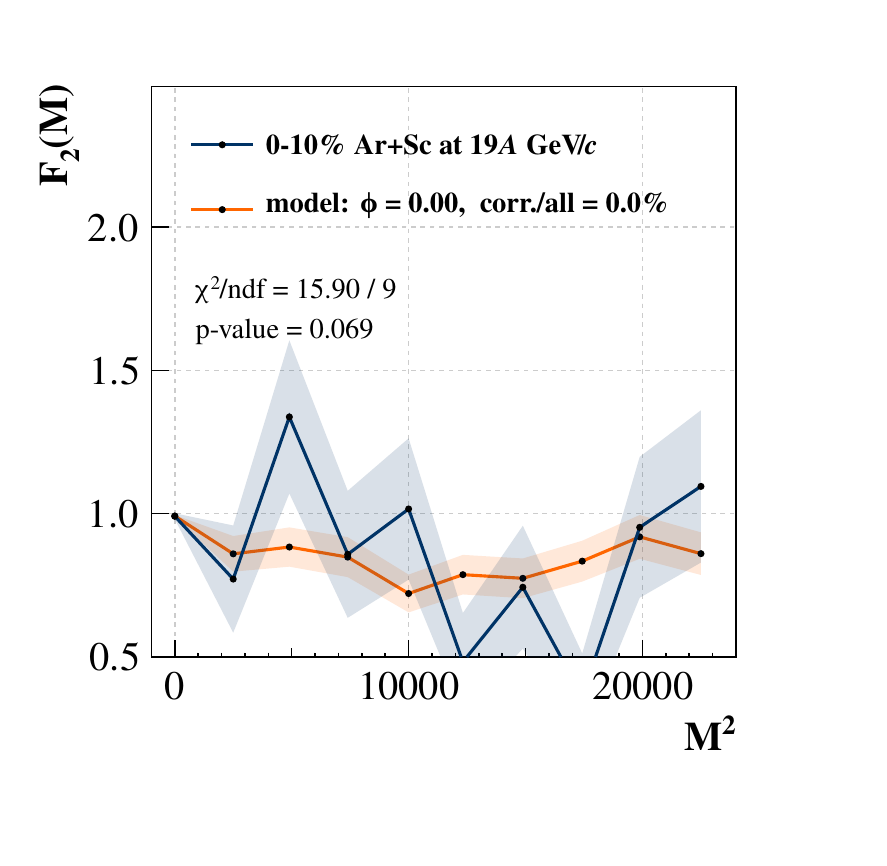}\qquad
    \includegraphics[width=.45\textwidth]{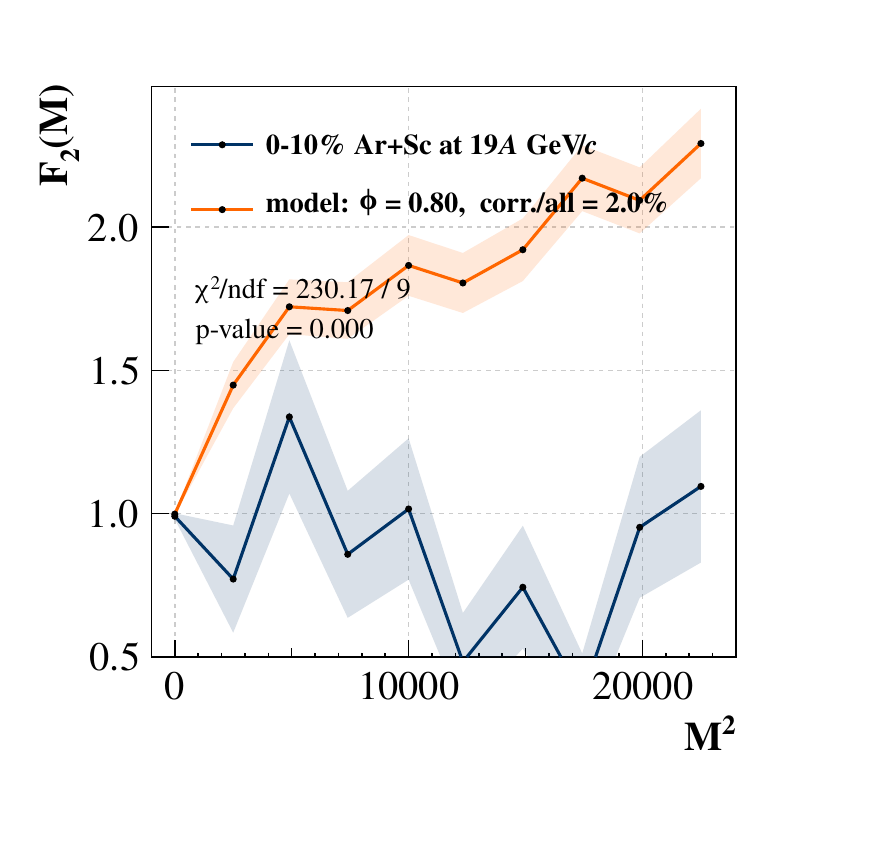}\\
    \includegraphics[width=.45\textwidth]{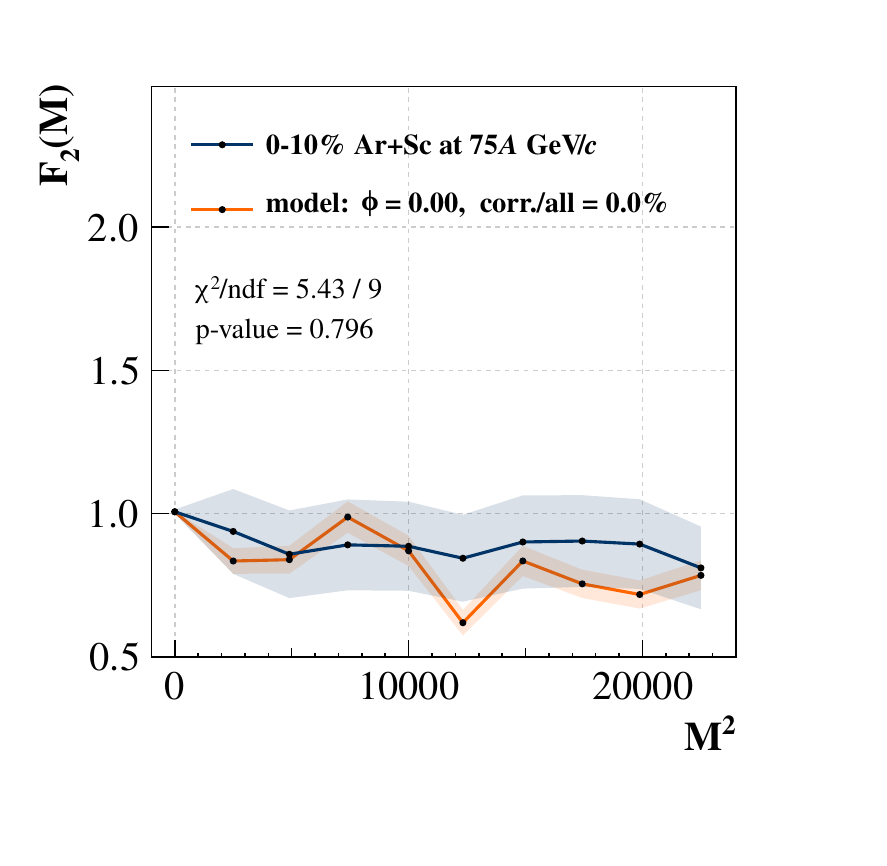}\qquad
    \includegraphics[width=.45\textwidth]{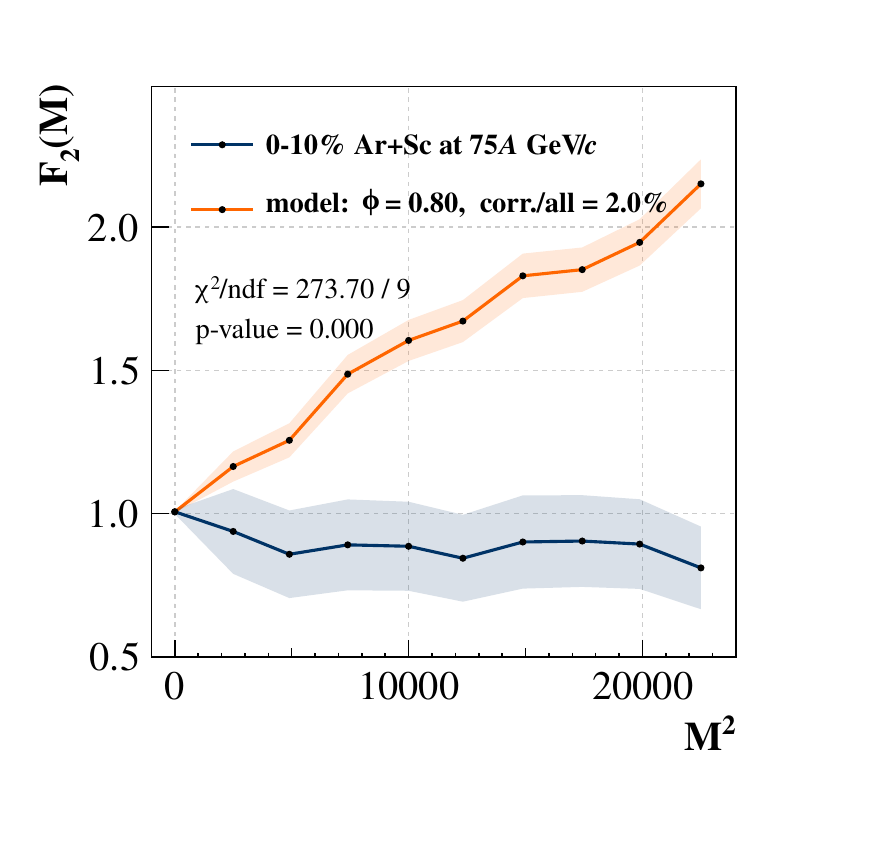}
    \rule{42em}{0.5pt}
    \caption{
        Examples of comparison of results for two Power-law Model data sets with the experimental data for central \ArSc collisions at 19\AGeVc (\textit{top}) and 75\AGeVc (\textit{bottom}).
        The \textit{left} panel includes model predictions assuming only
        uncorrelated protons, whereas the \textit{right} one shows predictions
        for 2.0\% of correlated protons with power-law exponent $\phi = 0.80$.
    }
    \label{fig:model-results}
\end{figure}

Next, all generated data sets with all the above effects included have been analyzed the same
way as the experimental data. Obtained $F_{2}(M)$ results have been compared
with the corresponding experimental results and
$\chi^{2}$ and a \textit{p}-value were calculated. Statistical uncertainties from the
model with similar statistics to the data were used for the calculation.
Examples of such comparisons for central \ArSc collisions at 19\AGeVc (\textit{left}) and 75\AGeVc (\textit{right}) are presented in Fig.~\ref{fig:model-results}.

\begin{figure}
    \centering
    \includegraphics[width=.8\textwidth]{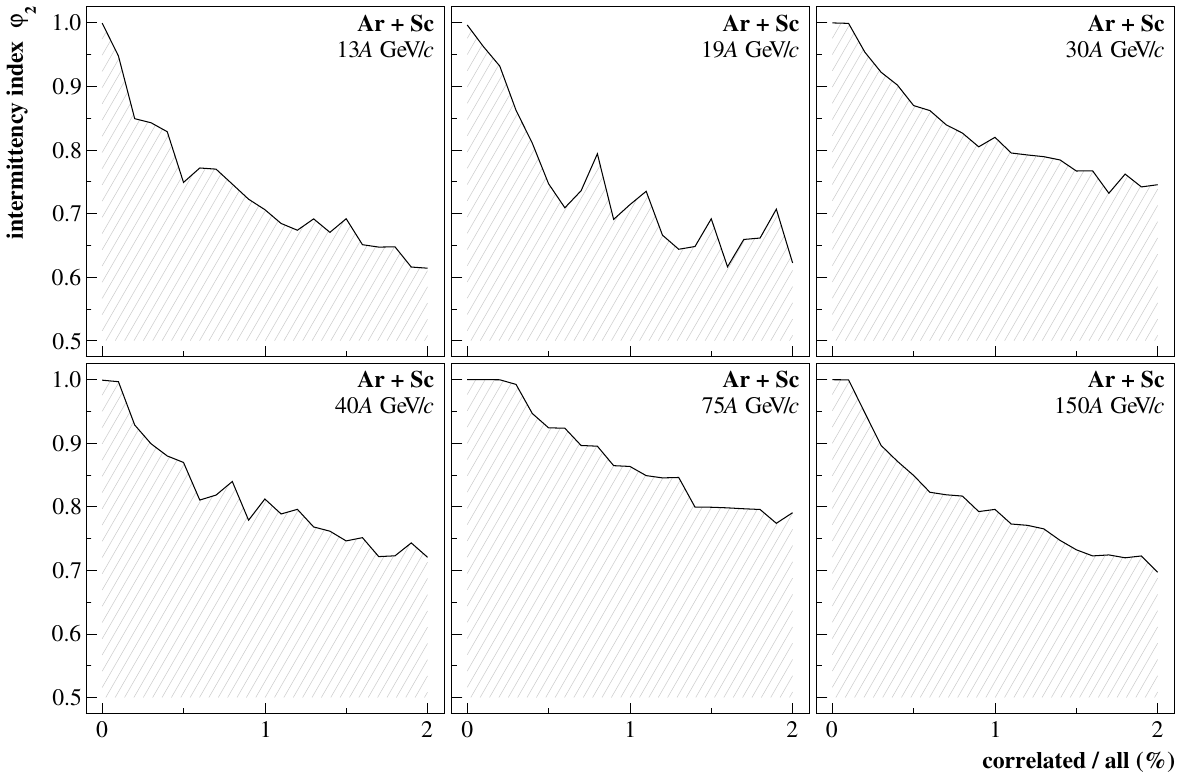}
    \rule{42em}{0.5pt}
    \caption{
    Exclusion plots for the Power-law Model~\cite{Czopowicz:2023xcu} parameters -- the fraction of correlated protons and the power-law exponent for central \ArSc collisions at \mbox{13\A--75\AGeVc} and 150\AGeVc. The white areas above the line correspond to \textit{p}-values less than 5\%. The exclusion plots were obtained using data (see Fig.~\ref{fig:results-cum-full}) for the fine subdivisions ($1^{2} \leq M^{2} \leq 150^{2}$).}
    

    \label{fig:exclusion-plot}
\end{figure}

Figure~\ref{fig:exclusion-plot} shows obtained exclusion plots 
as a function of
the fraction of correlated protons and the intermittency index (calculated from power-law exponent) for central \ArSc collisions at \mbox{13\A--75\AGeVc} and 150\AGeVc.\footnote{Note, that in the previously published paper on \ArSc at 150\AGeVc~\cite{NA61SHINE:2023gez}, the same notation was used for intermittency index ($\varphi_{2}$) and power-law exponent ($\phi$)}
White areas above the line correspond to a \textit{p}-value of less than 5\% and may be considered excluded (for this
particular model). Fluctuations of the exclusion lines in Fig.~\ref{fig:exclusion-plot} are due to limited statistics of the experimental data.

Results for the coarse subdivision have low statistical uncertainties (see Fig.~\ref{fig:results-cum-small}), thus
small deviations from the behavior expected for uncorrelated particle production due to non-critical
correlations (conservation laws, resonance decays, quantum statistics, and others), as well as possible experimental biases, may lead to a significant decrease of the \textit{p}-values. Thus, exclusion plots for the coarse subdivisions were not calculated.


The intermittency index $\varphi_{2}$ for an infinite system at QCD critical point is expected
to be $\varphi_{2}$ = 5/6~\cite{Antoniou:2006zb}, assuming that the latter belongs to the 3-D Ising universality class.
If this value is set as the power-law exponent of the Power-law Model with fine subdivisions
(see Fig.~\ref{fig:exclusion-plot}), for the \NASixtyOne data on \mbox{0--10\%} central \ArSc collisions at 13\A--75\AGeVc, an upper limit on the fraction of correlated protons is of the order of 1\%. It should be underlined that these numbers are specific to the Power-law Model. Analyses made with other models, which could provide different values depending on the assumed scenarios, remain beyond the scope of this paper.

\clearpage
\section{Summary}
\label{sec:summary}

This paper reports
on the search for the critical point of strongly interacting matter in central \ArSc collisions at beam momenta of 13\A, 19\A, 30\A, 40\A, and 75\AGeVc.
Results on second-order scaled factorial moments of proton multiplicity distribution at mid-rapidity are presented.
Protons produced in strong and electromagnetic processes in \ArSc interactions and selected by the single- and two-particle acceptance maps, as well as the identification cuts, are used.

\begin{figure}[!ht]
    \centering
    \includegraphics[width=.47\textwidth]{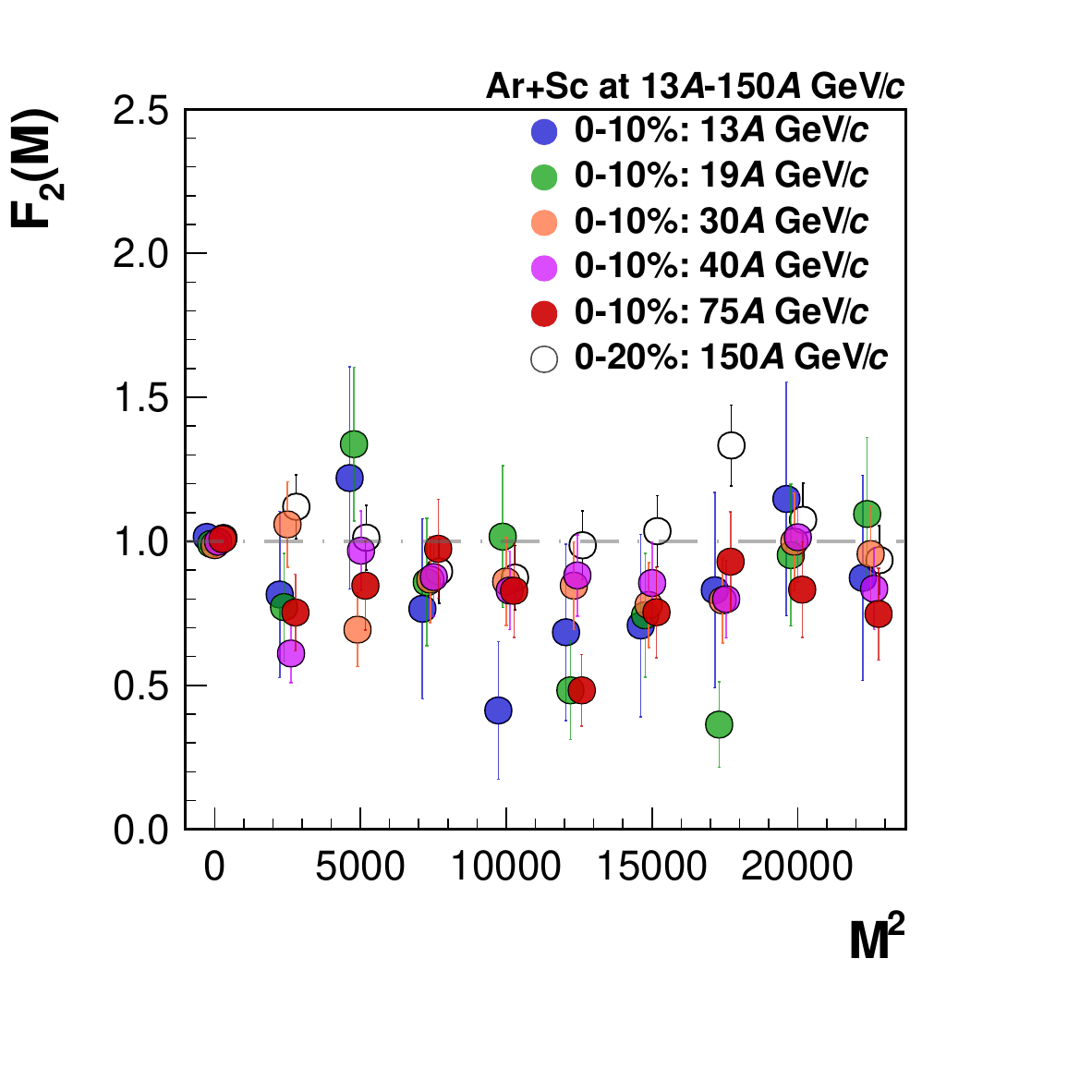}\hfill
    \includegraphics[width=.47\textwidth]{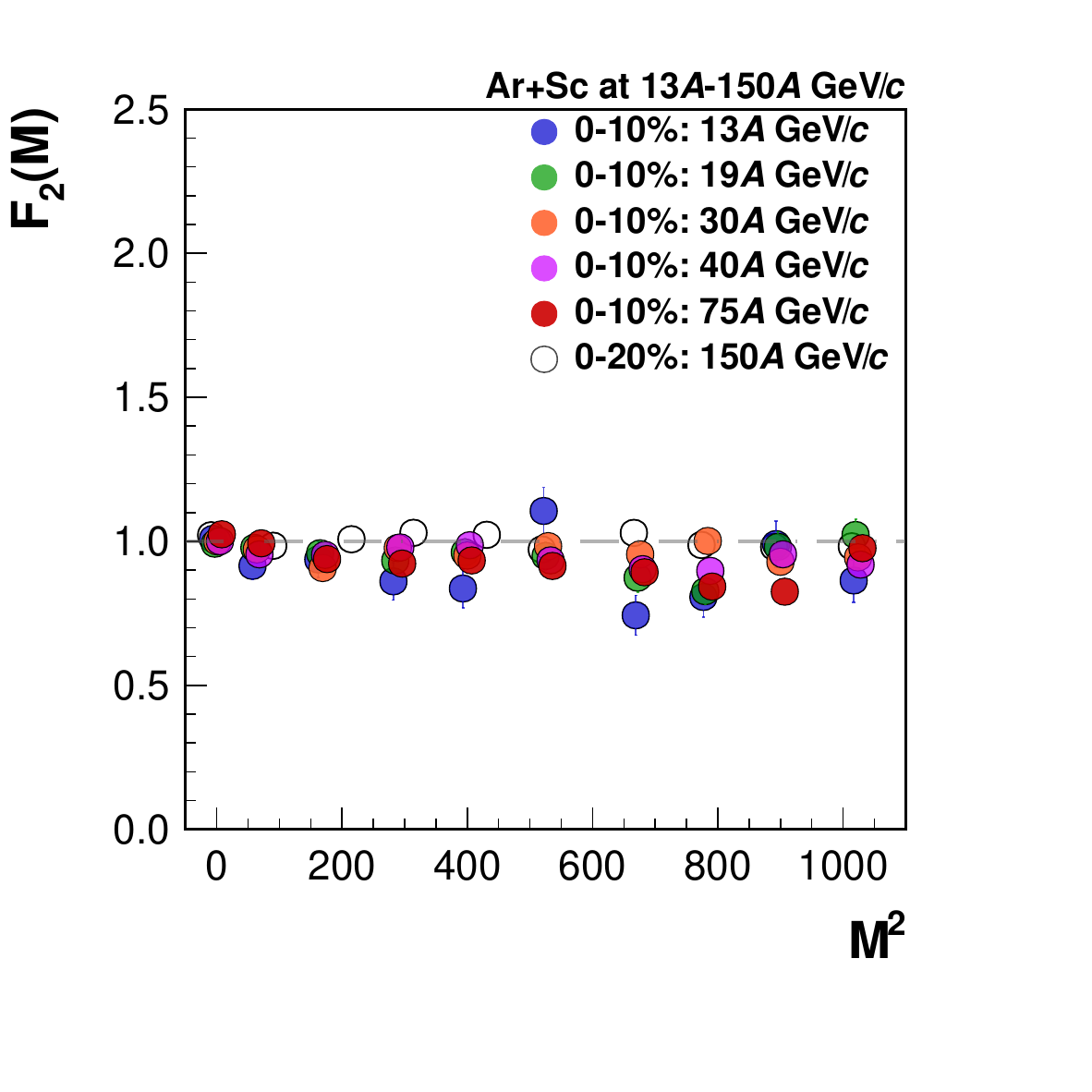}
    \rule{42em}{0.5pt}
    \caption{Summary of the proton intermittency results from the \NASixtyOne Ar+Sc energy scan. Results on the dependence of the scaled factorial moment of proton multiplicity distribution on the number of subdivisions in cumulative transverse momentum space $M^{2}$ for $1^{2} \leq M^{2} \leq 150^{2}$ (\textit{left}) and $1^{2} \leq M^{2} \leq 32^{2}$ (\textit{right}) are shown. The open circles represent results on 0--20\% central \ArSc collisions at 150\AGeVc~\cite{NA61SHINE:2023gez}. Closed circles indicate the experimental data results obtained within this work for 0--10\% central \ArSc collisions at 13\A, 19\A, 30\A, 40\A, and 75\AGeVc. Points for different energies are slightly shifted in horizontal axis to increase readability.}
    \label{fig:ArScScan}
\end{figure}

The scaled factorial moments are shown as a function of the number of subdivisions of transverse momentum space -- the so-called intermittency analysis.
The analysis was performed for cumulative and non-cumulative transverse momentum components.
Independent data sets were used to calculate results for each subdivision.
The influence of several experimental effects was discussed and quantified.
The results show no intermittency signal. A summary of the proton intermittency from the Ar+Sc energy scan results is shown in Fig.~\ref{fig:ArScScan}.

The experimental data are consistent with the mixed events and the \Epos model predictions. An upper limit on the fraction of correlated protons of the order of 1\% was obtained based on a comparison with the Power-law Model.

\begin{figure}
    \centering
    \includegraphics[width=.65\textwidth]{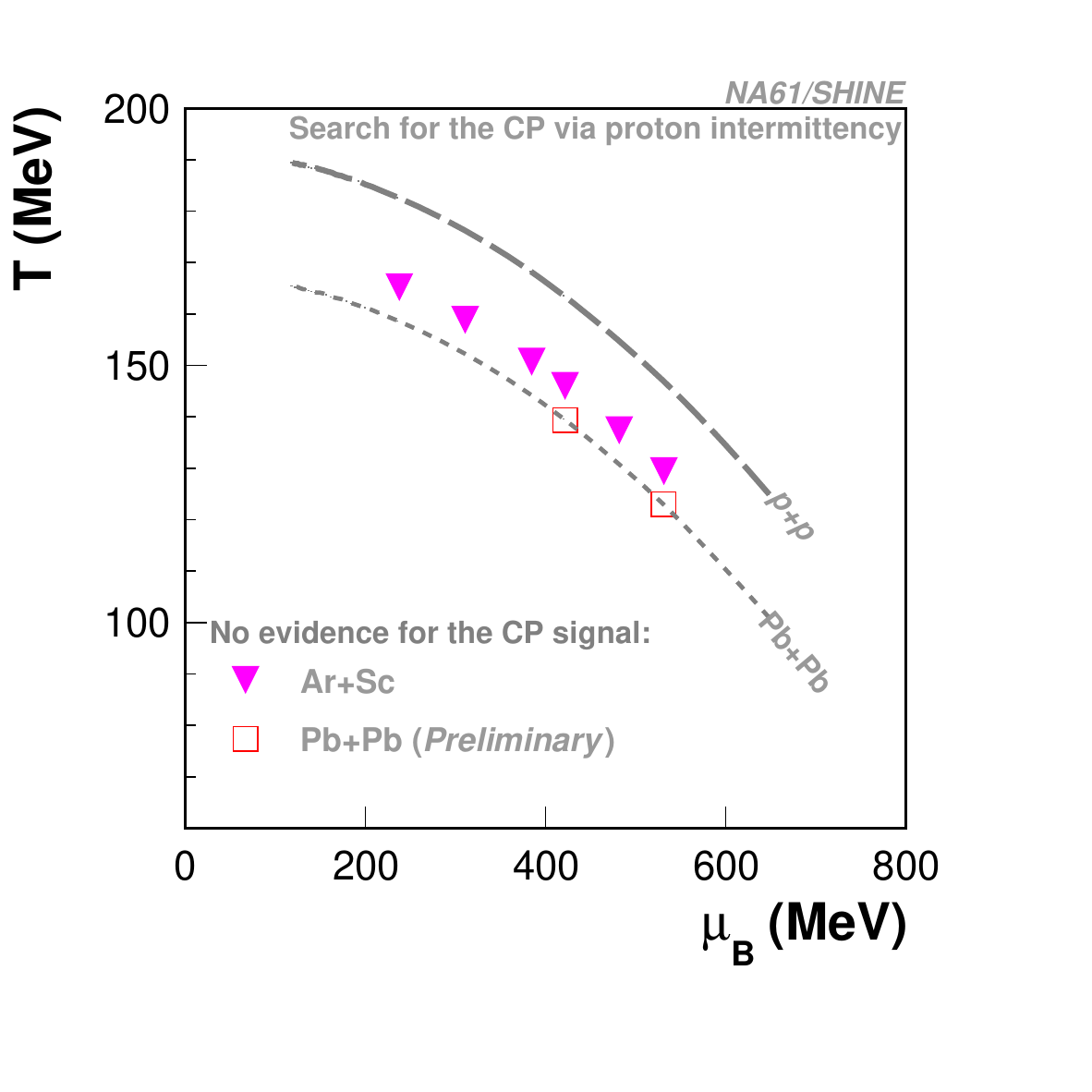}
     \rule{42em}{0.5pt}
    \caption{ Diagram of chemical freeze-out temperature and baryon-chemical potential. The dashed line indicates parameters in \pp interactions and the dotted line in the central Pb+Pb collisions; points estimated and extrapolated ($\mu_{B}$ for \pp) based on Ref.~\cite{Becattini:2005xt}. The colored points mark reactions (Ar+Sc and Pb+Pb~\cite{Adhikary:2022sdh}) in the $T-\mu_{B}$ phase diagram for which the search for the critical point was conducted, and no evidence for the critical point was found.}
    \label{fig:CEPSummary}
\end{figure}
The ongoing critical point search studies via proton intermittency are summarized on the diagram of chemical freeze-out temperature and chemical potential (estimated based on Ref.~\cite{Becattini:2005xt}) and shown in Fig.~\ref{fig:CEPSummary}.
The intermittency analysis of other reactions recorded within the \NASixtyOne program on strong interactions is well advanced, and new results should be expected soon.

\clearpage

\section*{Acknowledgements}
We would like to thank the CERN EP, BE, HSE and EN Departments for the
strong support of \NASixtyOne.

This work was supported by
the Hungarian Scientific Research Fund (grant NKFIH 138136\slash138152),
the Polish Ministry of Science and Higher Education
(DIR\slash WK\slash\-2016\slash 2017\slash\-10-1, WUT ID-UB), the National Science Centre Poland (grants
2014\slash 14\slash E\slash ST2\slash 00018, 
2016\slash 21\slash D\slash ST2\slash 01983, 
2017\slash 25\slash N\slash ST2\slash 02575, 
2018\slash 29\slash N\slash ST2\slash 02595, 
2018\slash 30\slash A\slash ST2\slash 00226, 
2018\slash 31\slash G\slash ST2\slash 03910, 
2019\slash 33\slash B\slash ST9\slash 03059, 
2020\slash 39\slash O\slash ST2\slash 00277), 
the Norwegian Financial Mechanism 2014--2021 (grant 2019\slash 34\slash H\slash ST2\slash 00585),
the Polish Minister of Education and Science (contract No. 2021\slash WK\slash 10),
the Russian Science Foundation (grant 17-72-20045),
the Russian Academy of Science and the
Russian Foundation for Basic Research (grants 08-02-00018, 09-02-00664
and 12-02-91503-CERN),
the Russian Foundation for Basic Research (RFBR) funding within the research project no. 18-02-40086,
the Ministry of Science and Higher Education of the Russian Federation, Project "Fundamental properties of elementary particles and cosmology" No 0723-2020-0041,
the European Union's Horizon 2020 research and innovation programme under grant agreement No. 871072,
the Ministry of Education, Culture, Sports,
Science and Tech\-no\-lo\-gy, Japan, Grant-in-Aid for Sci\-en\-ti\-fic
Research (grants 18071005, 19034011, 19740162, 20740160 and 20039012),
the German Research Foundation DFG (grants GA\,1480\slash8-1 and project 426579465),
the Bulgarian Ministry of Education and Science within the National
Roadmap for Research Infrastructures 2020--2027, contract No. D01-374/18.12.2020,
Ministry of Education
and Science of the Republic of Serbia (grant OI171002), Swiss
Nationalfonds Foundation (grant 200020\-117913/1), ETH Research Grant
TH-01\,07-3 and the Fermi National Accelerator Laboratory (Fermilab), a U.S. Department of Energy, Office of Science, HEP User Facility managed by Fermi Research Alliance, LLC (FRA), acting under Contract No. DE-AC02-07CH11359 and the IN2P3-CNRS (France).\\

The data used in this paper were collected before February 2022.

\clearpage
\bibliographystyle{include/na61Utphys}
\bibliography{include/na61References.bib}

\clearpage
{\Large The \NASixtyOne Collaboration}
\bigskip
\begin{sloppypar}

\noindent
\mbox{H.\;Adhikary\href{https://orcid.org/0000-0002-5746-1268}{\includegraphics[height=1.7ex]{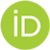}}\textsuperscript{\,14}},
\mbox{P.\;Adrich\href{https://orcid.org/0000-0002-7019-5451}{\includegraphics[height=1.7ex]{figures/orcid-logo.png}}\textsuperscript{\,16}},
\mbox{K.K.\;Allison\href{https://orcid.org/0000-0002-3494-9383}{\includegraphics[height=1.7ex]{figures/orcid-logo.png}}\textsuperscript{\,27}},
\mbox{N.\;Amin\href{https://orcid.org/0009-0004-7572-3817}{\includegraphics[height=1.7ex]{figures/orcid-logo.png}}\textsuperscript{\,5}},
\mbox{E.V.\;Andronov\href{https://orcid.org/0000-0003-0437-9292}{\includegraphics[height=1.7ex]{figures/orcid-logo.png}}\textsuperscript{\,23}},
\mbox{T.\;Anti\'ci\'c\href{https://orcid.org/0000-0002-6606-0191}{\includegraphics[height=1.7ex]{figures/orcid-logo.png}}\textsuperscript{\,3}},
\mbox{I.-C.\;Arsene\href{https://orcid.org/0000-0003-2316-9565}{\includegraphics[height=1.7ex]{figures/orcid-logo.png}}\textsuperscript{\,13}},
\mbox{M.\;Bajda\href{https://orcid.org/0009-0005-8859-1099}{\includegraphics[height=1.7ex]{figures/orcid-logo.png}}\textsuperscript{\,17}},
\mbox{Y.\;Balkova\href{https://orcid.org/0000-0002-6957-573X}{\includegraphics[height=1.7ex]{figures/orcid-logo.png}}\textsuperscript{\,19}},
\mbox{M.\;Baszczyk\href{https://orcid.org/0000-0002-2595-0104}{\includegraphics[height=1.7ex]{figures/orcid-logo.png}}\textsuperscript{\,18}},
\mbox{D.\;Battaglia\href{https://orcid.org/0000-0002-5283-0992}{\includegraphics[height=1.7ex]{figures/orcid-logo.png}}\textsuperscript{\,26}},
\mbox{A.\;Bazgir\href{https://orcid.org/0000-0003-0358-0576}{\includegraphics[height=1.7ex]{figures/orcid-logo.png}}\textsuperscript{\,14}},
\mbox{S.\;Bhosale\href{https://orcid.org/0000-0001-5709-4747}{\includegraphics[height=1.7ex]{figures/orcid-logo.png}}\textsuperscript{\,15}},
\mbox{M.\;Bielewicz\href{https://orcid.org/0000-0001-8267-4874}{\includegraphics[height=1.7ex]{figures/orcid-logo.png}}\textsuperscript{\,16}},
\mbox{A.\;Blondel\href{https://orcid.org/0000-0002-1597-8859}{\includegraphics[height=1.7ex]{figures/orcid-logo.png}}\textsuperscript{\,4}},
\mbox{M.\;Bogomilov\href{https://orcid.org/0000-0001-7738-2041}{\includegraphics[height=1.7ex]{figures/orcid-logo.png}}\textsuperscript{\,2}},
\mbox{Y.\;Bondar\href{https://orcid.org/0000-0003-2773-9668}{\includegraphics[height=1.7ex]{figures/orcid-logo.png}}\textsuperscript{\,14}},
\mbox{N.\;Bostan\href{https://orcid.org/0000-0002-1129-4345}{\includegraphics[height=1.7ex]{figures/orcid-logo.png}}\textsuperscript{\,26}},
\mbox{A.\;Brandin\textsuperscript{\,23}},
\mbox{W.\;Bryli\'nski\href{https://orcid.org/0000-0002-3457-6601}{\includegraphics[height=1.7ex]{figures/orcid-logo.png}}\textsuperscript{\,22}},
\mbox{J.\;Brzychczyk\href{https://orcid.org/0000-0001-5320-6748}{\includegraphics[height=1.7ex]{figures/orcid-logo.png}}\textsuperscript{\,17}},
\mbox{M.\;Buryakov\href{https://orcid.org/0009-0008-2394-4967}{\includegraphics[height=1.7ex]{figures/orcid-logo.png}}\textsuperscript{\,23}},
\mbox{A.F.\;Camino\textsuperscript{\,29}},
\mbox{P.\;Christakoglou\href{https://orcid.org/0000-0002-4325-0646}{\includegraphics[height=1.7ex]{figures/orcid-logo.png}}\textsuperscript{\,7}},
\mbox{M.\;\'Cirkovi\'c\href{https://orcid.org/0000-0002-4420-9688}{\includegraphics[height=1.7ex]{figures/orcid-logo.png}}\textsuperscript{\,24}},
\mbox{M.\;Csan\'ad\href{https://orcid.org/0000-0002-3154-6925}{\includegraphics[height=1.7ex]{figures/orcid-logo.png}}\textsuperscript{\,9}},
\mbox{J.\;Cybowska\href{https://orcid.org/0000-0003-2568-3664}{\includegraphics[height=1.7ex]{figures/orcid-logo.png}}\textsuperscript{\,22}},
\mbox{T.\;Czopowicz\href{https://orcid.org/0000-0003-1908-2977}{\includegraphics[height=1.7ex]{figures/orcid-logo.png}}\textsuperscript{\,14}},
\mbox{C.\;Dalmazzone\href{https://orcid.org/0000-0001-6945-5845}{\includegraphics[height=1.7ex]{figures/orcid-logo.png}}\textsuperscript{\,4}},
\mbox{N.\;Davis\href{https://orcid.org/0000-0003-3047-6854}{\includegraphics[height=1.7ex]{figures/orcid-logo.png}}\textsuperscript{\,15}},
\mbox{F.\;Diakonos\href{https://orcid.org/0000-0003-0142-9098}{\includegraphics[height=1.7ex]{figures/orcid-logo.png}}\textsuperscript{\,7}},
\mbox{A.\;Dmitriev\href{https://orcid.org/0000-0001-7853-0173}{\includegraphics[height=1.7ex]{figures/orcid-logo.png}}\textsuperscript{\,23}},
\mbox{P.~von\;Doetinchem\href{https://orcid.org/0000-0002-7801-3376}{\includegraphics[height=1.7ex]{figures/orcid-logo.png}}\textsuperscript{\,28}},
\mbox{W.\;Dominik\href{https://orcid.org/0000-0001-7444-9239}{\includegraphics[height=1.7ex]{figures/orcid-logo.png}}\textsuperscript{\,20}},
\mbox{P.\;Dorosz\href{https://orcid.org/0000-0002-8884-0981}{\includegraphics[height=1.7ex]{figures/orcid-logo.png}}\textsuperscript{\,18}},
\mbox{J.\;Dumarchez\href{https://orcid.org/0000-0002-9243-4425}{\includegraphics[height=1.7ex]{figures/orcid-logo.png}}\textsuperscript{\,4}},
\mbox{R.\;Engel\href{https://orcid.org/0000-0003-2924-8889}{\includegraphics[height=1.7ex]{figures/orcid-logo.png}}\textsuperscript{\,5}},
\mbox{G.A.\;Feofilov\href{https://orcid.org/0000-0003-3700-8623}{\includegraphics[height=1.7ex]{figures/orcid-logo.png}}\textsuperscript{\,23}},
\mbox{L.\;Fields\href{https://orcid.org/0000-0001-8281-3686}{\includegraphics[height=1.7ex]{figures/orcid-logo.png}}\textsuperscript{\,26}},
\mbox{Z.\;Fodor\href{https://orcid.org/0000-0003-2519-5687}{\includegraphics[height=1.7ex]{figures/orcid-logo.png}}\textsuperscript{\,8,21}},
\mbox{M.\;Friend\href{https://orcid.org/0000-0003-4660-4670}{\includegraphics[height=1.7ex]{figures/orcid-logo.png}}\textsuperscript{\,10}},
\mbox{M.\;Ga\'zdzicki\href{https://orcid.org/0000-0002-6114-8223}{\includegraphics[height=1.7ex]{figures/orcid-logo.png}}\textsuperscript{\,14,6}},
\mbox{O.\;Golosov\href{https://orcid.org/0000-0001-6562-2925}{\includegraphics[height=1.7ex]{figures/orcid-logo.png}}\textsuperscript{\,23}},
\mbox{V.\;Golovatyuk\href{https://orcid.org/0009-0006-5201-0990}{\includegraphics[height=1.7ex]{figures/orcid-logo.png}}\textsuperscript{\,23}},
\mbox{M.\;Golubeva\href{https://orcid.org/0009-0003-4756-2449}{\includegraphics[height=1.7ex]{figures/orcid-logo.png}}\textsuperscript{\,23}},
\mbox{K.\;Grebieszkow\href{https://orcid.org/0000-0002-6754-9554}{\includegraphics[height=1.7ex]{figures/orcid-logo.png}}\textsuperscript{\,22}},
\mbox{F.\;Guber\href{https://orcid.org/0000-0001-8790-3218}{\includegraphics[height=1.7ex]{figures/orcid-logo.png}}\textsuperscript{\,23}},
\mbox{S.N.\;Igolkin\textsuperscript{\,23}},
\mbox{S.\;Ilieva\href{https://orcid.org/0000-0001-9204-2563}{\includegraphics[height=1.7ex]{figures/orcid-logo.png}}\textsuperscript{\,2}},
\mbox{A.\;Ivashkin\href{https://orcid.org/0000-0003-4595-5866}{\includegraphics[height=1.7ex]{figures/orcid-logo.png}}\textsuperscript{\,23}},
\mbox{A.\;Izvestnyy\href{https://orcid.org/0009-0009-1305-7309}{\includegraphics[height=1.7ex]{figures/orcid-logo.png}}\textsuperscript{\,23}},
\mbox{K.\;Kadija\textsuperscript{\,3}},
\mbox{A.\;Kapoyannis\href{https://orcid.org/0000-0002-7732-8552}{\includegraphics[height=1.7ex]{figures/orcid-logo.png}}\textsuperscript{\,7}},
\mbox{N.\;Kargin\textsuperscript{\,23}},
\mbox{N.\;Karpushkin\href{https://orcid.org/0000-0001-5513-9331}{\includegraphics[height=1.7ex]{figures/orcid-logo.png}}\textsuperscript{\,23}},
\mbox{E.\;Kashirin\href{https://orcid.org/0000-0001-6062-7997}{\includegraphics[height=1.7ex]{figures/orcid-logo.png}}\textsuperscript{\,23}},
\mbox{M.\;Kie{\l}bowicz\href{https://orcid.org/0000-0002-4403-9201}{\includegraphics[height=1.7ex]{figures/orcid-logo.png}}\textsuperscript{\,15}},
\mbox{V.A.\;Kireyeu\href{https://orcid.org/0000-0002-5630-9264}{\includegraphics[height=1.7ex]{figures/orcid-logo.png}}\textsuperscript{\,23}},
\mbox{H.\;Kitagawa\textsuperscript{\,11}},
\mbox{R.\;Kolesnikov\href{https://orcid.org/0009-0006-4224-1058}{\includegraphics[height=1.7ex]{figures/orcid-logo.png}}\textsuperscript{\,23}},
\mbox{D.\;Kolev\href{https://orcid.org/0000-0002-9203-4739}{\includegraphics[height=1.7ex]{figures/orcid-logo.png}}\textsuperscript{\,2}},
\mbox{Y.\;Koshio\textsuperscript{\,11}},
\mbox{V.N.\;Kovalenko\href{https://orcid.org/0000-0001-6012-6615}{\includegraphics[height=1.7ex]{figures/orcid-logo.png}}\textsuperscript{\,23}},
\mbox{S.\;Kowalski\href{https://orcid.org/0000-0001-9888-4008}{\includegraphics[height=1.7ex]{figures/orcid-logo.png}}\textsuperscript{\,19}},
\mbox{B.\;Koz{\l}owski\href{https://orcid.org/0000-0001-8442-2320}{\includegraphics[height=1.7ex]{figures/orcid-logo.png}}\textsuperscript{\,22}},
\mbox{A.\;Krasnoperov\href{https://orcid.org/0000-0002-1425-2861}{\includegraphics[height=1.7ex]{figures/orcid-logo.png}}\textsuperscript{\,23}},
\mbox{W.\;Kucewicz\href{https://orcid.org/0000-0002-2073-711X}{\includegraphics[height=1.7ex]{figures/orcid-logo.png}}\textsuperscript{\,18}},
\mbox{M.\;Kuchowicz\href{https://orcid.org/0000-0003-3174-585X}{\includegraphics[height=1.7ex]{figures/orcid-logo.png}}\textsuperscript{\,21}},
\mbox{M.\;Kuich\href{https://orcid.org/0000-0002-6507-8699}{\includegraphics[height=1.7ex]{figures/orcid-logo.png}}\textsuperscript{\,20}},
\mbox{A.\;Kurepin\href{https://orcid.org/0000-0002-1851-4136}{\includegraphics[height=1.7ex]{figures/orcid-logo.png}}\textsuperscript{\,23}},
\mbox{A.\;L\'aszl\'o\href{https://orcid.org/0000-0003-2712-6968}{\includegraphics[height=1.7ex]{figures/orcid-logo.png}}\textsuperscript{\,8}},
\mbox{M.\;Lewicki\href{https://orcid.org/0000-0002-8972-3066}{\includegraphics[height=1.7ex]{figures/orcid-logo.png}}\textsuperscript{\,21}},
\mbox{G.\;Lykasov\href{https://orcid.org/0000-0002-1544-6959}{\includegraphics[height=1.7ex]{figures/orcid-logo.png}}\textsuperscript{\,23}},
\mbox{V.V.\;Lyubushkin\href{https://orcid.org/0000-0003-0136-233X}{\includegraphics[height=1.7ex]{figures/orcid-logo.png}}\textsuperscript{\,23}},
\mbox{M.\;Ma\'ckowiak-Paw{\l}owska\href{https://orcid.org/0000-0003-3954-6329}{\includegraphics[height=1.7ex]{figures/orcid-logo.png}}\textsuperscript{\,22}},
\mbox{Z.\;Majka\href{https://orcid.org/0000-0003-3064-6577}{\includegraphics[height=1.7ex]{figures/orcid-logo.png}}\textsuperscript{\,17}},
\mbox{A.\;Makhnev\href{https://orcid.org/0009-0002-9745-1897}{\includegraphics[height=1.7ex]{figures/orcid-logo.png}}\textsuperscript{\,23}},
\mbox{B.\;Maksiak\href{https://orcid.org/0000-0002-7950-2307}{\includegraphics[height=1.7ex]{figures/orcid-logo.png}}\textsuperscript{\,16}},
\mbox{A.I.\;Malakhov\href{https://orcid.org/0000-0001-8569-8409}{\includegraphics[height=1.7ex]{figures/orcid-logo.png}}\textsuperscript{\,23}},
\mbox{A.\;Marcinek\href{https://orcid.org/0000-0001-9922-743X}{\includegraphics[height=1.7ex]{figures/orcid-logo.png}}\textsuperscript{\,15}},
\mbox{A.D.\;Marino\href{https://orcid.org/0000-0002-1709-538X}{\includegraphics[height=1.7ex]{figures/orcid-logo.png}}\textsuperscript{\,27}},
\mbox{H.-J.\;Mathes\href{https://orcid.org/0000-0002-0680-040X}{\includegraphics[height=1.7ex]{figures/orcid-logo.png}}\textsuperscript{\,5}},
\mbox{T.\;Matulewicz\href{https://orcid.org/0000-0003-2098-1216}{\includegraphics[height=1.7ex]{figures/orcid-logo.png}}\textsuperscript{\,20}},
\mbox{V.\;Matveev\href{https://orcid.org/0000-0002-2745-5908}{\includegraphics[height=1.7ex]{figures/orcid-logo.png}}\textsuperscript{\,23}},
\mbox{G.L.\;Melkumov\href{https://orcid.org/0009-0004-2074-6755}{\includegraphics[height=1.7ex]{figures/orcid-logo.png}}\textsuperscript{\,23}},
\mbox{A.\;Merzlaya\href{https://orcid.org/0000-0002-6553-2783}{\includegraphics[height=1.7ex]{figures/orcid-logo.png}}\textsuperscript{\,13}},
\mbox{{\L}.\;Mik\href{https://orcid.org/0000-0003-2712-6861}{\includegraphics[height=1.7ex]{figures/orcid-logo.png}}\textsuperscript{\,18}},
\mbox{A.\;Morawiec\href{https://orcid.org/0009-0001-9845-4005}{\includegraphics[height=1.7ex]{figures/orcid-logo.png}}\textsuperscript{\,17}},
\mbox{S.\;Morozov\href{https://orcid.org/0000-0002-6748-7277}{\includegraphics[height=1.7ex]{figures/orcid-logo.png}}\textsuperscript{\,23}},
\mbox{Y.\;Nagai\href{https://orcid.org/0000-0002-1792-5005}{\includegraphics[height=1.7ex]{figures/orcid-logo.png}}\textsuperscript{\,9}},
\mbox{T.\;Nakadaira\href{https://orcid.org/0000-0003-4327-7598}{\includegraphics[height=1.7ex]{figures/orcid-logo.png}}\textsuperscript{\,10}},
\mbox{M.\;Naskr\k{e}t\href{https://orcid.org/0000-0002-5634-6639}{\includegraphics[height=1.7ex]{figures/orcid-logo.png}}\textsuperscript{\,21}},
\mbox{S.\;Nishimori\href{https://orcid.org/~0000-0002-1820-0938}{\includegraphics[height=1.7ex]{figures/orcid-logo.png}}\textsuperscript{\,10}},
\mbox{V.\;Ozvenchuk\href{https://orcid.org/0000-0002-7821-7109}{\includegraphics[height=1.7ex]{figures/orcid-logo.png}}\textsuperscript{\,15}},
\mbox{A.D.\;Panagiotou\textsuperscript{\,7}},
\mbox{O.\;Panova\href{https://orcid.org/0000-0001-5039-7788}{\includegraphics[height=1.7ex]{figures/orcid-logo.png}}\textsuperscript{\,14}},
\mbox{V.\;Paolone\href{https://orcid.org/0000-0003-2162-0957}{\includegraphics[height=1.7ex]{figures/orcid-logo.png}}\textsuperscript{\,29}},
\mbox{O.\;Petukhov\href{https://orcid.org/0000-0002-8872-8324}{\includegraphics[height=1.7ex]{figures/orcid-logo.png}}\textsuperscript{\,23}},
\mbox{I.\;Pidhurskyi\href{https://orcid.org/0000-0001-9916-9436}{\includegraphics[height=1.7ex]{figures/orcid-logo.png}}\textsuperscript{\,14,6}},
\mbox{R.\;P{\l}aneta\href{https://orcid.org/0000-0001-8007-8577}{\includegraphics[height=1.7ex]{figures/orcid-logo.png}}\textsuperscript{\,17}},
\mbox{P.\;Podlaski\href{https://orcid.org/0000-0002-0232-9841}{\includegraphics[height=1.7ex]{figures/orcid-logo.png}}\textsuperscript{\,20}},
\mbox{B.A.\;Popov\href{https://orcid.org/0000-0001-5416-9301}{\includegraphics[height=1.7ex]{figures/orcid-logo.png}}\textsuperscript{\,23,4}},
\mbox{B.\;P\'orfy\href{https://orcid.org/0000-0001-5724-9737}{\includegraphics[height=1.7ex]{figures/orcid-logo.png}}\textsuperscript{\,8,9}},
\mbox{M.\;Posiada{\l}a-Zezula\href{https://orcid.org/0000-0002-5154-5348}{\includegraphics[height=1.7ex]{figures/orcid-logo.png}}\textsuperscript{\,20}},
\mbox{D.S.\;Prokhorova\href{https://orcid.org/0000-0003-3726-9196}{\includegraphics[height=1.7ex]{figures/orcid-logo.png}}\textsuperscript{\,23}},
\mbox{D.\;Pszczel\href{https://orcid.org/0000-0002-4697-6688}{\includegraphics[height=1.7ex]{figures/orcid-logo.png}}\textsuperscript{\,16}},
\mbox{S.\;Pu{\l}awski\href{https://orcid.org/0000-0003-1982-2787}{\includegraphics[height=1.7ex]{figures/orcid-logo.png}}\textsuperscript{\,19}},
\mbox{J.\;Puzovi\'c\textsuperscript{\,24}\textsuperscript{\dag}},
\mbox{R.\;Renfordt\href{https://orcid.org/0000-0002-5633-104X}{\includegraphics[height=1.7ex]{figures/orcid-logo.png}}\textsuperscript{\,19}},
\mbox{L.\;Ren\href{https://orcid.org/0000-0003-1709-7673}{\includegraphics[height=1.7ex]{figures/orcid-logo.png}}\textsuperscript{\,27}},
\mbox{V.Z.\;Reyna~Ortiz\href{https://orcid.org/0000-0002-7026-8198}{\includegraphics[height=1.7ex]{figures/orcid-logo.png}}\textsuperscript{\,14}},
\mbox{D.\;R\"ohrich\textsuperscript{\,12}},
\mbox{E.\;Rondio\href{https://orcid.org/0000-0002-2607-4820}{\includegraphics[height=1.7ex]{figures/orcid-logo.png}}\textsuperscript{\,16}},
\mbox{M.\;Roth\href{https://orcid.org/0000-0003-1281-4477}{\includegraphics[height=1.7ex]{figures/orcid-logo.png}}\textsuperscript{\,5}},
\mbox{{\L}.\;Rozp{\l}ochowski\href{https://orcid.org/0000-0003-3680-6738}{\includegraphics[height=1.7ex]{figures/orcid-logo.png}}\textsuperscript{\,15}},
\mbox{B.T.\;Rumberger\href{https://orcid.org/0000-0002-4867-945X}{\includegraphics[height=1.7ex]{figures/orcid-logo.png}}\textsuperscript{\,27}},
\mbox{M.\;Rumyantsev\href{https://orcid.org/0000-0001-8233-2030}{\includegraphics[height=1.7ex]{figures/orcid-logo.png}}\textsuperscript{\,23}},
\mbox{A.\;Rustamov\href{https://orcid.org/0000-0001-8678-6400}{\includegraphics[height=1.7ex]{figures/orcid-logo.png}}\textsuperscript{\,1,6}},
\mbox{M.\;Rybczynski\href{https://orcid.org/0000-0002-3638-3766}{\includegraphics[height=1.7ex]{figures/orcid-logo.png}}\textsuperscript{\,14}},
\mbox{A.\;Rybicki\href{https://orcid.org/0000-0003-3076-0505}{\includegraphics[height=1.7ex]{figures/orcid-logo.png}}\textsuperscript{\,15}},
\mbox{K.\;Sakashita\href{https://orcid.org/0000-0003-2602-7837}{\includegraphics[height=1.7ex]{figures/orcid-logo.png}}\textsuperscript{\,10}},
\mbox{K.\;Schmidt\href{https://orcid.org/0000-0002-0903-5790}{\includegraphics[height=1.7ex]{figures/orcid-logo.png}}\textsuperscript{\,19}},
\mbox{A.Yu.\;Seryakov\href{https://orcid.org/0000-0002-5759-5485}{\includegraphics[height=1.7ex]{figures/orcid-logo.png}}\textsuperscript{\,23}},
\mbox{P.\;Seyboth\href{https://orcid.org/0000-0002-4821-6105}{\includegraphics[height=1.7ex]{figures/orcid-logo.png}}\textsuperscript{\,14}},
\mbox{U.A.\;Shah\href{https://orcid.org/0000-0002-9315-1304}{\includegraphics[height=1.7ex]{figures/orcid-logo.png}}\textsuperscript{\,14}},
\mbox{Y.\;Shiraishi\textsuperscript{\,11}},
\mbox{A.\;Shukla\href{https://orcid.org/0000-0003-3839-7229}{\includegraphics[height=1.7ex]{figures/orcid-logo.png}}\textsuperscript{\,28}},
\mbox{M.\;S{\l}odkowski\href{https://orcid.org/0000-0003-0463-2753}{\includegraphics[height=1.7ex]{figures/orcid-logo.png}}\textsuperscript{\,22}},
\mbox{P.\;Staszel\href{https://orcid.org/0000-0003-4002-1626}{\includegraphics[height=1.7ex]{figures/orcid-logo.png}}\textsuperscript{\,17}},
\mbox{G.\;Stefanek\href{https://orcid.org/0000-0001-6656-9177}{\includegraphics[height=1.7ex]{figures/orcid-logo.png}}\textsuperscript{\,14}},
\mbox{J.\;Stepaniak\href{https://orcid.org/0000-0003-2064-9870}{\includegraphics[height=1.7ex]{figures/orcid-logo.png}}\textsuperscript{\,16}},
\mbox{M.\;Strikhanov\textsuperscript{\,23}},
\mbox{H.\;Str\"obele\textsuperscript{\,6}},
\mbox{T.\;\v{S}u\v{s}a\href{https://orcid.org/0000-0001-7430-2552}{\includegraphics[height=1.7ex]{figures/orcid-logo.png}}\textsuperscript{\,3}},
\mbox{L.\;Swiderski\href{https://orcid.org/0000-0001-5857-2085}{\includegraphics[height=1.7ex]{figures/orcid-logo.png}}\textsuperscript{\,16}},
\mbox{J.\;Szewi\'nski\href{https://orcid.org/0000-0003-2981-9303}{\includegraphics[height=1.7ex]{figures/orcid-logo.png}}\textsuperscript{\,16}},
\mbox{R.\;Szukiewicz\href{https://orcid.org/0000-0002-1291-4040}{\includegraphics[height=1.7ex]{figures/orcid-logo.png}}\textsuperscript{\,21}},
\mbox{A.\;Taranenko\href{https://orcid.org/0000-0003-1737-4474}{\includegraphics[height=1.7ex]{figures/orcid-logo.png}}\textsuperscript{\,23}},
\mbox{A.\;Tefelska\href{https://orcid.org/0000-0002-6069-4273}{\includegraphics[height=1.7ex]{figures/orcid-logo.png}}\textsuperscript{\,22}},
\mbox{D.\;Tefelski\href{https://orcid.org/0000-0003-0802-2290}{\includegraphics[height=1.7ex]{figures/orcid-logo.png}}\textsuperscript{\,22}},
\mbox{V.\;Tereshchenko\textsuperscript{\,23}},
\mbox{A.\;Toia\href{https://orcid.org/0000-0001-9567-3360}{\includegraphics[height=1.7ex]{figures/orcid-logo.png}}\textsuperscript{\,6}},
\mbox{R.\;Tsenov\href{https://orcid.org/0000-0002-1330-8640}{\includegraphics[height=1.7ex]{figures/orcid-logo.png}}\textsuperscript{\,2}},
\mbox{L.\;Turko\href{https://orcid.org/0000-0002-5474-8650}{\includegraphics[height=1.7ex]{figures/orcid-logo.png}}\textsuperscript{\,21}},
\mbox{T.S.\;Tveter\href{https://orcid.org/0009-0003-7140-8644}{\includegraphics[height=1.7ex]{figures/orcid-logo.png}}\textsuperscript{\,13}},
\mbox{M.\;Unger\href{https://orcid.org/0000-0002-7651-0272~}{\includegraphics[height=1.7ex]{figures/orcid-logo.png}}\textsuperscript{\,5}},
\mbox{M.\;Urbaniak\href{https://orcid.org/0000-0002-9768-030X}{\includegraphics[height=1.7ex]{figures/orcid-logo.png}}\textsuperscript{\,19}},
\mbox{F.F.\;Valiev\href{https://orcid.org/0000-0001-5130-5603}{\includegraphics[height=1.7ex]{figures/orcid-logo.png}}\textsuperscript{\,23}},
\mbox{M.\;Vassiliou\textsuperscript{\,7}},
\mbox{D.\;Veberi\v{c}\href{https://orcid.org/0000-0003-2683-1526}{\includegraphics[height=1.7ex]{figures/orcid-logo.png}}\textsuperscript{\,5}},
\mbox{V.V.\;Vechernin\href{https://orcid.org/0000-0003-1458-8055}{\includegraphics[height=1.7ex]{figures/orcid-logo.png}}\textsuperscript{\,23}},
\mbox{V.\;Volkov\href{https://orcid.org/0000-0002-4785-7517}{\includegraphics[height=1.7ex]{figures/orcid-logo.png}}\textsuperscript{\,23}},
\mbox{A.\;Wickremasinghe\href{https://orcid.org/0000-0002-5325-0455}{\includegraphics[height=1.7ex]{figures/orcid-logo.png}}\textsuperscript{\,25}},
\mbox{K.\;W\'ojcik\href{https://orcid.org/0000-0002-8315-9281}{\includegraphics[height=1.7ex]{figures/orcid-logo.png}}\textsuperscript{\,19}},
\mbox{O.\;Wyszy\'nski\href{https://orcid.org/0000-0002-6652-0450}{\includegraphics[height=1.7ex]{figures/orcid-logo.png}}\textsuperscript{\,14}},
\mbox{A.\;Zaitsev\href{https://orcid.org/0000-0003-4711-9925}{\includegraphics[height=1.7ex]{figures/orcid-logo.png}}\textsuperscript{\,23}},
\mbox{E.D.\;Zimmerman\href{https://orcid.org/0000-0002-6394-6659}{\includegraphics[height=1.7ex]{figures/orcid-logo.png}}\textsuperscript{\,27}},
\mbox{A.\;Zviagina\href{https://orcid.org/0009-0007-5211-6493}{\includegraphics[height=1.7ex]{figures/orcid-logo.png}}\textsuperscript{\,23}}, and
\mbox{R.\;Zwaska\href{https://orcid.org/0000-0002-4889-5988}{\includegraphics[height=1.7ex]{figures/orcid-logo.png}}\textsuperscript{\,25}}
\\\rule{2cm}{.5pt}\\[-.5ex]\textit{\textsuperscript{\dag} \footnotesize deceased}

\end{sloppypar}

\noindent
\textsuperscript{1}~National Nuclear Research Center, Baku, Azerbaijan\\
\textsuperscript{2}~Faculty of Physics, University of Sofia, Sofia, Bulgaria\\
\textsuperscript{3}~Ru{\dj}er Bo\v{s}kovi\'c Institute, Zagreb, Croatia\\
\textsuperscript{4}~LPNHE, University of Paris VI and VII, Paris, France\\
\textsuperscript{5}~Karlsruhe Institute of Technology, Karlsruhe, Germany\\
\textsuperscript{6}~University of Frankfurt, Frankfurt, Germany\\
\textsuperscript{7}~University of Athens, Athens, Greece\\
\textsuperscript{8}~Wigner Research Centre for Physics, Budapest, Hungary\\
\textsuperscript{9}~E\"otv\"os Lor\'and University, Budapest, Hungary\\
\textsuperscript{10}~Institute for Particle and Nuclear Studies, Tsukuba, Japan\\
\textsuperscript{11}~Okayama University, Japan\\
\textsuperscript{12}~University of Bergen, Bergen, Norway\\
\textsuperscript{13}~University of Oslo, Oslo, Norway\\
\textsuperscript{14}~Jan Kochanowski University, Kielce, Poland\\
\textsuperscript{15}~Institute of Nuclear Physics, Polish Academy of Sciences, Cracow, Poland\\
\textsuperscript{16}~National Centre for Nuclear Research, Warsaw, Poland\\
\textsuperscript{17}~Jagiellonian University, Cracow, Poland\\
\textsuperscript{18}~AGH - University of Science and Technology, Cracow, Poland\\
\textsuperscript{19}~University of Silesia, Katowice, Poland\\
\textsuperscript{20}~University of Warsaw, Warsaw, Poland\\
\textsuperscript{21}~University of Wroc{\l}aw,  Wroc{\l}aw, Poland\\
\textsuperscript{22}~Warsaw University of Technology, Warsaw, Poland\\
\textsuperscript{23}~Affiliated with an institution covered by a cooperation agreement with CERN\\
\textsuperscript{24}~University of Belgrade, Belgrade, Serbia\\
\textsuperscript{25}~Fermilab, Batavia, USA\\
\textsuperscript{26}~University of Notre Dame, Notre Dame, USA\\
\textsuperscript{27}~University of Colorado, Boulder, USA\\
\textsuperscript{28}~University of Hawaii at Manoa, Honolulu, USA\\
\textsuperscript{29}~University of Pittsburgh, Pittsburgh, USA\\

\end{document}